\documentclass[a4paper,11pt]{article}
\usepackage{jheppub} % for details on the use of the package, please see the JINST-author-manual
\usepackage{lineno}
\usepackage{lmodern}
\usepackage{lipsum}
\usepackage{etoolbox}
\usepackage{tocloft} % For customizing Table of Contents
\usepackage{xpatch}
\usepackage{mathrsfs}
\usepackage{bm}
\usepackage[english]{babel}
\usepackage{amsfonts}
\usepackage{graphicx}
\usepackage[T1]{fontenc}
\usepackage[scr]{rsfso}
\usepackage{mathrsfs}
\usepackage{euscript}
\usepackage{mathalfa}

\usepackage{bbm} % to get covariant phase space style field space calculus
\usepackage{physics} % for usual physics stuff
\usepackage{comment} % to comment out stuff

\usepackage{amsmath}
\usepackage{amsthm}

\usepackage{multirow}
\usepackage[dvipsnames]{xcolor}
\usepackage[normalem]{ulem}

\newcommand\Lie{\cL}

\newcommand{\beq}{\begin{eqnarray}}
\newcommand{\eeq}{\end{eqnarray}}
\newcommand{\beqn}{\begin{eqnarray}}
\newcommand{\eeqn}{\end{eqnarray}}
\newcommand{\pa}{\partial}

\newcommand{\cL}{{\cal L}}

\usepackage{tikz-cd}
\usepackage{pict2e}
\makeatletter
\newcommand{\variable@rule}[1]{%
  \fontdimen8  
  \ifx#1\displaystyle\textfont3\else
    \ifx#1\textstyle\textfont3\else
      \ifx#1\scriptstyle\scriptfont3\else
        \scriptscriptfont3\relax
  \fi\fi\fi
}

\usepackage{mathtools}

\usepackage{tocloft}

\newcommand{\ve}{\varepsilon}

\newcommand{\nn}{\nonumber}

\newcommand{\cN}{\cal{N}}

\newcommand{\rd}{\text{d}}

\newcommand{\chkM}{{\color{red} \,\checkmark\kern-5pt{}_{M}}}

\def\be#1\ee{\begin{align}#1\end{align}} 
\newcommand{\bea}{\begin{eqnarray}}
\newcommand{\eea}{\end{eqnarray}}

\def\pa{\partial}

\def\k{k}

\def\s{\sigma}
\def\rd{\mathrm{d}}
\def\e{\mathrm{e}}
\def\lb{\label}
\def\N{\mathcal{N}}
\def\S{\mathcal{S}}
\def\form{\bm}
\def\ac{\varphi}
\def\vor{\varpi}
\def\sss{\scriptscriptstyle}

\newcommand{\sD}{\mathscr{D}}
\DeclareMathAlphabet{\esstix}{U}{esstixcal}{m}{n}
\DeclareMathAlphabet{\boondox}{U}{BOONDOX-cal}{m}{n}
\def\Carr{\esstix{C}}
\def\RCarr{\esstix{rC}}
\def\SCarr{\esstix{sC}}
\def\CGamma{\gamma}
\def\volN{{\nu}_{\sss \N}}

\def\volS{{\nu}_{\sss \mathrm{S}}}
\def\vol{{\nu}}
\def\M{\mathcal{M}}
\def\H{\mathcal{H}}

\def\btheta{\bar{\theta}}

\def\Neq{\stackrel{\sss \N}{=}}
\def\sR{\mathscr{R}}
\def\mr{\mathring}

\newcommand{\perimeter}[1]{
	\centerline{
		\begin{minipage}[c]{0.7\textwidth}
			\begin{center}
			$^a$ Perimeter Institute for Theoretical Physics,\\
			 31 Caroline St. N., Waterloo ON, Canada, N2L 2Y5
			\end{center}
		\end{minipage}
		}
	}
 \newcommand{\riken}[1]{
	\centerline{
		\begin{minipage}[c]{0.7\textwidth}
			\begin{center}
			$^b$ RIKEN iTHEMS, Wako, Saitama 351-0198, Japan
			\end{center}
		\end{minipage}
		}
	}
\newcommand{\basq}[1]{
	\centerline{
		\begin{minipage}[c]{0.8\textwidth}
			\begin{center}
		$^c$ Departamento de Fısica, Facultad de Ciencia y Tecnologıa, \\
  Universidad del Pais Vasco UPV/EHU, Apartado 644, 48080 Bilbao, Spain
			\end{center}
		\end{minipage}
		}
	}

\begin{document}

\begin{center}
    \vspace*{0.25cm}
\textbf{\LARGE{Foundations of Carrollian Geometry}}
\end{center}
\vspace{0.3cm}
\begin{center}
Luca Ciambelli$^a$ and Puttarak Jai-akson$^b$\\
\vspace{0.4cm}
\textit{\perimeter{}}\\
\textit{\riken{}}
\end{center}
\vspace{0.25cm}
\begin{abstract}

Carrollian physics provides the natural framework for describing null hypersurfaces. This review explores the geometry of Carrollian manifolds -- spaces endowed with a degenerate metric. We begin with an algebraic overview of the Carroll group, its conformal extension, and its relation to the BMS group. Then, in the core of the review, we follow the standard pseudo-Riemannian narrative: metric $\to$ connection $\to$ curvature. We first introduce the modern, general definition of a Carrollian structure, the analogue of the metric on such manifolds, reviewing the historical developments, symmetries, and link with the algebraic groups. The second part concerns connections. We show the breakdown of the Levi-Civita theorem in the Carrollian setting and construct the most general intrinsic Carrollian connection. The standard connection is then identified intrinsically and later shown to coincide with the rigged connection induced by embedding a null hypersurface in an ambient spacetime. The third part develops the associated curvature tensors. We include novel results presented here for the first time.

Two advanced topics highlight the broader scope of this framework. The first treats null hypersurfaces via the rigging technique, deriving the induced geometry from the ambient space. This provides a unified language for spacelike, timelike, and null hypersurfaces, and shows that the induced rigged connection exactly reproduces the intrinsic Carrollian one. From this, the Gauss and Codazzi-Mainardi equations follow, and the Einstein equations emerge as conservation laws for the null Brown-York stress tensor. The second topic extends the Carrollian setup to generic, non-null hypersurfaces, enabling a smooth null limit and completing the unified geometric description of hypersurfaces of all causal characters.
\end{abstract}
    \vspace{0.8cm}
\begin{center}{\small{For corrections, typos, and suggestions, please write to \\ \href{mailto:ciambelli.luca@gmail.com}{ciambelli.luca@gmail.com}, \ 
\href{mailto:puttarak.jaiakson@gmail.com}{puttarak.jaiakson@gmail.com}}}\end{center}
\thispagestyle{empty}

\newpage
\addtocontents{toc}{\protect\thispagestyle{empty}}
\tableofcontents
\thispagestyle{empty}
\newpage
\clearpage
\pagenumbering{arabic} 

\paragraph{Disclaimer}  
This review is intended for students already versed in differential geometry and general relativity, for researchers approaching null physics for the first time, and for experts in this and adjacent areas seeking a unified perspective. It is not conceived as an exhaustive survey of the state of the art, that is, a literature review on recent trends and applications. Instead, this review is a coherent collection of results organized and presented as teaching material. Our objective is to construct a unified and self-contained framework that renders the subject both accessible and conceptually transparent.

The need for such a synthesis has become increasingly pressing. Despite the growing relevance of null and Carrollian structures throughout high-energy physics, the essential tools and results remain scattered across the literature and are often formulated in technically intricate ways that obscure their geometric meaning. This recognition has shaped both the content and the pedagogical design of this review, which aims to offer a clear, systematic, and approachable entry point to the geometry of null manifolds and its Carrollian reformulation. We are grateful to the many colleagues whose contributions have shaped this rich and evolving field, and whose insights continue to inspire its development.\\

\section{Introduction}

The Lorentzian spacetime we inhabit is defined by the existence of a universal maximal speed of propagation, $c=299792458\,\mathrm{m/s}$, the speed of light. This property has profound implications for the geometric and mathematical structure of spacetime. Among the most significant is the emergence of null hypersurfaces, which delineate regions that are causally connected from those that are causally disconnected. These hypersurfaces form the natural stage for the physics of particles propagating at the speed of light. 

We start by describing null hypersurfaces through a simple example. Consider 4-dimensional Minkowski space, with metric in spherical coordinates
\beq
\rd s^2=-c^2\rd t^2+\rd r^2+r^2 \rd \Omega_2^2 \ ,
\eeq
where $\rd\Omega_2^2$ is the metric on the unit sphere. The induced metric at $r=r_0$ is
\beq\label{time}
\rd s^2=-c^2\rd t^2+r_0^2 \rd \Omega_2^2 \ ,
\eeq
which describes the physics of an observer living on a sphere of fixed radius. This is a timelike hypersurface, with induced metric of signature $(-,+,+)$.

Now let us move to null coordinates. Defining $u=c \ t-r$ and $v=c \ t+r$, the metric reads
\beq
\rd s^2=-2 \rd u \rd v+\frac{(v-u)^2}{4} \rd \Omega_2^2 \ .
\eeq
Now consider the induced metric at $u=u_0$, and set $u_0=0$ for simplicity, 
\beq\label{sphere}
\rd s^2=\frac{v^2}{4} \rd \Omega_2^2 \ .
\eeq
This describes a sphere expanding at the speed of light, since $u=0$ implies $r=c \ t$. The other null coordinate, $v$, is part of the intrinsic description, but the induced metric has no $\rd v^2$ term, signaling a degeneracy. This is a null hypersurface, with induced metric of signature $(0,+,+)$. 

From this simple example, we can already extract several key insights into the peculiar nature of null hypersurfaces. First, their intrinsic metric is degenerate, endowing them with an unconventional geometric structure: in particular, no unique inverse metric exists. How, then, can we meaningfully describe their intrinsic geometry?

Second, specifying a degenerate metric alone is insufficient to characterize a null hypersurface. For instance, how do we account for the null time coordinate $v$ introduced in the previous example? Consider the vector field $\pa_v$: its norm on the hypersurface satisfies $g_{vv}=0$. If one were to define the null hypersurface solely through the metric \eqref{sphere}, nothing would distinguish this degenerate metric from a perfectly regular two-dimensional Euclidean metric on a sphere. This observation indicates that the vector field generating the null direction must be supplied as an additional ingredient -- precisely the element that appears in the definition of a Carrollian structure, which provides the intrinsic geometric framework for null hypersurfaces.

\vspace{0.5cm}
\hrule
\vspace{0.5cm}

The purpose of this review is to present a fully intrinsic account of null manifolds -- manifolds equipped with a degenerate metric. As illustrated by the preceding example, their geometric properties are so peculiar that many of the standard tools developed for pseudo-Riemannian manifolds cease to apply, rendering null manifolds a distinct category of their own. The intrinsic study of such manifolds, whether or not they are embedded in a higher-dimensional ambient space, falls under the broad domain of Carrollian physics. Interestingly, Carrollian physics also possesses an independent and purely algebraic origin, stemming from a contraction of the Poincaré group. We shall return to this perspective later; for now, one may think of this contraction heuristically as taking the limit of vanishing light speed.\footnote{We emphasize that this is heuristic, as $c$ is a dimensionful parameter. The proper formulation of this limiting procedure will be discussed in the next section.} Consider the metric \eqref{time} and formally take the limit $c\to 0$:
\beq
\rd s^2=-c^2\rd t^2+r_0^2 \rd \Omega_2^2 \ \stackrel{c\to 0}\longrightarrow \ r_0^2 \rd \Omega_2^2 \ .
\eeq
We thus see that, starting from a metric of signature $(-,+,+)$, we reach a degenerate metric of signature $(0,+,+)$. The intrinsic geometric description of Carrollian degenerate spacetimes was first addressed in the seminal work of Henneaux \cite{Henneaux:1979vn}.\footnote{There are other important and physically relevant examples of degenerate spacetimes, such as Galilean spacetimes.} Prior to that, these manifolds have been the subject of intense study from an ambient spacetime perspective, prominently by Penrose \cite{Penrose:1962ij} et al, as we will carefully review in  section \ref{7}. At the algebraic level, the Inönü-Wigner contraction associated with $c\to 0$ of the Poincaré group led Lévy-Leblond and Sen Gupta to the discovery of the Carroll group \cite{LevyLeblond1965, SenGupta:1966qer}. Subsequent developments extended this structure to its conformal counterparts, linking them to the conformal isometries of Carrollian manifolds. A remarkable outcome of this line of research was the realization in \cite{Duval:2014uva} that, when the spatial manifold is a sphere, a particular conformal Carroll group is isomorphic to the Bondi-van der Burg-Metzner-Sachs (BMS) group \cite{Bondi, Sachs:1961zz}, the asymptotic symmetry group of four-dimensional asymptotically flat spacetimes. This correspondence is geometrically natural, since null infinity is itself a null hypersurface. This established a profound bridge between Carrollian geometry and asymptotic symmetry analysis, thereby laying the foundations for the modern study of flat-space holography through the framework of Carrollian physics. There are by now many fascinating applications of Carrollian physics, in a plethora of topics. We provide a survey of recent applications and various approaches to Carrollian physics in section \ref{7}.

\vspace{0.5cm}
\hrule
\vspace{0.5cm}

The technical challenges and mathematical subtleties inherent to Carrollian manifolds are what motivated us to write this review. Our goal is to normalize null geometry, placing it on the same conceptual footing as Riemannian and pseudo-Riemannian geometries. To this end, we adopt the pedagogical approach typically used to introduce pseudo-Riemannian geometry to physicists, structured around three main guiding principles:

\begin{itemize}
    \item \textbf{Metric $\to$ Carrollian structure.} A Carrollian structure is the set of geometric data defining the fundamental properties of null manifolds. As briefly illustrated in the example above, it consists of a degenerate metric together with a vector field lying in its kernel. The latter represents the arrow of null time and must be provided as additional structure whenever the metric is degenerate.
    
    \item \textbf{Levi-Civita connection $\to$ Carrollian connections.} The Levi-Civita theorem determines the Christoffel symbols uniquely from a non-degenerate metric in pseudo-Riemannian geometry. This theorem, however, fails for Carrollian structures, where no unique preferred connection exists. After introducing the most general intrinsic Carrollian connection, we will select a specific torsionless yet non-metric-compatible one and analyze its properties. This canonical connection will later be shown to naturally arise from the embedding of a null hypersurface. The explicit derivation of the connection symbols for this  Carrollian connection is a novel result derived here.
    
    \item \textbf{Curvature $\to$ Carrollian curvature.} The lack of metric compatibility in the Carrollian connection gives rise to a more intricate notion of curvature, which we will examine in detail. Since various connections and corresponding curvature tensors appear in the literature, we will clarify their interrelations and demonstrate that the Carrollian connection adopted here subsumes all previous ones as special cases.
\end{itemize}

As stated, the central goal of this review is to revisit these three points with the aim of making Carrollian geometry accessible. We adhere to two guiding principles. First, every computation presented here is carefully derived step by step to ensure clarity and reproducibility. Second, except for a few explicitly indicated sections, the entire review is formulated in a fully covariant language, employing both abstract index notation and adapted frames. Consequently, all results discussed herein are independent of any specific choice of coordinates or parametrization, and readily suitable to describe any Carrollian manifold and thus any null hypersurface. 

Describing null manifolds in full generality may seem like a daunting and overly technical endeavor at first. Yet this effort is indispensable: the intrinsic Carrollian formalism reveals its true power precisely when one studies hypersurfaces embedded in an ambient spacetime. The second part of this review is therefore devoted to two more advanced developments aimed at embedding our intrinsic constructions. This serves a dual purpose -- first, to demonstrate the practical utility of the formalism, and second, to justify some of the intrinsic constructions from an ambient, embedding perspective.

The first advanced topic revisits the rigging procedure originally introduced in \cite{Mars:1993mj}, designed to describe null hypersurfaces embedded in a pseudo-Riemannian manifold. The rigged projector enables one to induce a Carrollian structure from the bulk metric and to construct the corresponding rigged connection. On the null hypersurface, this connection is torsionless but not metric-compatible. We show explicitly that it reproduces the Carrollian connection derived earlier from purely intrinsic considerations. We then evaluate the Gauss and Codazzi-Mainardi equations for the null hypersurface. To the best of our knowledge, only the latter has appeared previously in the literature, while the former represents a novel result. We conclude by showing how the Codazzi-Mainardi equation naturally leads to the construction of the null Brown-York stress tensor \cite{Chandrasekaran:2021hxc}, whose conservation law is equivalent to the projection of Einstein’s equations onto the null hypersurface.

The second advanced topic discusses how to endow a generic hypersurface -- of any causal character -- with a Carrollian structure. Although this construction is not required for non-null hypersurfaces, extending it there provides a unified geometric framework that smoothly interpolates between non-null and null cases. In this sense, the intrinsic Carrollian language emerges as a universal description of all hypersurfaces, allowing for continuous transitions in causal character. This proves particularly useful for analyzing foliations that asymptote to null hypersurfaces and, ultimately, for treating hypersurfaces that contain intrinsically distinct causal regions. 

\vspace{0.5cm}
\hrule
\vspace{0.5cm}

Given the pedagogical nature of this review, we have gathered and reorganized results from previous works into a coherent and unified presentation. We provide below a list of key  papers from which several results have been drawn. 
\begin{itemize}
    \item The algebraic account of Carroll algebras in section~\ref{2} is inspired by the original papers by Lévy-Leblond and Sen Gupta \cite{LevyLeblond1965, SenGupta:1966qer}, while we also utilized the recent paper \cite{Afshar:2024llh} for our account on  conformal Carroll algebras.
    \item The subsection~\ref{zerosig} on zero-signature spacetimes is directly adapted from the original paper \cite{Henneaux:1979vn}.
    \item Carrollian manifolds, their conformal isometries, and the link with the BMS group in subsection~\ref{carrman} follow closely the derivation in \cite{Duval:2014uoa, Duval:2014uva}.
    \item Section~\ref{sec:modern} is mainly influenced by the fibre-bundle construction presented in \cite{Ciambelli:2019lap}, and the subsequent analyses in \cite{Freidel:2022bai, Ciambelli:2023mir}, except for subsection~\ref{334} on Carrollian diffeomorphisms, which were introduced in \cite{Ciambelli:2018wre}.
    \item Section~\ref{sec:Carr-connection} on Carrollian connections builds upon ideas from the literature, starting from \cite{Bekaert:2015xua}, and continuing with \cite{Ciambelli:2023mir}, appendix A in \cite{Ciambelli:2023xqk}, and the intrinsic Carrollian version of results discussed in \cite{Chandrasekaran:2021hxc}. The relationship to metric hypersurface connections follows \cite{Manzano:2023wxx}. 
    \item The Levi-Civita-Carroll  derivative discussed in subsection \ref{423} and the Riemann-Carroll tensor used in \ref{curva} first appeared in \cite{Ciambelli:2018wre}.
    \item The discussion of Carrollian connections at null infinity presented in subsection~\ref{422} is influenced by the original treatment in \cite{Ashtekar:1981hw} and the more recent work \cite{Ashtekar:2024bpi}.
    \item The null Brown-York stress tensor has been discussed in \cite{Chandrasekaran:2021hxc}. It features in the Codazzi-Mainardi equation for null hypersurfaces, analyzed in \cite{Gourgoulhon:2005ng}.
    \item Section~\ref{rigg} presents the rigging construction from the original paper \cite{Mars:1993mj}, and is inspired by the more recent works \cite{Chandrasekaran:2021hxc} and \cite{Freidel:2022vjq}.
    \item The Carrollian structure on stretched horizons discussed in section~\ref{sec:sCarr} was derived and examined in \cite{Freidel:2022vjq}, with earlier related observations appearing in  \cite{Donnay:2019jiz}.
\end{itemize}

The present review reflects the authors’ perspective on Carrollian geometry and its role in null physics. A historical and conceptual excursus, placed near the end of the manuscript (section \ref{7}), situates this perspective within the broader development of null geometry and Carrollian structures, and discusses complementary approaches in the literature.

\bigskip

Unless explicitly stated, the conventions in the manuscript are as follows. We use $a,b,\dots$ as abstract indices on the null manifold with $a=(0,i)$, $i$ being the spatial index. Similarly, we use coordinates $x^a=(x^0,x^i)$. The algebraic generators are the time translation $P_0$, space translation $P_i$, Carroll boost $B_i$, spatial special conformal transformation $K_i$, temporal special conformal transformation $K$, dilatation $D$, rotations $J_{ij}$, generic supertranslation $M_f$. The Carrollian geometric data in abstract index notation are the null, aka Carrollian, manifold $\N$, degenerate metric $q_{ab}$, Carrollian vector $\ell^a$, Ehresmann connection, aka ruling $k_a$, expansion tensor, aka second fundamental tensor $\theta_{ab}$, acceleration $\varphi_a$, and vorticity $\varpi_{ab}$. The Lie derivative along a vector $\xi$ is denoted by $\cL_\xi$. The manifold $\N$ has dimension $d+1$, such that there are $d$ spatial directions. In later sections, we will introduce an ambient space, which is assumed to be Lorentzian, with dimension $d+2$.

\newpage 

\section{Carroll Algebra and Conformal Extensions}\label{2}

Carrollian physics has group theoretic origin. The Carroll group was firstly described by Lévy-Leblond and Sen Gupta \cite{LevyLeblond1965, SenGupta:1966qer}, arising as the Inönü-Wigner contraction of the Poincaré group. Although this is a review on Carroll geometry, for historical rigor and completeness, we review here the salient features of the Carroll group and its conformal extensions. \\

\subsection{Carroll Algebra}

To contextualize the Carrollian contraction of the Poincaré group, we begin with a brief recap of the Galilean contraction. Calling $J_{ij}, B_i, P_i, P_0$ the generators of rotations, boosts, spatial translations, and time translations, respectively,  the $d+1$-dimensional Galilean algebra $\mathfrak{gal}(d+1)$ can be obtained by suitably rescaling the Poincaré generators (see the table below for the $4d$ Poincaré algebra)\footnote{Contrary to \cite{LevyLeblond1965}, our generators are anti-Hermitian. Moreover, we denote the boost $B_i$ instead of $K_i$.}
\beq
B_i=\ve B_i\qquad P_i=\ve P_i\,,
\eeq
and by performing the $\ve\to 0$ limit. Conversely, Lévy-Leblond and Sen Gupta \cite{LevyLeblond1965, SenGupta:1966qer} noticed  that, rescaling the generators as
\beq
B_i=\eta B_i\qquad P_0=\eta P_0\,,
\eeq
and sending $\eta\to 0$, leads to a different -- somehow opposite -- contraction, giving rise to the Carroll algebra $\mathfrak{carr}(d+1)$. This is summarized in  table \ref{fig1}, taken (with conventions adapted) directly from the 1965 paper of Lévy-Leblond \cite{LevyLeblond1965}.

\begin{table}[h!]
\renewcommand{\arraystretch}{1.3}
\centering
\begin{tabular}{| c | c | c | }
\hline 
\textbf{Carroll} $\mathfrak{carr}(3)$ & \textbf{Poincaré} $\mathfrak{iso}(1,3)$ & \textbf{Galilei} $\mathfrak{gal}(3)$   \\ \hline 
$
\begin{aligned}
[J_i,J_j] &= \ve_{ijk}J_k \\
[J_i,B_j] &= \ve_{ijk}B_k\\
[B_i,B_j] &= 0\\
[J_i,P_j] &= \ve_{ijk}P_k \\
[B_i,P_j] &= \delta_{ij} P_0 \\
[J_i,P_0] &= 0\\
[B_i,P_0] &= 0\\
[P_i,P_j] &= 0 \\
[P_i,P_0] &= 0
\end{aligned}
$
& 
$
\begin{aligned}
[J_i,J_j] &= \ve_{ijk}J_k \\
[J_i,B_j] &= \ve_{ijk}B_k\\
[B_i,B_j] &= -\ve_{ijk}J_k\\
[J_i,P_j] &=  \ve_{ijk}P_k \\
[B_i,P_j] &=  \delta_{ij} P_0 \\
[J_i,P_0] &= 0\\
[B_i,P_0] &= P_i\\
[P_i,P_j] &= 0 \\
[P_i,P_0] &= 0
\end{aligned}
$
& 
$
\begin{aligned}
[J_i,J_j] &=  \ve_{ijk}J_k \\
[J_i,B_j] &= \ve_{ijk}B_k\\
[B_i,B_j] &= 0\\
[J_i,P_j] &=  \ve_{ijk}P_k \\
[B_i,P_j] &= 0 \\
[J_i,P_0] &= 0\\
[B_i,P_0] &= P_i\\
[P_i,P_j] &= 0 \\
[P_i,P_0] &= 0
\end{aligned}
$
\\ \hline
\end{tabular}
\caption{Table of comparison of the Carroll, Poincaré, and Galilei algebras in \cite{LevyLeblond1965}, in the specific case of $3+1$ spacetime dimensions, in which one defines $J_i=\frac12 \ve_{ijk}J_{jk}$. }
\label{fig1}
\end{table}
%%%%%%%%

This Carrollian contraction can be understood at the group level as arising from the vanishing speed-of-light limit. Indeed, consider a Poincaré transformation, composed of a rotation $R$, a Lorentz transformation with velocity $\beta^i$, and a translation $(a_0,a^i)$:
\beq
\begin{cases}
x'^0=\gamma(x^0+\beta^i R_{ij}x^j)+a^0
  \\
x'^i=R^i{}_jx^j+\frac{\gamma^2}{\gamma+1}(\beta^jR_{jk} x^k)\beta^i+\gamma \beta^i x^0+a^i
\end{cases}\,,
\eeq
where $\gamma=\frac{1}{\sqrt{1-\beta^2}}$.

From this, the Galilean limit is performed setting
\beq\label{c}
t=\frac{x^0}{c}\qquad v^i=c\beta^i\qquad b=\frac{a^0}{c}\,,
\eeq
and sending $c\to \infty$, leading to
\beq
\begin{cases}
t'=t+b
  \\
x'^i=R^i{}_jx^j+v^it+a^i
\end{cases}\,.
\eeq
This is a Galilean transformation: time is absolute, whereas space undergoes the Galilean relativity, that is, the fundamental laws of physics are the same in all inertial reference frames moving at constant velocity relative to each other. In this limit, the speed of light has been formally sent to infinity, and thus, as expected, information propagates instantaneously between events. This is formal, as the speed of light is a dimensionful parameter. More rigorously, the limit is performed setting $\Delta x/\Delta t \ll c$, such that the light cone opens up into the whole plane. 

Conversely, the Carrollian limit is performed setting
\beq\label{C}
t=C x^0\qquad v^i=C\beta^i\qquad b=C a^0\,,
\eeq
and sending $C\to \infty$,\footnote{Comparing the first equation in \eqref{C} with the first one in \eqref{c}, we see that $C \leftrightarrow 1/c $, and thus $C\to \infty \leftrightarrow c\to 0$. Note that, however, the velocity is rescaled differently, in order to be compatible with $\Delta x/\Delta t \gg c$.} leading to
\beq\label{carr}
\begin{cases}
t'=t+v^iR_{ij}x^j+b
  \\
x'^i=R^i{}_jx^j+a^i
\end{cases}\,.
\eeq
This is a Carrollian transformation: space is absolute, whereas time is not, a situation specular yet opposite to the Galilean one. In this limit, the speed of light has been formally sent to zero or, more properly, this is the limit $\Delta x/\Delta t \gg c$. This means that the light cone shrinks to a line: information does not propagate between neighboring events. Such a feature is oftentimes referred to as ultra-locality.

The fate of the light cone in these contractions is depicted in the figure below,
\begin{center}
    \includegraphics[scale=0.4]{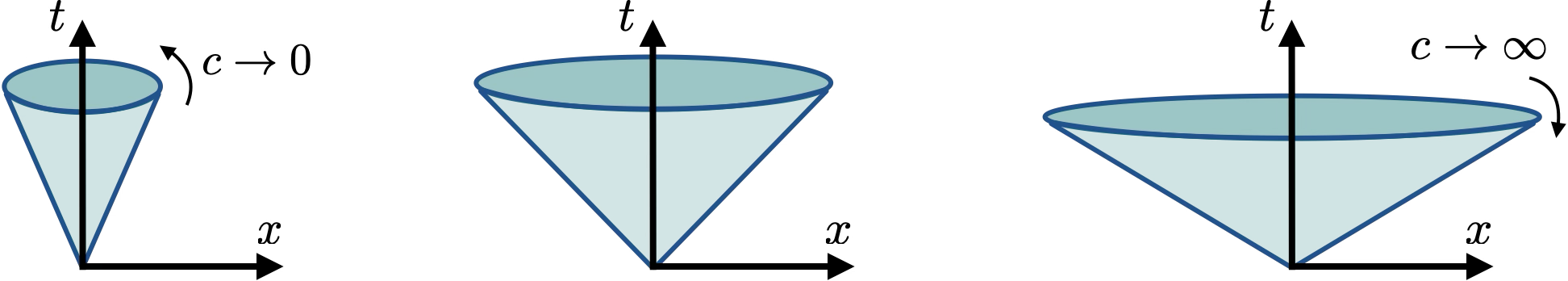}
\end{center}
In particular, the image in the center describes the Poincaré light cone, the right image its Galilean limit, while the left image the Carrollian one.\\

\subsection{Conformal Carroll Algebra}\label{sec:CC-algebra}

Just as the conformal algebra provides the conformal enhancement of the Poincaré algebra, it is natural to inquire whether the algebras discussed above admit similar conformal enhancements. Specifically, one may ask whether a conformal Carroll algebra exists, and how it is constructed.

To describe the conformal Carroll algebra, we first write the generators of the Carroll algebra $\mathfrak{carr}(d+1)$ inducing the transformation \eqref{carr} as\footnote{The notation here coincides with \cite{Afshar:2024llh}, except for the Hamiltonian, here denoted $P_0$ instead of $H$.}
\beq\label{gencarr}
P_0=\partial_0\qquad P_i=\pa_i\qquad B_i=x_i\pa_0 \qquad J_{ij}=x_i\pa_j-x_j\pa_i\,,
\eeq
such that the algebra $\mathfrak{carr}(d+1)$ reads\footnote{Conventions throughout the whole review: $A_{[ij]}=\frac12 (A_{ij}-A_{ji})$ and $A_{(ij)}=\frac12 (A_{ij}+A_{ji})$.}
\beq
[P_i,B_j]=\delta_{ij}P_0\qquad [P_i,J_{jk}]=2\delta_{i[j}P_{k]}\qquad [B_i,J_{jk}]=2\delta_{i[j}B_{k]}\qquad [J_{ij},J_{kl}]=4\delta_{[i[k}J_{l]j]}\,.
\eeq

While there are many possible conformal extensions of this  algebra, we confine our attention to the so-called level $2/z$ conformal Carroll algebra $\mathfrak{ccarr}_{2/z}(d+1)$. This algebra is readily realized as the algebra of conformal isometries of a Carrollian manifold, as we will reveal in section \ref{carrman}. Here, we describe it entirely algebraically. One notes that we can introduce the dilatation, temporal, and spatial special conformal  operators
\beq\label{Dila}
D=z x^0\pa_0+x^i\pa_i\qquad K=x^2\pa_0 \qquad K_i=x^2 \pa_i-2x_i(zx^0\pa_0+x^j\pa_j)\, ,
\eeq
in which we introduced the dynamical exponent $z$, since space and time dilate differently a priori. This finite-dimensional system is actually an algebra only when $z=1$.\footnote{In this case, one also has that this algebra generates the Poincaré algebra in one higher dimension.} Indeed, in
\beq
&[P_i,B_j]=\delta_{ij}P_0\qquad [P_i,J_{jk}]=2\delta_{i[j}P_{k]}\qquad [B_i,J_{jk}]=2\delta_{i[j}B_{k]}\qquad [J_{ij},J_{kl}]=4\delta_{[i[k}J_{l]j]}&\nonumber\\
&[P_i,K]=2B_i\qquad [P_0,D]=z P_0\qquad [P_i,D]=P_i\qquad [B_i,D]=(z-1)B_i&\nonumber\\
&[D,K]=(2-z)K\qquad [P_0,K_i]=-2z B_i\qquad [P_i,K_j]=2J_{ij}-2\delta_{ij}D\qquad [D,K_i]=K_i&\nonumber\\
&[B_i,K_j]=-\delta_{ij}K+(1-z)x_ix_j\pa_0\qquad [K_i,J_{jk}]=2\delta_{i[j}K_{k]}\qquad [K_i,K]=2x_i(z-1)x^2\pa_0\,,\nonumber&
\eeq
the commutators $[B_i,K_j]$ and $[K_i,K]$ do not lead to a closed algebra, unless $z=1$. 

However, one remarks that the non-closure is given by higher $x$-polynomials in $\partial_0$. This suggests that the infinite-dimensional enhancement in which the generators $P_0$, $B_i$, and $K$ are enhanced to the supertranslation generators
\beq
M_f=f(x^i)\pa_0\,,
\eeq
do form an algebra. 

This is indeed the case, and one obtains in this way the infinite-dimensional enhancement of the level $2/z$ conformal Carroll algebra $\mathfrak{ccarr}_{2/z}(d+1)$ generated by
\beq
&M_f=f(x^i)\pa_0 \qquad D=z x^0\pa_0+x^i\pa_i\qquad K_i=x^2 \pa_i-2x_i(zx^0\pa_0+x^j\pa_j)&\nonumber\\
&P_i=\pa_i \qquad J_{ij}=x_i\pa_j-x_j\pa_i\,,&\label{ccarr}
\eeq
and given by
\beq
&[P_i,J_{jk}]=2\delta_{i[j}P_{k]}\qquad [J_{ij},J_{kl}]=4\delta_{[i[k}J_{l]j]}\qquad [P_i,D]=P_i\qquad [D,K_i]=K_i&\nonumber\\
& [P_i,K_j]=2J_{ij}-2\delta_{ij}D\qquad  [K_i,J_{jk}]=2\delta_{i[j}K_{k]}\qquad [J_{ij},M_f]=M_{J(f)}&\\
& [P_i,M_f]=M_{P(f)}\qquad [D,M_f]=M_{D(f)}\qquad [K_i,M_f]=M_{K(f)}\,,\nonumber&
\eeq
with 
\beq
&J(f)=(x_i\pa_j-x_j\pa_i)f\qquad P(f)=\pa_i f& \nonumber \\
&D(f)=(x^i\pa_i -z)f \qquad K(f)=(x^2\pa_i -2 x_i( x^j\pa_j-z))f&\,.
\eeq
This is the conformal extension of the Carroll algebra that we will consider in the following. One of the major breakthroughs in Carrollian physics was the realization by Duval, Gibbons, and Horvathy \cite{Duval:2014uva}, that this infinite-dimensional algebra, for $z=1$ and thus level $2$, is isomorphic to the BMS algebra derived as the asymptotic symmetry algebra of four-dimensional asymptotically flat spacetimes. Note that the level-2 conformal Carroll algebra corresponds to the case where space and time dilate equally, a situation often referred to as the "relativistic scaling". This can be seen by inserting $z=1$ in \eqref{Dila}.

This concludes our brief overview of the salient algebraic features of the Carroll group and its conformal extension. It retraces the historical derivation of the Carroll group, which constitutes the first account of Carrollian physics. We can now move to the geometric description of Carrollian manifolds -- the core of the present review. After introducing a proper geometric description of a Carrollian manifold, we will recover the Carroll group and its conformal extension from the isometries and conformal isometries of a background structure, respectively.

\newpage

\section{Carrollian Structures}

This section marks the beginning of our geometric exploration of Carrollian physics. We introduce the notion of a Carrollian structure -- the analogue of a pseudo-Riemannian geometry in the Carrollian setting. By Carrollian structure we refer to the geometric data that naturally arise on a null manifold. Throughout this review, we will use the terms null manifold, zero-signature spacetime, and Carrollian manifold interchangeably. The coexistence of these names reflects the historical evolution of the field, which we  review in section \ref{7}, after we introduce the intrinsic Carrollian data and show how they can be induced from an ambient space. What matters is that the modern terminology, i.e., Carrollian structure, has come to unify and generalize all these earlier formulations.

We begin with the earliest purely intrinsic formulation, namely the zero-signature spacetimes described by Henneaux \cite{Henneaux:1979vn}.\footnote{It is important to recall the works of Penrose \cite{Penrose:1962ij}, Geroch \cite{Geroch:1977big}, and Ashtekar-Hansen \cite{Ashtekar:1978zz}, which dates prior to Henneaux's work, where many aspects of Carrollian geometry are discussed at length in the context of asymptotic infinity. We will discuss these works and further earlier developments in section \ref{7}.} We then turn to the key result of Duval, Gibbons, and Horvathy \cite{Duval:2014uva}, who discovered that the conformal isometries of a Carrollian manifold -- which we will define in detail below -- are precisely described by the conformal Carroll algebra introduced in section \ref{sec:CC-algebra}. As shown by Duval, Gibbons, and Horvathy \cite{Duval:2014uva}, this algebra is isomorphic to the BMS group, thereby establishing a deep link between two major developments in the field.

We next review the modern formulation of Carrollian structures, emphasizing their underlying geometric framework. After presenting the various internal symmetries, we introduce the essential notions of Carrollian acceleration, vorticity, and expansion. The remainder of the section is devoted to the discussion of adapted coordinates and to the transformation properties of these quantities under diffeomorphisms.\\

\subsection{Zero-signature Spacetimes}\label{zerosig}

The first description of a Carrollian structure can be traced back to the zero-signature spacetime defined by Henneaux \cite{Henneaux:1979vn}. Remarkably, this precursory paper contains already many of the elements needed to define a general Carrollian structure. 

A zero-signature spacetime of dimension $(d+1)$ consists of a degenerate metric $q_{ab}$ and a positive density $\Omega$ of weight +1. As a matrix, $q_{ab}$ is assumed to have rank $d$, inferring that its kernel, $\mathrm{ker}(q)$, spans a 1-dimensional null subspace. We also assume that the norm of any vector $V^a$ is always non-negative, meaning that $q_{ab} V^a V^b \geq 0$. The determinant of the metric is given by 
\begin{align}
\det q = \frac{1}{d+1}C^{ab} q_{ab}\,, \qquad \text{or equivalently} \qquad (\det q) \delta_a^b = C^{bc} q_{ca} \,,
\end{align}
where $C^{ab}$, called the minor of $q_{ab}$, is a symmetric contravariant tensor density of weight\footnote{A tensor density $T_a{}^b$ has weight $w$ if it transforms under general coordinate transformations $x \to x'(x)$ as $T'_{a}{}^{b} = (\det J)^{-w}(J^{-1})_a{}^{c}T_c{}^d J_d{}^b $ where $J_a{}^b = \frac{\pa x'^b}{\pa x^a}$ is the Jacobian matrix of the transformation.} +2 defined as 
\begin{align}
C^{ab} = \frac{1}{d!} \epsilon^{aa_1a_2 \dots a_d} \epsilon^{bb_1b_2\dots b_d} q_{a_1b_1}q_{a_2b_2} \dots q_{a_db_d} \, ,
\end{align}
where $\epsilon^{a a_1 a_2 \dots a_d}$ is the Levi-Civita, totally antisymmetric, symbol. 

By construction, $\det q = 0$, which implies $C^{ac} q_{cb} = 0$. This condition suggests that each column of the cofactor $C^{ab}$, when viewed as a matrix, lies in the null subspace $\mathrm{ker}(q)$ of the metric. Denoting the vector spanning $\mathrm{ker}(q)$ by $\ell^a$, the cofactor can be written as
\begin{align}
C^{ab} = \Omega^2 \ell^a \ell^b\,,
\end{align}
together with the condition that $\ell^a \in \mathrm{ker}(q)$,  
\begin{align}
q_{ab} \ell^b = 0\,. 
\end{align}
Therefore, the weight +1 density $\Omega$ plays two roles here: it corrects the weight of the tensor density $C^{ab}$ and fixes the scale of the vector $\ell^a$. Hence, the pair $(q_{ab}, \Omega)$, or equivalently $(q_{ab}, \ell^a)$, defines the geometry of the zero-signature spacetime. Later, in section \ref{sec:modern}, $(q_{ab}, \ell^a)$ will indeed be the starting point to define a general Carrollian structure.

Since the metric is degenerate, there is no non-degenerate inverse metric $q^{ab}$ such that $q^{ac}q_{cb} = \delta^a_b$. Instead, the contravariant symmetric tensor $q^{ab}$ is defined by\footnote{Comparison with the notation employed in \cite{Henneaux:1979vn}: $\ell^a \rightarrow n^\alpha$, $k_a\rightarrow \theta_\alpha$, $q_{ab}\rightarrow g_{\alpha\beta}$, $q^{ab}\rightarrow G^{\alpha\beta}$. We also present it in general dimension, whereas the analysis in \cite{Henneaux:1979vn} is performed in $4$ dimensions.}
\begin{align}
q^{ac} q_{cb} = \delta^a_b - k_b \ell^a \,,
\end{align}
for an auxiliary covector $k_a$ satisfying $\ell^a k_a =1$. It immediately follows that 
\begin{align}
q^{ab} q_{ac} q_{bd} = q_{cd} \,, \qquad \text{and} \qquad q^{ab}q_{ab} = d\,.
\end{align}
The above equation defines a family of tensors $q^{ab}$, whose members are related by the transformation
\begin{align}
q^{ab} \to q^{ab} - 2\lambda^{(a} \ell^{b)}\,,
\end{align}
for an arbitrary vector $\lambda^a$. This transformation corresponds to the shift of the covector $k_a \to k_a + q_{ab} \lambda^b$. The structure $(q_{ab}, \ell^a, k_a)$ endows the spacetime with a ruled Carrollian structure, and the above transformation represents one of its internal symmetries. We will discuss this in detail in section \ref{sec:modern}.\\

It is useful to introduce adapted coordinates $x^a = (x^0, x^i)$ in which the components of the vector $\ell^a$ are $(1, 0, 0, \dots)$. In these adapted coordinates, the components of the degenerate metric and the minor are expressed as
\beq
q_{ab}=\begin{pmatrix}
    0 & 0\\
    0 & q_{ij}
\end{pmatrix} \,, \qquad \text{and} \qquad
C^{ab}=\begin{pmatrix}
    q & 0\\
    0 & 0
\end{pmatrix} \,,
\eeq
where we denote the determinant of the spatial metric $q_{ij}$ with $q := \det q_{ij}$. In these coordinates, the scale $\Omega$ is given by $\Omega = \sqrt{q}$, and $q^{ij}$ is given by the inverse of the spatial metric.  

We note that the adapted coordinates are defined up to the coordinate transformations $x^a \to x'^a (x)$,
\begin{align}
x'^0 = x^0 + f(x^i) \,, \qquad \text{and} \qquad x'^i = f^i(x^j)\,,
\end{align}
for arbitrary functions $f(x^i)$ and $f^i(x^j)$ of the spatial coordinates. One easily recognizes that $f(x^i)$ generates supertranslations, that is, space-dependent time transformations.\footnote{Here and throughout, we use the term "supertranslation" in the standard broad sense of space-dependent time translations. In section \ref{sec:CC-algebra}, these transformations are introduced within the conformal framework: they are not arbitrary functions, but carry a definite conformal weight (equal to 1) and transform accordingly under the conformal group.} Similarly, the functions $f^i(x^j)$ parametrize diffeomorphisms of the spatial subspace. To the best of our knowledge, this is the first instance in which the infinite-dimensional enhancement of the Carroll group appears  in the literature from geometric considerations. 

The special case where the zero-signature spacetime is flat occurs when there exists a coordinate system such that $q_{0a} =0$, $q_{ij} = \delta_{ij}$, and $\Omega =1$. As we will carefully review below, the isometries of this flat spacetime still leads to an infinite-dimensional group. Indeed, Duval, Gibbons and Horvathy \cite{Duval:2014uva} realized that this group becomes exactly the Carroll group given in \eqref{carr} (for the case $d=3$) only upon further requiring that the connection is preserved. 

The second fundamental tensor is defined as\footnote{This is defined with opposite sign with respect to $K_{\alpha\beta}$ in \cite{Henneaux:1979vn}.}
\beq
\theta_{ab}=\frac12 \cL_{\ell}q_{ab}\,.\label{Kmh}
\eeq
Spaces for which $\theta_{ab}=0$ are called static or completely reducible since, if working in adapted coordinates, one has $\pa_0q_{ij}=0$ and thus the metric is completely time independent. The tensor $\theta_{ab}$ will be called the expansion tensor in \eqref{theta}. Already in this original paper (and reference [10] therein), it is understood that this tensor is what controls the failure of the Levi-Civita theorem on a null manifold. Therefore, completely reducible spaces admit a unique connection. If the space under consideration is not completely reducible, there still persists a limited notion of intrinsic parallel transport
\beq
\nabla_\ell q_{ab}=0=\nabla_\ell \ell^a\,.\label{neweq}
\eeq
We will extensively talk about the connection on a null manifold in section \ref{sec:connection}, in which we will derive these equations in full generality.\\

\subsection{Carrollian Manifolds}\label{carrman}

Carrollian manifolds are defined by Duval, Gibbons, and Horvathy \cite{Duval:2014uva}  as the quadruple $({\cN},q,\ell,\nabla)$, in which $\cN$ is the smooth manifold, endowed with a symmetric covariant tensor field $q$, with kernel generated by the nowhere vanishing vector field $\ell$. This has been called a strong Carrollian structure, since in its definition enters a choice of connection $\nabla$, that parallel-transports $q$ and $\ell$, that is, $\nabla q=0=\nabla \ell$. Conversely, a weak Carrollian structure is defined by $({\cN},q,\ell)$, with no reference to a preferred connection. With coordinates $x^a=(x^0,x^i)$,\footnote{The conventions in this subsection differ from \cite{Duval:2014uva} as follows: $x^i \rightarrow x^A$, $q \rightarrow g$, $\ell \rightarrow \xi$, $x^0\rightarrow u$, $\N\rightarrow C$.} a strong flat Carrollian structure is defined as
\beq\label{flat}
q=\delta_{ij} \rd x^i \rd x^j\qquad \ell =\pa_0\qquad  \Gamma^a_{bc}=0\,.
\eeq

The Carroll algebra is defined here as the algebra of vector fields $X$ such that
\beq\label{ciso}
\cL_X q=0\qquad \cL_X\ell=0\,,
\eeq
and also 
\beq\label{cisoextra}
\cL_X\nabla=0\,.
\eeq
In the flat case, it is instructive to first solve the two conditions in \eqref{ciso} alone, giving
\beq
X=(\omega^i{}_jx^j+\gamma^i)\pa_i+f(x^i)\pa_0\,,
\eeq
with $\omega$ skew symmetric. It is interesting to note the appearance of supertranslations $f(x^i)$ directly from the isometries of the Carroll structure: they are not associated with the conformal nature. In other words, the algebra of isometries is already infinite-dimensional. Imposing the extra condition \eqref{cisoextra}, using $\cL_{X}\Gamma^a_{bc}=\pa_b\pa_c X^a$, tells us that $f$ is at best linear in $x^i$, leading to
\beq
X=(\omega^i{}_jx^j+\gamma^i)\pa_i+(\varphi-\beta_i x^i)\pa_0\,.\label{isoca}
\eeq
This is the generator of the Carroll algebra $\mathfrak{carr}(d+1)$ in \eqref{gencarr}. In particular, $\omega^{ij}$ is the parameter for rotations $J_{ij}$, $\gamma^i$ the one for spatial translations $P_i$, $\varphi$ for temporal translations $P_0$, and $-\beta^i$ for the Carroll boosts $B_i$.\\

The level-$2/z$ conformal Carroll algebra is instead found studying the conformal isometries
\beq
\cL_X q=\lambda q\qquad \cL_X\ell=\mu\ell \qquad \lambda+\frac{2}{z}\mu=0\,.
\eeq
For the flat Carrollian structure \eqref{flat}, the solution of these equations is
\beq
X=(\omega^i{}_jx^j+\gamma^i+\chi x^i+k^i x^2-2k_j x^j x^i )\pa_i
+\left(z(\chi-2k_ix^i)x^0+f(x^j)\right)\pa_0\,.
\eeq
This is exactly the algebra $\mathfrak{ccarr}_{2/z}(d+1)$ as reported in \eqref{ccarr}. This establishes the profound link between the abstract algebraic discussion of section \ref{2} and the geometric construction of this section.\\

Let us then solve the conformal isometries equation for a generic metric $q=\hat q_{ij}(x) \rd x^i \rd x^j$. Calling $X=\hat X+X^0\pa_0$ with $\hat X(x)=\hat X^i\pa_i$, the equation $\cL_X\ell=\mu\ell$ gives
\beq\label{pasX}
\pa_0 \hat X^i=0\qquad \pa_0 X^0=-\mu\,.
\eeq
Moreover, the equation $\cL_X q=\lambda q$ reinforces $\pa_0 \hat X^i=0$ and gives
\beq
\cL_{\hat X}\hat q = \lambda \hat q \qquad \lambda=\frac{2}{d} \hat\nabla_i\hat X^i\,,
\eeq
where $\hat\nabla_i$ is the connection compatible with $\hat q_{ij}$.
Finally, using $\lambda+\frac{2}{z}\mu=0$ and \eqref{pasX}, we learn that 
\beq
\pa_0^2 X^0=-\pa_0 \mu=\frac{z}{2}\pa_0 \lambda=\frac{z}{d}\pa_0 (\hat\nabla_i\hat X^i)=0\,.
\eeq
Therefore, putting things together, we found
\beq
X=\hat X(x^i)+\left(\frac{z\lambda}{2}x^0+f(x^i)\right)\pa_0\,,
\eeq
with $\hat X$ a conformal Killing vector of $\hat q$. 

To connect more explicitly to BMS symmetries, consider $q$ to be the metric of the $2$-sphere, with thus $d=2$. Then, we have that the generators of $\mathfrak{ccarr}_{2/z}(3)$ read
\beq
X=\hat X(x^i)+\left(\frac{z}{2}\hat\nabla_i\hat X^ix^0+f(x^i)\right)\pa_0\,,
\eeq
where $\hat X$ is now an element of $\mathfrak{sl}(2,\mathbb{C})$, the conformal group of the $2$-sphere. The level-$2$ conformal Carroll algebra dilates space and time equally, and is obtained setting $z=1$, leading to 
\beq\label{ccarralg}
X=\hat X(x)+\left(\frac12\hat\nabla_i\hat X^i x^0+f(x^i)\right)\pa_0\,.
\eeq
This generates the algebra
\beq
\mathfrak{ccarr}_{2}(3)=\mathfrak{sl}(2,\mathbb{C}) \ltimes \mathbb{R}^S\,,
\eeq
in which by $\mathbb{R}^S$ we mean functions of the $2$-sphere forming an Abelian algebra. 

This is exactly the BMS algebra obtained studying asymptotically flat $4$-dimensional spacetimes. Therefore, if $q$ is a $2$-sphere,
\beq
\mathfrak{ccarr}_{2}(3)\cong \mathfrak{bms}_4\,.
\eeq
This result, originally obtained by Duval, Gibbons, and Horvathy \cite{Duval:2014uva}, has been the starting point driving tremendous advancement in Carrollian physics and flat-space holography. It truly put Carroll at the center stage for discussing the geometric structure of null hypersurfaces, providing a unified framework thereof.\\

%%%%%%
\subsection{General Carrollian Structure} \lb{sec:modern}

The zero-signature spacetime and the Carrollian manifold described in sections \ref{zerosig} and \ref{carrman}, respectively, can both be viewed as specific realizations of a general Carrollian structure. In this subsection, we reformulate the notion of Carrollian structure from a more geometric perspective, emphasizing its underlying fibre-bundle description and discussing the role of adapted coordinates and symmetries. A detailed analysis of Carrollian connections and their associated curvature will be deferred to section \ref{sec:connection}.\\

\subsubsection{Geometry}

A $d+1$-dimensional manifold $\N$ is a Carrollian manifold when equipped with a Carrollian structure $\Carr := (\N, q, \ell)$, where $q = q_{ab} \rd x^a \otimes \rd x^b$ is the (corank-1) degenerate metric in the direction of a nowhere-vanishing Carrollian vector field $\ell = \ell^a \pa_a$, that is $q(\ell, \cdot)  = 0 = \ell^a q_{ab}$. It is important that $\ell^a\neq 0$, otherwise the condition $\ell^a q_{ab}=0$ does not imply that the metric is degenerate. In the following, we denote with $\{x^a\}_{a = 0}^{d}$ general coordinates on the Carrollian manifold $\N$, such that $a,b,c,\dots$ are abstract indices. Our construction will therefore be completely invariant under diffeomorphisms of $\N$, unless explicitly stated.

The Carrollian structure can be elegantly described as a line bundle, $\pi: \N \to \S$ over a codimension-1 base manifold $\S$. In what follows, we will denote with $\{\s^A\}_{A = 1}^{d}$ the local coordinates on $\S$. Stemming from the fiber bundle structure is a Carrollian vector, $\ell \in \text{ker}(\pi_*)$, that lies in the 1-dimensional kernel of the pushforward map $\pi_*: T\N \to T\S$. For a 1-form $\eta = \eta_A \rd \sigma^A \in T^*\S$, this implies that its interior product with the Carrollian vector yields
\begin{align}
0 = \eta \left(\pi_* \ell\right) = (\pi^*\eta) (\ell) = \eta_A (\pi^*\rd \sigma^A)(\ell)\,,
\end{align}
where the first equality follows from the definition of $\ell$, and the second equality follows from the duality between the pushforward and the pullback, $\pi^*: T^*\S \to T^*\N$.
Therefore, a choice of coordinates on $\S$ defines the closed forms 
\begin{align}
e^A = \rd x^a e_a{}^A := \pi^*\rd \s^A \,, \quad \text{such that} \quad \iota_\ell e^A=\ell^a e_a{}^A=0 \quad \text{and} \quad \rd e^A = 0\,. \lb{coframe-hor}
\end{align}

A degenerate Carrollian metric $q$ on $\N$ that obeys the corank-1 condition, $q (\ell, \cdot) = 0$, is given by 
\begin{align}
q = q_{AB} e^A \otimes e^B\,, \lb{metric}
\end{align} 
where $q_{AB}(x)$, viewed as the components of the spatial metric, are the components of $q$ in the bases $e^A$. \\

Let us mention a special case where the metric $q$ is induced from a Riemannian metric $\mr{q}$ on the base $\S$, defined as 
\beq
\mr{q} = \mr{q}_{AB} \rd \s^A \otimes \rd \s^B\,, \qquad \text{with inverse}\qquad \mr{q}^{-1} = \mr{q}^{AB}\pa_{\s^A} \otimes \pa_{\s^B}\,.\label{qS}
\eeq 
Using $e^A$ as in \eqref{coframe-hor}, we define the "static" Carrollian metric $\hat{q}$ as the pullback of the base metric,
\begin{align}
\hat{q} = \pi^* \mr{q} = \mr{q}_{AB} (x) e^A \otimes e^B\,,
\end{align}  
which satisfies the corank-1 condition $\hat{q}(\ell,\cdot) =0$. This is a special case because, from the properties of the fibre bundle, $\hat{q}$ does not depend on time, that is, on the points along the fibre. We can verify this by computing the Lie derivative of $\hat{q}$ along the Carrollian vector
\begin{equation}
\begin{aligned}
\Lie_\ell \hat{q} = \Lie_\ell (\pi^* \mr{q}) = \pi^* (\Lie_{\pi_* \ell} \mr{q}) =0\,,\label{lqh}
\end{aligned}
\end{equation}
since $\pi_* \ell =0$. Alternatively, we compute
\begin{align}
\Lie_\ell \hat{q} = \ell[\mr{q}_{AB}(x)]  e^A \otimes e^B +2 \mr{q}_{AB}(x)  (\Lie_\ell e^A) \otimes e^B = \ell[\mr{q}_{AB}(x)]  e^A \otimes e^B \, ,
\end{align}
where we used $\Lie_\ell e^A = \rd (\iota_\ell e^A) + \iota_\ell \rd e^A =0$, following from \eqref{coframe-hor}. Then, equation \eqref{lqh} implies
\begin{align}
\ell[\mr{q}_{AB}(x)] = 0 \,.
\end{align}
We clearly see that the choice $\hat{q} = \pi^* \mr{q}$ leads to a degenerate and time-independent metric. Therefore, $(\N,\hat{q},\ell)$ gives rise to a static Carrollian structure. While certainly relevant as a subcase, in the following we will consider the general case in which the degenerate metric $q$ given in \eqref{metric} is fully general, and thus time-dependent.  \\

To proceed, one would like to introduce a projector to the space orthogonal to $\ell$, and to be able to describe tensors with mixed indices. For that, we need to introduce a one form dual to $\ell$. However, the naive object one would define by "lowering" the index of $\ell$ using the metric is, by the very virtue of the degenerate metric of a Carrollian structure, identically vanishing,
\beq\label{kaq}
k_a\stackrel{?}{=}q_{ab}\ell^b=0\,.
\eeq
Nonetheless, we can exploit the fact that the tangent bundle $T\cN$ admits a dual bundle $T^*\cN$, and thus one has a natural pairing of forms and vectors, such that there exists a 1-form $k$ satisfying
\beq
k_a\ell^a=\iota_\ell k=1\,.
\eeq

Given the underlying fibre bundle structure, such a 1-form $k$ is geometrically given by a choice of Ehresmann connection.
Indeed, if one further supplements an Ehresmann connection on the fiber bundle, the tangent bundle $T\N$ admits a local decomposition,
\begin{align}
T\N \simeq  V \oplus H\,, 
\end{align}
where $V = \mathrm{ker}(\pi_*)$ is the  universal vertical subbundle, while $H$ is the horizontal subbundle. The latter is ambiguous and thus requires a choice of Ehresmann connection, such that a vector field is horizontal if its action -- that is, interior product -- on the connection is zero. Given the definition of $V$, $\ell$ defines a direction in the $1d$ vertical subbundle $V$, and serves as its local basis. The Ehresmann connection -- also called ruling -- is the 1-form $\k = \k_a d x^a \in \Omega^1(\N)$ introduced earlier, the tangent-bundle dual of $\ell$. We stress that $k$ is not the metric-dual of $\ell$, since by degeneracy of the metric, $q(\ell,\cdot)=0$. 

The Ehresmann connection is a linear map, $\k: T\N \to \mathbb{R}$, whose kernel defines the $d$-dimensional horizontal subbundle, $H := \{ X \in T\N | \iota_X \k =0 \}$. Furthermore, $H$ is spanned by the horizontal frame fields $e_A = e_A{}^a \pa_a$ obeying, by definition, the condition $\iota_{e_A} \k=0$, and are chosen, without loss of generality, such that they are dual to the horizontal coframe fields \eqref{coframe-hor}, $e^A = \rd x^a e_a{}^A$.
Overall, the frame fields $(\ell, e_A)$ and their duals $(\k, e^A)$ serve as a complete basis for the tangent bundle $T\N$ and the cotangent bundle $T^*\N$, respectively, and satisfy the pairing conditions,
\begin{align}
\ell^a \k_a = 1=\iota_\ell k \quad \ell^a e_a{}^A =0=\iota_\ell e^A \quad e_A{}^a \k_a = 0=\iota_{e_A}k \quad e_A{}^a e_a{}^B= \delta_A^B=i_{e_A}e^B\,. \label{orthogonal}
\end{align}
This basis, derived from a fibre bundle description, will henceforth be called a Carrollian frame. If this structure pertains to the  holonomic coordinate basis, this is the adapted coordinate systems introduced by Henneaux \cite{Henneaux:1979vn}, with $\ell\rightarrow n=\pa_{x^0}$.\\

The Carrollian structure together with the choice of an Ehresmann connection is called a ruled Carrollian structure, $\RCarr = (\Carr, k)$. In addition, introducing the inverse $q^{AB}$ such that $q^{AC} q_{CB} = \delta^A_B$, we define the corank-1 tensor $q^{ab} = q^{AB} e_A{}^a e_B{}^b$, satisfying $q^{ab} k_b = 0$, such that its contraction with the degenerate metric produces the horizontal projector
\begin{align}
q_a{}^b = e_a{}^A e_A{}^b\,,\qquad \text{such that} \qquad q_{ac}q^{cb} = q_a{}^b = \delta_a^b - k_a \ell^b \qquad q_a{}^b q_b{}^c=q_a{}^c\,.\label{proj}
\end{align}
This tensor projects to the space orthogonal to $\ell$ and $k$: $q_a{}^b k_b =0$ and $\ell^a q_a{}^b =0$. Thus, when acting on vectors, it projects to $H$, whereas when acting on forms it projects to the dual of $H$, $H^*$. Note that if we restrict to horizontal quantities only, the degenerate tensor $q^{ab}$ can be used to "raise" indices and take traces, that is, given a 1-form $j_a$,
\beq
\text{if} \quad \ell^a j_a=0 \ \quad\Rightarrow\quad q^{ab}j_b=j^a\,, \quad \text{with} \quad j^ak_a=0\,.\label{lj}
\eeq
Conversely, note that if $j_a$ is not horizontal, then acting with $q^{ab}$ projects to its horizontal part only, thereby losing the information about its vertical component. 

A vector field $X \in T\N$ and a 1-form $\omega\in \Omega^1(\N)$ can be uniquely decomposed in the Carrollian frame as
\begin{align}
X= (\iota_X \k) \ell + (\iota_X e^A) e_A\,, \qquad \text{and} \qquad \omega = (\iota_{\ell} \omega) \k + (\iota_{e_A} \omega) e^A\,.
\end{align}
In components, these equations can be written using the projector as
\beq\label{deco}
X^a=X^b k_b \ \ell^a+X^b q_b{}^a\,,\qquad \text{and} \qquad \omega_a=\ell^b \omega_b \ k_a+q_a{}^b\omega_b\,.
\eeq

We will also denote the components of a horizontal tensor in the horizontal frame with the uppercase Latin indices ($A,B, C, ...$). For instance, $q_{AB} = q(e_A,e_B)$ are components of the degenerate metric in the Carrollian frame, see \eqref{metric}. A tensor $T_a{}^b$ is horizontal if $T_a{}^b = q_a{}^c T_c{}^d q_d{}^b$ whereas $\ell^a T_a{}^b = 0=  T_a{}^b k_b =0$. The degenerate metric is a perfect example of a horizontal tensor, as it trivially satisfies $q_a{}^c q_{cd}q_b{}^d=q_{ab}$ and $\ell^a q_{ab}=0$. Similarly, the differential of a function $F$ on $\N$ can be expressed as  
\begin{align}
\rd F = \ell [F]  \k + e_A[F] \e^A\qquad \pa_a F=(\ell^b\pa_b F)\k_a+q_a{}^b\pa_b F\,,
\end{align}
where $V[\cdot] = V^a\pa_a (\cdot)$ denotes a directional derivative along a vector $V$\,.\\

In the following, we will often decompose quantities both in the Carrollian frame and in the abstract index notation. One should be careful in translating between the two, especially in expressions involving derivatives. Let us show this with a simple explicit example. Consider the specific  vector field with $Y^B=\ell[y^B]$,
\beq
Y=Y^\ell \ell+Y^Be_B=Y^\ell \ell+\ell[y^B]e_B\,.
\eeq
We would like to express it in abstract index notation as $Y=Y^a\pa_a$. To do so, we use $\ell=\ell^a\pa_a$ and $e_B=e_B{}^a\pa_a$ and write
\beq
Y=(Y^\ell \ell^a+\ell[y^B]e_B{}^a)\pa_a\,.\label{YYY}
\eeq
To process the horizontal part, we use
\beq
\ell[y^B]e_B{}^a=\ell[y^b e_b{}^B]e_B{}^a=\ell[y^b]e_b{}^B e_B{}^a+y^b \ell[e_b{}^B]e_B{}^a=q_b{}^a \ell[y^b]+y^b \ell[e_b{}^B]e_B{}^a\,,\label{ly}
\eeq
where we recall the identity \eqref{proj}, $e_a{}^Ae_A{}^b=q_a{}^b$. To further process the last term, one notes that 
\beq
\iota_\ell e^A=0\qquad \rd e^A=0\qquad \Rightarrow \qquad \cL_\ell e^A=\iota_\ell \rd e^A+\rd \iota_\ell e^A=0\,,
\eeq
and thus, in components
\beq
0=\cL_\ell e_a{}^A=\ell[e_a{}^A]+e_b{}^A\pa_a\ell^b \qquad \Rightarrow\qquad \ell[e_a{}^A]=-e_b{}^A \pa_a \ell^b\,.
\eeq
Therefore, eq. \eqref{ly} becomes\footnote{Later, we will define the horizontal derivative $\overline{\pa}_a=q_a{}^b\pa_b$, see \eqref{ol}. Then, this simply reads $\ell[y^B]e_B=\cL_\ell y^b\overline{\pa}_b$.}
\beq
\ell[y^B]e_B{}^a=q_b{}^a \ell[y^b]-y^b e_c{}^Be_B{}^a \pa_b\ell^c=q_b{}^a \ell[y^b]-q_b{}^a y[\ell^b]=q_b{}^a \cL_\ell y^b\,.
\eeq
In conclusion, \eqref{YYY} reads
\beq\label{Yfin}
Y=Y^\ell \ell+\ell[y^B]e_B=(Y^\ell \ell^a+q_b{}^a \cL_\ell y^b)\pa_a\,.
\eeq
This kind of manipulations are important in jogging between the Carrollian frame and the abstract index notations. Since both have their utility and merits, we will keep using them both in what follows.

The ruled Carrollian structure $\RCarr$ completely captures the intrinsic geometry of a null manifold, without the need to embed the latter in an ambient space. This provides a modern take and a universal framework encompassing Carrollian manifolds as discussed in subsection \ref{carrman} and zero-signature spacetimes, introduced in subsection \ref{zerosig}.\\

\subsubsection{Internal Symmetries} \lb{sec:C-symm}

The various quantities introduced to define a ruled Carrollian structure are generally not uniquely determined, as the conditions they satisfy leave a residual freedom. The appropriate strategy is to select a representative for each of these ambiguities and subsequently verify how the physically meaningful expressions transform under such residual transformations. There is nothing problematic in working with quantities that are, in this sense, ambiguous, provided one consistently tracks the transformation properties of the tensors involved. This situation is entirely analogous to General Relativity, where one often manipulates objects built from the Christoffel symbols -- quantities that do not transform tensorially under diffeomorphisms -- while ultimately ensuring that all physically relevant expressions remain covariant or invariant under the full diffeomorphism group.

The main quantities involved in the definition of a ruled Carrollian structure are $\ell$, $q$, and $k$. Their defining properties are
\beq\label{lqk}
\ell^a q_{ab}=0\,,\qquad \text{and} \qquad \ell^a k_a=1\,.
\eeq
The first equation is left invariant if we rescale $\ell$ and/or $q$ by some nowhere vanishing function, whereas in the second equation one can shift $k$ by any 1-form that is annihilated by $\ell$. These transformations preserve the vertical subbundle, in the sense that they at best rescale its basis. There is however another transformation, which does not preserve the vertical subbundle: this is a shift of $\ell$ by a horizontal vector accompanied by a shift of the degenerate metric $q$. In summary, there are four types of symmetry transformations preserving eqs. \eqref{lqk}: rescaling of $\ell$, rescaling of $q$, shifts of $k$, and shifts of $\ell$ combined with shifts of $q$. 

\paragraph{$\ell$-rescaling} The Carrollian vector $\ell$ can be viewed as a preferred representative of the equivalence class under rescaling, $[\ell] := \{ \Phi \ell | \ \Phi : \N \to \mathbb{R} \}$, by a smooth, nowhere-vanishing function $\Phi$.

To preserve the condition $\ell^a k_a=1$, this local rescaling acts on the Ehresmann connection as $k \to \frac{k}{\Phi}$. For the infinitesimal parameter $\phi$ associated with $\Phi$, the transformation is given by 
\begin{align}
\delta_\phi \ell^a = \phi \ell^a \qquad
\delta_\phi k_a = -\phi k_a \qquad \delta_\phi q_a{}^b=0 \qquad
\delta_\phi q_{ab} = 0 \qquad \delta_\phi q^{ab}= 0\,. \label{rescaling}
\end{align}
A tensor $O_s$ on $\N$ is said to have weight $s$ under the rescaling if it transforms as $\delta_\phi O_s=s\phi O_s$.

\paragraph{$q$-rescaling} The degenerate metric is defined up to rescaling, thereby giving rise to the equivalence class $[q] := \{ \Psi q | \ \Psi : \N \to \mathbb{R} \}$, by a smooth, nowhere-vanishing function $\Psi$. 

Since this does not act on $\ell$, the condition $\ell^a k_a=1$ tells us that $k$ is also invariant. Infinitesimally, we have
\begin{align}
\delta_\psi \ell^a = 0 \qquad
\delta_\psi k_a = 0 \qquad \delta_\psi q_a{}^b=0 \qquad
\delta_\psi q_{ab} = \psi q_{ab} \qquad \delta_\psi q^{ab}= -\psi q^{ab}\,.
\end{align}

\paragraph{$k$-shift}

This symmetry represents a local shift of the Ehresmann connection. This comes about since the condition $\ell^ak_a=1$ is left invariant under a shift of $k\to k+\zeta$ as long as $\iota_\ell\zeta=0$. Defining the horizontal vector $\zeta^a=q^{ab}\zeta_b$, we thus have, infinitesimally,
\begin{align} 
\delta_\zeta \ell^a = 0 \quad
\delta_\zeta k_a = \zeta_a \quad \delta_\zeta q_a{}^b=-\zeta_a\ell^b \quad
\delta_\zeta q_{ab} = 0 \quad
\delta_\zeta q^{ab}= -\left(\ell^a \zeta^b + \ell^b \zeta^a\right)\,. \label{shift}
\end{align}
As usual, some internal symmetries can be realized by isometries, when the background is flat. This is indeed the case for this shift, realized as an isometry in \eqref{isoca}, where it is associated to the Carroll boost $-\beta_A$. The Carroll boost symmetry is an important ingredient of Carrollian physics.

\paragraph{$\ell \& q$-shift}

The condition $\ell^a q_{ab}=0$ is preserved when combining a horizontal shift of $\ell$ and of the metric $q$, generated by a vector $\eta$. Infinitesimally, this is achieved by
\begin{align}
\delta_\eta \ell^a = \eta^a \quad
\delta_\eta k_a = 0 \quad \delta_\eta q_a{}^b=-k_a \eta^b \quad
\delta_\eta q_{ab} = -\left( k_a \eta_b + k_b \eta_a \right) \quad \delta_\eta q^{ab}= 0\,,
\end{align}
where $\eta^a$ is horizontal and $\eta_a = q_{ab} \eta^b$. As we remarked, this symmetry does not preserve the vertical subbundle, which is now spanned by $\ell+\eta$.\\

Depending on the explicit problem at hand, note that some of these transformations may not be allowed. In other words, these ambiguities can be lifted by external input on the problem. For instance, once a null hypersurface is embedded at finite distance in a pseudo-Riemannian manifold subject to Einstein equations, its degenerate metric is dictated by the induced metric from the bulk. Then, rescaling $q$ leads to either rescaling the projector to the hypersurface, which would make it a projector no more, or to rescale the bulk metric itself, which is not a symmetry of General Relativity. Therefore, in this specific example, the $q$-rescaling ambiguity is lifted by the embedding. Similar ambiguity resolutions may occur when applying this geometric framework in other contexts.\\

\subsubsection{Acceleration, Vorticity, Expansion}

We can define various geometric objects from the ruled Carrollian structure $\RCarr$. First, the field strength of the Ehresmann connection admits the general decomposition,\footnote{The vorticity here is minus the vorticity defined in \cite{Ciambelli:2019lap}.}
\begin{equation}
\begin{aligned}
\rd \k &:= - \left( \k \wedge \ac + \vor \right)\,, \lb{d-k}
\end{aligned}
\end{equation}
where the 1-form $\ac$ is the Carrollian acceleration, which is horizontal $\iota_\ell \varphi=0$, and the 2-form $\vor$ is the Carrollian vorticity, also horizontal $\iota_\ell \varpi=0$. 

The Carrollian acceleration expresses how the Ehresmann connection is Lie-dragged along $\ell$
\beq
\cL_{\ell}k=\iota_\ell \rd k+\rd(\iota_\ell k)=-\iota_{\ell}\left( \k \wedge \ac + \vor \right)=-\varphi\,,
\eeq
where we used $\iota_\ell k=1$. The vorticity controls the non-integrability of the horizontal subbundle $H$, which in general is a distribution. Indeed, Frobenius theorem states that
\beq
k\wedge \rd k=-k\wedge \varpi
\eeq
vanishes if and only if the distribution $H$ is integrable. Therefore, the Carrollian vorticity $\varpi$ is responsible for the non-integrability of the horizontal subbundle.\\

Recalling that $[\iota_X, \cL_Y ]\gamma = \iota_{[X,Y]} \gamma$, $\cL_X = \rd \iota_X + \iota_X \rd$, and $\iota_X\iota_Y \gamma=0$, an important identity for what follows is
\begin{equation}
\iota_X \iota_Y \rd \gamma = \iota_{[X,Y]}\gamma + \cL_Y (\iota_X \gamma) -\cL_X (\iota_Y \gamma)\,, \lb{duality-com}
\end{equation} 
for any vector fields $X, Y \in T\N$ and any 1-form $\gamma \in \Omega^1(\N)$. Indeed, we will repeatedly apply this equation to the Carrollian coframes. If $\gamma$ is a coframe, since it must satisfy \eqref{orthogonal}, one has
$\cL_Y (\iota_X \gamma)=0=\cL_X (\iota_Y \gamma)$. For instance, if $\gamma=e^A$, $X=\ell$ and $Y=e_B$, one has $\cL_{e_B} (\iota_\ell e^A)=0$ and $\cL_\ell (\iota_{e_B} e^A)=\cL_\ell \delta_B^A=0$.

We then proceed and first apply eq. \eqref{duality-com} to $\gamma=e^A$. Since the curvature of the horizontal coframe vanishes, $\rd e^A =0$, specifying $X=e_B$ and $Y=e_C$ we have
\beq
0 = \iota_{[e_B,e_C]}e^A=C_{BC}{}^D \iota_{e_D}e^A+C_{BC}{}^\ell \iota_{\ell}e^A=C_{BC}{}^A\,,
\eeq
where we used $\iota_{e_B}e^A=\delta_B^A$ and $\iota_{\ell}e^A=0$. If instead one uses $X=\ell$ and $Y=e_B$, the result is
\beq
0 = \iota_{[\ell,e_B]}e^A=C_{\ell B}{}^A\,.
\eeq
Similarly, let us now choose $\gamma=k$. Now the left hand side of eq. \eqref{duality-com} is non-vanishing due to \eqref{d-k}. Choosing then $X=e_A$ and $Y=e_B$, the left hand side indeed gives
\beq
\iota_{e_A}\iota_{e_B}\rd k=-\iota_{e_A}\iota_{e_B}\varpi=-\varpi_{BA}=\varpi_{AB}\,,
\eeq
while the right hand side of \eqref{duality-com} is
\beq
\iota_{[e_A,e_B]}k=C_{AB}{}^\ell\,.
\eeq
So we have found 
\beq
C_{AB}{}^\ell=\varpi_{AB}\,.
\eeq
Using instead $X=\ell$ and $Y=e_A$, we eventually get
\beq
\iota_{\ell}\iota_{e_A}\rd k=-\iota_{\ell}\iota_{e_A}(k\wedge \varphi)=\varphi_A=\iota_{[\ell,e_A]}k=C_{\ell A}{}^\ell\,.
\eeq

Given that 
\beq
[e_B,e_C]=C_{BC}{}^A e_A+C_{BC}{}^\ell \ell\,, \qquad \text{and}\qquad [\ell,e_C]=C_{\ell C}{}^A e_A+C_{\ell C}{}^\ell \ell\,,
\eeq
we thus have derived the Carrollian Lie brackets
\begin{align}
[e_A,e_B] = \vor_{AB}\ell
\qquad \text{and} \qquad
[\ell,e_A]= \ac_A \ell\,. \lb{C-comm}
\end{align}
Therefore, we confirm that, in the presence of the Carrollian vorticity $\vor_{AB}$, the Lie bracket between horizontal frames, $[e_A,e_B]$, lies outside the horizontal subbundle $H$, another manifestation of the fact that $\varpi$ controls the non-integrability of $H$. 

The Jacobi identity of the Lie brackets constrains the evolution of the Carrollian vorticity. Using
\begin{equation}
    \begin{aligned}
        0 & =[\ell,[e_A,e_B]]+[e_B,[\ell,e_A]]+[e_A,[e_B,\ell]]\\
& =[\ell,\varpi_{AB}\ell]+[e_B,\varphi_A \ell]-[e_A,\varphi_B \ell]\\
&= (\ell[\varpi_{AB}]+e_B[\varphi_A]-\varphi_A\varphi_B-e_A[\varphi_B]+\varphi_A\varphi_B)\ell\,,
    \end{aligned}
\end{equation}
we find the equation
\begin{align}
\ell[\vor_{AB}]= e_A\big[\ac_B\big] -e_B\big[\ac_A\big]\,. 
\end{align}\\

We now turn our attention to the expansion tensor, $\theta_{ab}$, defined as the change of the Carrollian metric along the vertical direction,\footnote{This tensor is the second fundamental tensor defined in \eqref{Kmh}.}
\begin{align}
\theta_{ab} = \frac{1}{2}\cL_\ell q_{ab}\,.\label{theta}
\end{align}
Using that $\cL_{\ell}\ell=0$, one has 
\beq
\ell^a\theta_{ab}=0\,,
\eeq
therefore, this tensor is fully horizontal and symmetric. Moreover, by design, in the adapted Carrollian frame,
\begin{align}
\theta_{AB} =  \frac{1}{2}\ell \big[ q_{AB} \big]\,
. \label{adtheta}
\end{align}

Since this tensor is horizontal, we can take its trace by contracting with $q^{ab}$. This defines the expansion
\begin{align}
 \theta = q^{ab} \theta_{ab}\,. \lb{expansion-scalar}
\end{align} 
This is a central quantity, as it measures the rate of change of the area element\footnote{With a slight abuse of notation, we denote the area element of the spatial metric with $\sqrt{q} := \sqrt{\det q_{AB}}$, which is the determinant of $q_{AB}$ when viewed as a matrix. This is different from the standard notation $\sqrt{q} = \sqrt{\det q_{ab}}$, which gives zero due to the degeneracy of the metric in a Carrollian structure.\lb{detq}} $\sqrt{q} = \sqrt{\det q_{AB}}$ of the spatial metric along the Carrollian vector $\ell$. To see this, we can use \eqref{adtheta}, and recall that for horizontal tensors $q^{AB}$ is the inverse of $q_{AB}$
\begin{align}
 \theta=\frac{q^{AB}}{2}\ell[q_{AB}]=\frac12 \text{Tr}(\ell[q_{AB}])=\ell \left[\ln \sqrt{q}\right]\,. \label{thC}
\end{align}

The expansion tensor can be decomposed into its trace-full and trace-free parts
\beq
\theta_{ab}=\frac{1}{d} \theta q_{ab}+\s_{ab}\,,
\eeq
such that $q^{ab}\s_{ab}=0$. The symmetric and trace-free tensor
\begin{align}
\s_{ab} = \theta_{ab} - \frac{1}{d} \theta q_{ab}\,, 
\end{align}
is the Carrollian shear. All these quantities have been introduced from the intrinsic Carrollian viewpoint. They encode the decomposition of the first derivative of the Carrollian vector field $\ell$.

It is instructive to display how the expansion, acceleration, and vorticity transform under the shift \eqref{shift}. One has
\beq\label{newshift}
\delta_\zeta \theta_{ab}=0\qquad \delta_\zeta \varphi_a=-\cL_{\ell}\zeta_a   \qquad \delta_\zeta \varpi_{ab}=-(q_a{}^c \pa_c-\varphi_a)\zeta_b+(q_b{}^c \pa_c-\varphi_b)\zeta_a  \,.
\eeq

Note that, given a rank-2 covariant tensor $L_{ab}$, the tensor $L_a{}^b$ is uniquely specified requiring
\beq
L_a{}^b q_{bc}=L_{ac}\qquad \text{and} \qquad L_a{}^b k_b=0\,.
\eeq
The last condition is projecting to the horizontal part. Continuing, one can then define $L^{ab}$ by the conditions
\beq
L_c{}^b=q_{ac} L^{ab}\qquad \text{and}\qquad k_a L^{ab}=0\,.
\eeq
Therefore, as already mentioned in \eqref{lj}, if $L_{ab}$ is not fully horizontal, the tensor $L^{ab}$ is losing information about its non-horizontal part. Conversely, if $L_{ab}$ is fully horizontal, then $L^{ab}$ is simply its associated rank-2 contravariant tensor built by raising the indices with $q^{ab}$. Applying this to the fully horizontal tensor $\theta_{ab}$, one has
\beq
\theta_a{}^b=\frac{\theta}{d}q_a{}^b+\s_a{}^b\,,
\eeq
such that, recalling $q_a{}^a=d$, $\s_a{}^a=0$.\\

To emphasize the importance of these quantities, let us briefly anticipate that when this Carrollian structure is embedded in a one-dimension-higher pseudo-Riemannian manifold subject to Einstein gravity without matter, they combine to give the projected Einstein equation on the null hypersurface, the celebrated null Raychaudhuri and Damour equations
\beq\label{EE}
&(\cL_{\ell}+\theta)\theta-\mu\theta+\sigma_a{}^b\sigma_b{}^a=0& \\
&\left({\cal L}_\ell+\theta\right)\pi_a -q_a{}^b D_b \mu +(\theta-\mu)\varphi_a + q_b{}^c q_a{}^d(D_c+\varphi_c)\sigma_{d}{}^b=0&
\eeq
where we introduced the combination
\beq
\mu=\kappa+\frac{d-1}{d}\theta\,.
\eeq
The quantities $\kappa$ and $\pi_a$ entering these expressions will be discussed later (see section \ref{rigg}). From the intrinsic Carrollian standpoint, they parameterized the torsion-free but non metric-compatible Carrollian connection $D_a$, as we will review in section \ref{sec:connection}. This demonstrates how the different Carrollian objects introduced above find a relevant application to the physics of gravity induced on null hypersurfaces, which constitutes the primary application of Carrollian physics and geometry. We will focus on it in section \ref{rigg}.\\

%%%%%%%%%%%%%%%%%
\subsubsection{Adapted Coordinates}\label{334}
Although the geometric construction is presented in a coordinate-free manner, it is sometimes useful to discuss specific coordinates. Since the Carrollian manifold $\N$ is a line bundle over the base $\S$, we can choose a coordinate system\footnote{In the context of null surfaces, the “temporal” coordinate $x^0$ can be the retarded null time $u$ or the advanced null time $v$. This is oftentimes referred to as the clock.} $x^a = (x^0,x^i)$ such that open sets of the cuts at $x^0=\mathrm{constant}$, denoted $\S_{x^0}$, are identified with open sets of the base $\S$ through the bundle map, $\S_{x^0}  \to \S$, essentially mapping the spatial coordinates $x^i$ to the coordinates on $\S$\footnote{More rigorously, $\pi^A$ is a transition map, $\pi^A := (\s \circ \pi \circ x^{-1} (x^0,x^i))^A$, where $x: {\cN} \to {\mathbb{R}}^{D-1}$ and $\s: S \to {\mathbb{R}}^{D-2}$ provide, respectively, local coordinates on $\cN$ and $S$.}
\begin{align}
x^i \to \sigma^A= \pi^A(x^0,x^i)\,. 
\end{align}

We denote the Jacobian of the pushforward $T \S_{x^0}\to T\S$ by $J_i{}^B$, and it is explicitly given in coordinates by $J_i{}^B =\pa_i \pi^B$, where $\pa_i := \frac{\pa}{\pa x^i}$. It follows that $\pa_j J_i{}^A = \pa_i J_j{}^A$. In this coordinate system, the most general parametrization of the frame fields is given by 
 \begin{equation}
\lb{Cframe}
\begin{alignedat}{4}
&\ell &&= \e^{-\alpha} \mathrm{D}_0  \qquad \qquad  && k &&= \e^\alpha (\rd x^0  - \beta_A e^A) \\
& e_A &&= (J^{-1})_A{}^i \pa_i + \beta_A \mathrm{D}_0  \qquad \qquad && e^A &&= J_j{}^A (\rd x^j - V^j \rd x^0) 
\end{alignedat}
\end{equation}
where we defined $\mathrm{D}_0 := (\pa_0 + V^i \pa_i)$. The variables comprise a scale factor $\alpha$, a velocity field $V^i$, a Carrollian field $\beta_A$, the Jacobian $J_i{}^B$, and its inverse $J^{-1}$ such that $J_i{}^C (J^{-1})_C{}^j = \delta_i^j$ and $(J^{-1})_j{}^A J_B{}^j = \delta_B^A$.

Following from the definition of the horizontal coframe \eqref{coframe-hor} the velocity field $V^i$ is expressed in terms of the projection map as 
\begin{align}
V^i = - \pa_0 \pi^A  (J^{-1})_A{}^i\,, \qquad \text{such that} \qquad \mathrm{D}_0 \pi^A = 0\,,
\end{align}
where we used $\pa_i\pi^B=J_i{}^B$. In addition, the condition $\rd e^A =0$ imposes the following constraints on the variables,
\begin{align}\label{Carrollian}
\mathrm{D}_0 J_i{}^A  = -\pa_i V^j J_j{}^A\,, \qquad \text{and} \qquad \mathrm{D}_0 (J^{-1})_B{}^i  = (J^{-1})_B{}^j\pa_j V^i\,. 
\end{align}
We can evaluate the coordinate expressions for the Carrollian acceleration $\ac_A$ and the Carrollian vorticity $\vor_{AB}$ and obtain
\beq
\ac_A =  \mathrm{D}_0 \beta_A +e_A[\alpha]\qquad 
\vor_{AB} = \e^\alpha \left( e_A[\beta_B] - e_B[\beta_A] \right)\,.  
\eeq\\

So far, these expressions are simply a parametrization: they are fully diffeomorphism-covariant, provided the various quantities introduced transform judiciously. Often, it is convenient to break this covariance and work with the adapted coordinates $x^i=\sigma^A$ such that the action of the projection is trivial, $\pi:(x^0,\sigma^A) \to \sigma^A$.\footnote{In comparing with \cite{Ciambelli:2019lap}, one has $e^\alpha\to \Omega$, $\beta_A\to b_i$, $k\to \mathbf{e}$, $\varpi\to -\form{\varpi}$.} Here, the velocity field vanishes and the Jacobian is trivial, $V^i =0$ and $J_i{}^B = \delta_i{}^B$. The expressions for the frame fields simplify to
 \begin{equation}
\begin{alignedat}{4}
&\ell &&= \e^{-\alpha} \pa_0  \qquad \qquad  &&k &&= \e^\alpha (\rd x^0  - \beta_A \rd \s^A) \\
& e_A &&= \pa_{\s^A} + \beta_A \pa_0  \qquad \qquad && e^A &&= \rd \s^A\,.  \label{como}
\end{alignedat}
\end{equation}

The Carrollian vorticity becomes the curvature of the Carrollian field $\beta_A$, 
\begin{align}
\vor_{AB} = \e^\alpha\left(\pa_{\s^A} \beta_B -\pa_{\s^B} \beta_A + [\beta_A,\beta_B]_{\mathrm{W}}\right)\,,
\ \ \ \ \text{where} \ \ \ \
[a,b]_{\mathrm{W}} := a\pa_0 b-b\pa_0a
\end{align}
is the Witt bracket. 

As mentioned, the operation of adapting the coframes to these specific coordinates breaks part of the diffeomorphism invariance of the system. Indeed, the form of the frame fields written in \eqref{como} using the co-moving coordinate system is not preserved under a generic diffeomorphism. In particular, as soon as the spatial coordinates are transformed in a time-dependent way, the Carrollian vector field $\ell$ acquires a spatial component. Nevertheless, there exists a special residual subgroup of diffeomorphisms preserving the form of the frame fields. These are the so-called Carrollian diffeomorphisms
\beq\label{cadi}
{x^0}'={x^0}'(x^0,\sigma^A)\qquad \text{and}\qquad {\sigma^A}'={\sigma^A}'(\sigma^B)\,.
\eeq
Their action preserves \eqref{como}, provided the Carrollian field $\beta$ parametrizing the Ehresmann connection shifts accordingly, a manifestation of the latter being indeed a connection,
\beq
\beta_A\to (J^{-1})_A{}^B (J \beta_B+J_B)\,,\qquad \text{with}\qquad J=\frac{\pa {x^0}'}{\pa {x^0}}\qquad \text{and}\qquad J_A=\frac{\pa {x^0}'}{\pa {\sigma^A}}\,.
\eeq
The Carrollian diffeomorphisms \eqref{cadi} play a special role, as they are the automorphisms of the underlying Carrollian fiber bundle: given that $\pa_0$ defines the typical fibre, the automorphisms are by construction changes of the fibres that depend on the base point, while the base points transform among themselves, without mixing with the fibres. This realizes naturally the idea that, in Carrollian physics, space is absolute. One can therefore see that the Carrollian diffeomorphisms generalize the conformal Carroll algebra generated by \eqref{ccarralg}.

When applying this construction to finite-distance null hypersurfaces, one needs to account for the full diffeomorphism group. Then, this parametrization is not enough, and one has to generalize the co-moving coordinates. This is the reason why we presented the Carrollian structure in an abstract index notation: all our expressions before and after this subsection are fully diffeomorphism covariant. Note that the general parametrization \eqref{Cframe} can always be achieved and is diffeomorphism covariant. Indeed, the velocity $V^i$ can be generated via diffeomorphisms of the form ${x^i}'(x^0,x^j)$. In fact, the frame fields \eqref{Cframe} are simply the most general decomposition of vectors and forms on a basis. While this excursus aims at connecting with previous literature and common notation in Carrollian physics, in the rest of the review (except shortly in the coming subsection) we will refrain from using a specific form of the frame fields,  thus working in a manifestly covariant fashion.\\

%%%%%%%
\subsubsection{Diffeomorphisms}

Since we have introduced various notations and structures, it is useful to show how  the Carrollian structure changes under diffeomorphisms of the manifold $\N$, and in particular under their vertical and horizontal split. Under an infinitesimal diffeomorphism, generated by a vector field $\xi^a = f \ell^a + X^b q_b{}^a$ for a function $f(x)$ and a horizontal vector $X=X^a\pa_a=X^A e_A$, the ruled Carrollian objects $(\ell, k, q)$ change as follows:\footnote{In the Carrollian frame, using the discussion around eq. \eqref{Yfin}, these equations become
\begin{subequations}
\begin{align}
\cL_\xi \ell &= -\left( \ell[f] + X^A \ac_A \right) \ell - \ell[X^A] e_A \\
\cL_\xi k & = \left( \ell[f] + X^A \ac_A \right) k + \left( (e_A - \ac_A)[f] + \vor_{AB} X^B \right) e^A\\
\cL_\xi q &= 2\left(f \theta_{AB} + \sD_{(A}X_{B)} \right) e^A \otimes e^B + q_{AB} \ell[X^B] \left(k\otimes e^A + e^A \otimes k\right)\, ,
\end{align}
\end{subequations} where $2\sD_{(A}X_{B)}=X^Ce_C[q_{AB}]+q_{AC}e_B[X^C]+q_{BC}e_A[X^C]$. Here, we anticipated the notation for the horizontal covariant derivative that will be defined in \eqref{hder}.}
\begin{subequations}
\begin{align}
\cL_\xi \ell^a &= -\left(\ell[f]+X^b\varphi_b\right) \ell^a- q_b{}^a \cL_{\ell}X^b\\
\cL_\xi k_a & = \left( \ell[f] + X^b \ac_b \right) k_a +(\overline{\pa}_a - \ac_a)[f] + \vor_{ab} X^b\\
\cL_\xi q_{ab} &= 2 f \theta_{ab} + q_a{}^c q_b{}^d\cL_{X}q_{cd}+ (k_a q_{bc}+k_b q_{ac}) \cL_\ell X^c\,,
\end{align}
\end{subequations}
where we have used the notation $\overline{\pa}_a f=q_a{}^b\pa_b f$ -- see \eqref{ol}.

In this language, the Carrollian diffeomorphisms introduced above satisfy $\cL_\xi \ell \propto \ell$, that is,  $\ell[X^A] = 0$. This, in the co-moving coordinates, becomes $\pa_0 X^A = 0$. Carrollian isometries and conformal Carrollian isometries are then obtained by solving $\cL_\xi q =0$ and $\cL_\xi q \propto q$ plus $\cL_\xi \ell \propto \ell$, respectively. The solutions in three dimensions and their connection with the BMS group have already been provided in section \ref{carrman}. 

When working in the co-moving coordinates, the components $(\alpha, \beta_A, q_{AB})$ transform under Carrollian diffeomorphisms, $\xi = f(x^0, \s) \ell + X^A(\s)e_A$, as follows
\beq
&\delta_\xi \alpha = \ell[f] + X^A \ac_A  = \xi[\alpha] + \pa_0 \xi^0 & \nn  \\
&\delta_\xi \beta_A  =   \e^{-\alpha}(\ac_A - e_A)[f] - \e^{-\alpha}\vor_{AB} X^B = \xi[\beta_A] - \left( \pa_A + \beta_A \pa_0 \right) \xi^0 + \beta_B \pa_A \xi^B &\nn \\
&\delta_\xi q_{AB}= 2\left( f \theta_{AB} + \sD_{(A} X_{B)} \right) = \xi[q_{AB}] + q_{AC} \pa_B \xi^C + q_{CB}\pa_A \xi^C\,,\nn &
\eeq
where we used the components of the diffeomorphism vector in the adapted coordinates, $\xi^0 = \e^{-\alpha} f + X^A \beta_A$ and $\xi^A = X^A$. \\

This concludes our overview of the foundations of Carrollian geometry. We have reviewed the main historical developments in the field and presented a modern geometric formulation of Carrollian structures, emphasizing their symmetries and underlying geometric framework. Our discussion aimed to place Carrollian geometry on the same conceptual footing as pseudo-Riemannian geometry. In particular, we stressed that, owing to the degeneracy of the metric, one cannot dispense with the Carrollian vector field and its tangent-bundle dual 1-form in constructing the geometric data. A Carrollian structure should therefore be regarded as the analogue of a metric on non-degenerate manifolds. Following the usual pseudo-Riemannian development, the next natural step is to introduce the affine connections associated with a Carrollian structure.

\newpage

\section{Connection and Curvature} \lb{sec:connection}

The geometric study of a manifold traditionally unfolds in three successive steps: the specification of its geometric data, the definition of connections governing parallel transport, and the introduction of curvature tensors. Having discussed the geometric data of a Carrollian structure in the previous section, we now turn to the analysis of Carrollian connections and curvature.

A connection encodes the kinematics of fields on a manifold by prescribing their transport and differentiation. In the pseudo-Riemannian case, the well-known Levi-Civita theorem guarantees the existence of a unique, torsionless, and metric-compatible connection. Its Christoffel symbols are entirely determined by the metric, making it the canonical tool for studying parallel transport and curvature.

The situation is markedly different for a Carrollian structure, where the metric is degenerate. Imposing both torsionlessness and metricity no longer determines the connection uniquely; rather, it imposes non-trivial constraints on the metric itself. In particular, these conditions require the Carrollian metric to be time-independent. Hence, only time-independent Carrollian structures admit a unique torsionless and metric-compatible connection. In the general case, when time dependence is allowed, one must relax one of these conditions. A Carrollian structure may therefore admit either a torsionless or a metric-compatible connection, each option leading to distinct and interesting geometric consequences. 

In this section, we begin by briefly recalling the standard construction of connections and curvature tensors on pseudo-Riemannian manifolds. We then develop the Carrollian counterpart, presenting a detailed and self-contained discussion of the most general Carrollian connections. We then discuss how to intrinsically characterize a specific Carrollian connection, which, in section \ref{rigg}, will be shown to be compatible with embedding the Carrollian structure in a pseudo-Riemannian ambient space. Such Carrollian connection is torsionless and minimally non-metric compatible. We then perform a comparison of this connection with the so-called horizontal connection. Finally, we construct the curvature tensor associated with this Carrollian connection and derive simplified expressions valid in special geometric regimes.\\

\subsection{Connection on Pseudo-Riemannian Manifolds}\label{421}

We start with a pseudo-Riemannian manifold $\M$ as a warm up. The metric is a non-degenerate bilinear tensor, and we employ the notation
\beq
g:T\M\otimes T\M\to C^{\infty}(\M)\qquad g(X,Y)=g_{MN}X^M Y^N\,,
\eeq
for all $X,Y\in T\M$ vector fields. The indices $M,N$ refer to a coordinate basis $\partial_M$ of $T\M$, such that
\beq
X=X^M \pa_M,\quad Y=Y^M\pa_M\,.
\eeq
Since the basis is holonomic, the structure constants vanish
\beq
[\pa_M,\pa_N]=0\,.
\eeq
We introduce an affine connection $\nabla_X:T\M\to T\M$. Affinity means, for all $f\in C^{\infty}(\M)$,
\beq
\nabla_{fX}=f\nabla_X\qquad \nabla_X (fY)=f\nabla_X Y+X(f) Y\,,
\eeq
where $X(f)=X^M \pa_M(f)$. 

The key tensors are the torsion and the covariant derivative of the metric. The torsion is a skew symmetric bilinear tensor defined as
\beq\label{T}
T(X,Y)=\nabla_X Y-\nabla_Y X-[X,Y]\,,
\eeq
where the commutator ensures  $T(fX,Y)=fT(X,Y)$. The covariant derivative of the metric is given by
\beq\label{dg}
\nabla_X g(Y,Z)=X(g(Y,Z))-g(\nabla_X Y,Z)-g(Y,\nabla_X Z)\,.
\eeq
A connection satisfying $T(X,Y)=0$ for all $X,Y$ is called torsion-free (or torsionless), while the condition $\nabla_X g(Y,Z)=0$ for all $X,Y,Z$ is called metricity. The Christoffel symbols are defined by the action of the connection on the basis
\beq
\nabla_{M} \pa_N =\Gamma_{MN}^{P}\pa_P\,.
\eeq
The Levi-Civita theorem ensures that there exists a unique affine connection satisfying torsionless ($T(X,Y)=0$) and metricity ($\nabla_M g_{PQ}=0$). It is a non-trivial fact that these two equations can be entirely solved for the Christoffel symbols, without constraining the metric itself. As usual, one solves these conditions and gets the Christoffel symbols
\beq
\Gamma_{MN}^P&=&\frac12 g^{PQ}\Big(\pa_M g_{NQ}+\pa_N g_{MQ}-\pa_Q g_{MN})\,.
\eeq\\

We explained so far the holonomic case, where the tangent bundle basis is $\pa_M$. If the index $M$ refers to a general non-holonomic basis $e_M$, the structure constants are
\beq
[e_M,e_N]=C_{MN}{}^P e_P\,,
\eeq
and the Levi-Civita connection straightforwardly generalizes to ($C_{MNP}=C_{MN}{}^Q g_{QP}$)
\beq
\Gamma_{MN}^P=\frac{g^{PQ}}{2}\big(e_M(g_{NQ})+e_N(g_{MQ})-e_Q(g_{MN})+C_{MNQ}+C_{QNM}-C_{MQN}\big)\,.
\eeq
We emphasize that the connection is entirely solved by the structure constants and the metric, and no conditions are imposed on the latter from setting \eqref{T} and \eqref{dg} to zero. This will cease to be true for a Carrollian structure.

The Riemann tensor is defined as
\begin{align}
[\nabla_M, \nabla_N] X^P = R^P{}_{QMN} X^Q\,,\label{rdef}
\end{align}
for all $X\in T\M$. This tensor is built out of the connection symbols $\Gamma^A_{BC}$ as
\beq\label{Riem}
R^M{}_{NPQ}=e_P[\Gamma^M_{QN}]-e_Q[\Gamma^M_{PN}]+\Gamma^S_{QN}\Gamma^M_{PS}-\Gamma^S_{PN}\Gamma^M_{QS}-C_{PQ}^S\Gamma^M_{SN}\,.
\eeq\\

\subsection{Carrollian Connection} \lb{sec:Carr-connection}

Consider now the Carrollian structure discussed in Section \ref{sec:modern}, given by a Carrollian vector field $\ell^a$ and a degenerate metric $q_{ab}$. We will be as general as possible, and thus work in a non-holonomic basis
\beq
[e_a,e_b]=C_{ab}{}^c e_c\,.
\eeq
The affine connection symbols are defined by ($D_{e_a}=D_a$)
\beq
D_a e_b=\Gamma_{ab}^c e_c\,.
\eeq

For generic vector fields $X=X^ae_a$ and $Y=Y^ae_a$ in $T\N$, the torsion is
\beq
T(X,Y)=D_X Y-D_Y X-[X,Y]\,.
\eeq
For the basis, it reads
\beq
T(e_a,e_b)=(\Gamma_{ab}^c-\Gamma_{ba}^c-C_{ab}{}^c)e_c=T_{ab}{}^c e_c\,,
\eeq
whereas the covariant derivative of the degenerate metric
\beq
D_X q(Y,Z)=X(q(Y,Z))-q(D_X Y,Z)-q(Y,D_X Z)\,,
\eeq
gives for the basis ($q(e_b,e_c)=q_{bc}$)
\beq
D_a q_{bc}=e_a(q_{bc})-\Gamma_{ab}^dq_{dc}-\Gamma_{ac}^dq_{bd}=N_{abc}\,.
\eeq\\

In these expressions, we introduced the non-metricity tensor $N_{abc}$ and the torsion $T_{ab}{}^c$. Given $\ell^a q_{ab}=0$, and employing the notation $A_\ell=\ell^a A_a$ for all tensors $A$, the crucial property of the non-metricity is
\beq
N_{a\ell c}=\ell^b N_{abc}=\ell^b e_a(q_{bc})-\ell^b\Gamma_{ab}^d q_{dc}\,,
\eeq
and
\beq
N_{\ell bc}=\ell^a N_{abc}=\ell^a e_a(q_{bc})-\ell^a\Gamma_{ab}^d q_{dc}-\ell^a\Gamma_{ac}^d q_{bd}\,.
\eeq
Together with
\beq
T_{\ell ab}=\ell^c T_{ca}{}^d q_{db}=\ell^c\Gamma_{ca}{}^dq_{db}-\ell^c \Gamma_{ac}{}^dq_{db}-\ell^c C_{ca}{}^dq_{db}\,,
\eeq
we can use the previous results to write
\beq
\cL_{\ell}q_{ab}=N_{\ell ab}-2N_{(ab)\ell}+2 T_{\ell(ab)}+\ell^c C_{ca}{}^d q_{db}+\ell^c C_{cb}{}^d q_{da}\,.
\eeq

This is the main result, demonstrating how the non-metricity and the torsion combine to reconstruct the Lie derivative of the degenerate metric. If we furthermore impose $C_{ab}{}^c q_{cd}=0$,\footnote{This holds both in the coordinate basis, in which $C_{ab}{}^c=0$, and in the adapted frame, where $C_{ab}{}^c$ given in  \eqref{C-comm}, has only vertical components for the index $c$, and thus $C_{ab}{}^c q_{cd}=0$.} this reduces to
\beq
\cL_{\ell}q_{ab}=N_{\ell ab}-2N_{(ab)\ell}+2 T_{\ell(ab)}\,.\label{Ll}
\eeq
Thus, imposing metricity $N_{abc}=0$ and torsionless $T_{ab}{}^c=0$ leads to a constraint on the metric itself: $\cL_{\ell}q_{ab}=0$. This means that if the metric is time-independent, we can always impose the Levi-Civita conditions. Conversely, we remark that on a general background either the torsion or the metricity absorb the Lie derivative of the metric, and one is free to choose to work with a connection that is either torsionless or metric-compatible, but not both simultaneously. 

This is different from the pseudo-Riemannian case, where these two conditions have no restrictions on the underlying geometry. To recap, there are two options, either we are on a time-independent background, and then we can select a torsion-free and metric compatible connection $D$, or we need to give up one of these conditions. We noted that torsion-free or metricity can be separately imposed without leading to impositions on the background. Depending on the system under scrutiny, a choice could be more helpful than the other. Note that in either case the connection is still not entirely determined by the degenerate metric. Therefore, from an intrinsic Carrollian perspective, one must provide extra data on the Carrollian structure to introduce a connection.\\

\subsubsection{The Standard Carrollian Connection}\label{sec:CC}

As shown in section \ref{rigg}, for a null hypersurface embedded in a higher-dimensional manifold, a distinguished Carrollian connection is fixed by requiring it to coincide with the projection of the ambient Levi-Civita connection. From an ambient viewpoint, this object is known as the rigged connection. From an intrinsic perspective on a null manifold, however, we deliberately avoid this terminology, as it presupposes reference to an embedding (and in particular to a rigging structure). We therefore refer to this connection as the "standard Carrollian connection", with the understanding that this is identical to the induced rigged connection, when the null manifold is embedded in an ambient space.

Since the projection is a linear map, it automatically implies that the torsion must vanish -- as we will demonstrate in \eqref{tf}. Therefore, setting $T_{ab}{}^c=0$ in \eqref{Ll} selects the connection satisfying 
\beq
2\theta_{ab}=\cL_{\ell}q_{ab}=N_{\ell ab}-2 N_{(ab)\ell}\,,
\eeq
where we recalled the definition of the expansion tensor $\theta_{ab}$, \eqref{theta}. 

If one further assumes that $N_{abc}$ satisfies
\beq
N_{\ell ab}=0\qquad  
 q_d{}^b q_e{}^c N_{a b c}=0 \qquad N_{[ab]\ell}=0\,,\label{Ncond}
\eeq
one gets
\beq
N_{ab\ell}=N_{(ab)\ell}=-\theta_{ab}\,.
\eeq
Using $N_{a\ell\ell}=0$, and repeatedly applying the decomposition \eqref{deco}, one has
\beq
N_{abc}=N_{ab\ell}k_c + q_c{}^d N_{a\ell d} k_b + q_c{}^d q_b{}^e N_{a e d}=N_{ab\ell}k_c + N_{a c \ell} k_b=-\theta_{ab}k_c-\theta_{ac}k_b\,.
\eeq

This set of conditions for $N_{abc}$ is naturally realized when the Carrollian manifold is embedded in a pseudo-Riemannian bulk (as proven in \eqref{ppn} later), leading to the specific Carrollian connection satisfying torsionless and "minimal non-metricity", that is,
\beq
D_a q_{bc}=-k_b\theta_{ac}-k_c\theta_{ab}\,.
\eeq
We refer to this condition as minimal non-metricity since it relates some of the connection symbols to the expansion. \\

Given this we can derive how the connection acts on $\ell$,\footnote{This is the null analogue of the Brown-York identity for non-degenerate hypersurfaces $\theta_{\mu\nu}=\frac12 \cL_{n}g_{\mu\nu}=2\nabla_{(\mu}n_{\nu)}$, which requires the connection to be Levi-Civita and the metric non-degenerate such that $n_\mu=g_{\mu\nu}n^\nu$.}
\beq
D_a(\ell^b q_{bc})=0 \quad \Rightarrow \quad  \theta_{ac}=2q_{b(a} D_{c)}\ell^b\,,
\eeq
where we used the identities $\ell^a k_a=1$ and $\ell^a\theta_{ab}=0$. From this we can derive how the connection acts on $\ell$ up to terms proportional to $\ell$ itself:
\beq
D_a\ell^b=\theta_a{}^b+\omega_a\ell^b\,,
\eeq
where $\omega_a$ is unspecified, and $\theta_a{}^b=\theta_{ac}q^{cb}$, since $\theta_{ab}$ is horizontal. 

Thus, introducing new variables for the vertical and horizontal components of $\omega_a$,
\beq
\omega_a=\omega_\ell k_a +q_a{}^b\omega_b=\kappa k_a+\pi_a\,, 
\eeq
we see that the first pieces of the connection that are not dictated by the geometric data $\ell,q,k$ of the Carrollian structure are
\beq
\ell^a D_a\ell^b k_b=\kappa\,, \qquad q_c{}^aD_a\ell^bk_b=\pi_c\,.
\eeq

The remaining unspecified connection symbols are discussed observing that
\beq
0=D_a (\ell^b k_b)\,,\quad \Rightarrow \quad -\omega_a=\ell^b D_a k_b\,.
\eeq
this means that 
\beq
D_a k_b=-\omega_a k_b +J_{ab}\,,
\eeq
where $J_{ab}$ is an unspecified tensor satisfying $J_{a\ell}=0$.

We can constrain this tensor further, observing that \eqref{d-k} together with the torsionless condition imply
\beq
D_{[a} k_{b]}=\pa_{[a} k_{b]}=-k_{[a} \varphi_{b]}-\frac12 \varpi_{ab}\,.
\eeq
This imposes
\beq
J_{[ab]}=\omega_{[a}k_{b]}-k_{[a} \varphi_{b]}-\frac12 \varpi_{ab}\,,\label{Jvort}
\eeq
which, contracted with $\ell^a$, gives
\beq
J_{\ell b}=-\omega_b + \omega_{\ell} k_b-\varphi_b=-\pi_b-\varphi_b\,.
\eeq
Since this piece of $J_{ab}$ does not contain new independent data, we decompose $J_{ab}$ introducing the  horizontal tensor $\btheta_{ab}$ (with thus $\ell^a\btheta_{ab}=0=\btheta_{ab}\ell^b$),
\beq
J_{ab}=k_a J_{\ell b}+\btheta_{ab}=-k_{a}(\pi_b+\varphi_b)+\btheta_{ab}\,.\label{bthe}
\eeq
While the symmetric part of $\btheta_{ab}$ is left undetermined, from \eqref{Jvort} the skew part contains the vorticity
\beq
\btheta_{[ab]}=-\frac12 \varpi_{ab}\,. \label{btheta_[ab]}
\eeq
We have finally obtained how the covariant derivative acts on $k_a$
\beq
D_a k_b=-\omega_a k_b-k_a(\pi_b+\varphi_b)+\btheta_{ab}\,.
\eeq
Therefore, the remaining pieces of the connection that are not dictated by the Carrollian structure are
\beq
q_a{}^c D_{(c} k_{d)} q_b{}^d=\btheta_{(ab)}\,. 
\eeq\\

This concludes the construction of the standard Carrollian connection. To recap, we have found that the connection acts on the geometric data as
\beq\label{Dq}
D_a q_{bc}&=&-k_b\theta_{ac}-k_c\theta_{ab}\\
D_a \ell^b &=&\theta_a{}^b+\omega_a\ell^b \label{Dell} \\
D_a k_b&=& -\omega_a k_b-k_a(\pi_b+\varphi_b)+\btheta_{ab}\,.\label{Dk}
\eeq
In these expressions, the quantities
\beq
\omega_a = \kappa k_a + \pi_a \quad \text{and} \quad \btheta_{(ab)} \label{omtbar}
\eeq
encode the undetermined components of the connection and are independent of the ruled Carrollian structure $(\ell, q, k)$. They therefore represent external data that must be specified in order to fully define the Carrollian connection.

The special case in which these quantities vanish is often referred to as the metric hypersurface connection. Although this connection does not satisfy metricity, its symbols are entirely determined by the degenerate metric $q$. As we will see in section \ref{rigg}, the additional fields $\omega_a$ and $\btheta_{(ab)}$ naturally arise from the embedding of the Carrollian manifold into the higher-dimensional ambient spacetime.

It is crucial to emphasize that, in a generic background, the induced Carrollian connection is not completely fixed by the intrinsic geometric data defining the Carrollian structure. Furthermore, due to the degeneracy of the metric, such a connection cannot, in general, be metric-compatible.

The two equations \eqref{Dq} and \eqref{Dell} can be contracted with $\ell^a$ to give
\beq
\ell^a D_a q_{bc}=0\,\qquad \ell^a D_a\ell^b =\kappa \ell^b\,.
\eeq
Therefore, if $\kappa=0$, one recovers the equations \eqref{neweq}. This is a further justification for the condition imposed on $N_{abc}$ in \eqref{Ncond}.\\

\subsubsection{Excursus: Null Infinity}\label{422}

The standard connection we just described covers equally-well finite-distance embedded null hypersurfaces and null infinity in an asymptotically flat spacetime. With all the technical results derived, it is straightforward to cover this asymptotic case and recover the original description of the asymptotic equivalence class of connections due to Ashtekar \cite{Ashtekar:1981hw}.

An asymptotically flat metric displays a pole of order 2 in the radial direction as we approach the boundary. In covariant language, this means that there exists a function $\Omega$ such that the boundary is located at $\Omega\to 0$, and its conformal completion is then defined by stripping off this divergence. That is, null infinity ${\cal I}$ is defined as the boundary of the auxiliary manifold with conformal metric $\hat g_{\mu\nu}=\Omega^2 g_{\mu\nu}$. 

At asymptotic infinity, it turns out that the boundary degenerate metric and Carrollian vector field are background structures. Then, the ruled Carrollian structure can be taken to be as simple as possible: one can choose the vector field to be simply $\pa_u$, where $u$ is the null time at the boundary, and the metric to be completely time independent, such that the expansion tensor is zero. The Ehresmann connection can be further chosen to be an exact one form, such that both the acceleration and the vorticity vanish. Moreover, the bulk extension in the vicinity of the boundary of the Carrollian vector field and Ehresmann connection can be tuned such that $\omega_a=0$. While this is not the most general treatment, it allows us to reduce equations (\ref{Dq}-\ref{Dk}) to 
\beq\label{ccscri}
D_aq_{bc}=0 \qquad D_a\ell^b=0\qquad D_a k_b=\btheta_{ab}\,,
\eeq
with $\btheta_{[ab]}=0$. Without entering into details, not only the tensor $\btheta_{ab}$ cannot be set to zero arbitrarily, but it actually encodes the asymptotic shear 
\beq\label{nn}
D_a k_b=\btheta_{ab}=\frac12 C_{AB}\delta^A_a\delta^B_b\,,
\eeq
whose time derivative, the news tensor $N_{AB}=\pa_u C_{AB}$, encapsulate the gravitational radiation profile reaching null infinity. Note that, while in general $\btheta_{ab}$ has a trace, the equivalence class of connections introduced by Ashtekar \cite{Ashtekar:1981hw} is needed to eliminate this trace, singling out the radiative data, that is, the traceless shear. Indeed, the trace can always be set to
zero with a choice of conformal factor in the compactification (which means it is a pure gauge datum), hence what matters is only the
equivalence class of connections defined by Ashtekar. Therefore, the Carrollian connection \eqref{ccscri} coincides with Ashtekar’s connection (see in particular the discussion on page 5 of \cite{Ashtekar:2024bpi}). This is to be expected, since for a non-expanding null hypersurface, such as ${\cal I}$, the induced connection is unique.

This confirms that $\btheta_{(ab)}$ contains extrinsic data to the null hypersurface, parametrizing the independent pieces in the Carrollian connection.\\

\subsubsection{Levi-Civita-Carroll Covariant Derivative}\label{423}

While, as we just saw, introducing a connection on the Carrollian manifold $\N$ is a subtle point, given the horizontal frame $\left(e_A,e^A \right)$ one can introduce a notion of horizontal covariant derivative, which is metrical and acts on horizontal tensors in a canonical way. 

To do so, we first introduce the Christoffel-Carroll symbols, denoted $\CGamma^A_{BC}$, defined in the same manner as the standard Christoffel symbols but using the spatial metric $q_{AB}$ and the horizontal frame,
\begin{align}
\CGamma^A_{BC} := \frac{1}{2} q^{AD}\bigg( e_B[ q_{DC}] +e_C[ q_{BD}] - e_D [q_{BC}] \bigg)\,. \lb{Chris-Car}
\end{align}
The Levi-Civita-Carroll connection associated with these symbols is torsion-free, $\CGamma^A_{BC} = \CGamma^A_{CB}$, by definition. We then define the Levi-Civita-Carroll covariant derivative $\sD_A$ which acts on a horizontal tensor $T = T^A{}_B e_A \otimes e^B$ as 
\begin{align}\label{hder}
\sD_A T^B{}_C = e_A [T^B{}_C ]+ \CGamma^B_{DA} T^D{}_C - \CGamma^D_{CA} T^B{}_D\,.
\end{align}
This can straightforwardly be generalized to a tensor of any rank. Its defining property is that it is compatible with the  metric $q_{AB}$, that is, $\sD_C q_{AB} =0$. Therefore, this connection is torsion-free and metric-compatible. While it is often a useful tool to employ, its drawback is that it acts exclusively on horizontal tensors. Note that we deliberately call this the Levi-Civita-Carroll derivative to not confuse it with the horizontal projection of the general Carrollian covariant derivative introduced above, whose connection symbols are $\Gamma_{ab}^c$.\\

Indeed, one can relate the Levi-Civita-Carroll connection with the connection on the entire Carrollian structure $\Gamma_{ab}^c$ by projecting the latter to the horizontal subspace. To do so, we first recall the identities
\beq\label{useid}
q_{ab}=q_{AB}e_a{}^Ae_b{}^B\qquad q_a{}^b=e_a{}^Ae_A{}^b\qquad q_b{}^e e_e{}^F=e_b{}^F \qquad e_A{}^a\pa_a[ \ ]=e_A[ \ ]\,,
\eeq
and the property \eqref{Dq}
\beq
D_a q_{bc}=-k_b\theta_{ac}-k_c\theta_{ab}\,.
\eeq
Since $k_be_B{}^b=0$, the projection of this equation to the horizontal subspace vanishes
\beq
e_A{}^a e_B{}^be_C{}^c D_a q_{bc}=0\,.
\eeq
On the other hand, we have
\beq
e_A{}^a e_B{}^be_C{}^c D_a q_{bc}=e_A{}^a D_a q_{BC}-q_{bc} e_A{}^a D_a (e_B{}^be_C{}^c)=e_A[q_{BC}]-q_{bc} e_A{}^a  D_a (e_B{}^be_C{}^c)
\eeq
in which we used that $q_{BC}$ is a scalar with respect to $D_a$, and thus $D_a q_{BC}=\pa_a q_{BC}$.

Putting things together, we have
\begin{equation}
    \begin{aligned}
      0&=e_A[q_{BC}]-q_{bc} e_A{}^a  D_a (e_B{}^be_C{}^c)\\
&=e_A[q_{BC}]-q_{DC}e_b{}^D e_A{}^a (\pa_a e_B{}^b+\Gamma_{ad}^b e_B{}^d)-q_{BD}e_c{}^D e_A{}^a   (\pa_a e_C{}^c+\Gamma^c_{ad}e_C{}^d)\\
&=\gamma_{AB}^Dq_{DC}+\gamma_{AC}^Dq_{DB}-q_{DC}e_b{}^D e_A[e_B{}^b]-q_{DC} \Gamma_{AB}^D-q_{BD}e_c{}^D e_A[e_C{}^c]-q_{BD}  \Gamma^D_{AC} \\
&= q_{DC}\left( \gamma^D_{AB} - \Gamma^D_{AB} - e_A[e_B{}^b]e_b{}^D \right) + q_{DB}\left( \gamma^D_{AC} - \Gamma^D_{AC} - e_A[e_C{}^b]e_b{}^D \right)\,,  
    \end{aligned}
\end{equation}
where we introduced the notation $\Gamma^A_{BC}=e_a{}^A \Gamma^a_{bc}e_B{}^be_C{}^c$.
Taking the cyclic combination $ABC+BCA-CAB$,
we obtain
\beq\label{step}
2q_{CD}\left(\gamma^D_{AB}-\Gamma^D_{AB}-e_b{}^De_{(A}[e_{B)}{}^b]\right)-2q_{AD}e_b{}^De_{[B}[e_{C]}{}^b]-2q_{BD}e_b{}^De_{[A}[e_{C]}{}^b]=0\,.
\eeq
The last two terms vanish due to the Carrollian structure constants \eqref{C-comm}. Indeed
\beq
2e_b{}^De_{[B}[e_{C]}{}^b]=e_b{}^D[e_{B},e_{C}]^b=e_b{}^D \varpi_{BC}\ell^b=0\,,
\eeq
since $\ell^b e_b{}^B=0$. Therefore, from \eqref{step} we derive
\begin{align}\label{gG}
\gamma^C_{AB} = \Gamma^C_{AB} + e_A[e_B{}^a]e_a{}^C\,.
\end{align}
This explains how to relate the Levi-Civita-Carroll derivative with the horizontal projector of the Carrollian connection $D_a$. Note that \eqref{gG} turns out to be the well-known relationship between frames
\beq\label{de}
e_B{}^a \sD_A e_a{}^C=\Gamma^C_{AB}\,,
\eeq
Nonetheless, we preferred to offer an explicit derivation, which highlights the similarities and differences between the Levi-Civita-Carroll and the general Carrollian connections.\\

It is instructive to reconvert \eqref{de} to abstract index notation. To do so, here and in the following, we introduce a useful notation for the horizontal projector of tensors. For any tensor such as $T_b{}^a$, we define
\begin{align}
\overline{T}_b{}^a = q_b{}^c T_c{}^d q_d{}^a\,.
\end{align}
While it is in principle not needed, this compactly keeps track of the horizontal parts of tensors in lengthy computations. It is important to state how this notation is applied to derivations. We define the horizontal projection of the covariant derivative to be
\begin{align}
\overline{D}_a T_b{}^c = q_a{}^d q_e{}^cq_b{}^f D_d T_f{}^e\,,\label{ol}
\end{align}
and its straightforward generalization to a tensor of any rank. This equally applies to the partial derivative.\\

Using that $e_a{}^A f_A=q_a{}^b f_b=e_a{}^A e_A{}^b f_b$ for any $1$-form $f_A$, we can contract \eqref{gG} with the frames. On the LHS, this gives the horizontal tensor\footnote{Similarly, one has $e_c{}^C \gamma^c_{ab} e_A{}^a e_B{}^b =\gamma^C_{AB}$. Note that we do not need to use the overline notation for $\gamma_{ab}^c$, as this tensor is completely horizontal by definition, and thus $\overline{\gamma}_{ab}^c=\gamma_{ab}^c$.}
\beq
\gamma^c_{ab}=e_C{}^c \gamma^C_{AB}e_a{}^A e_b{}^B\,.
\eeq
Then, we obtain
\beq\label{gGbar}
\gamma^c_{ab}=\overline{\Gamma}^c_{ab} + e_b{}^B\overline{\pa}_a e_B{}^c\,,
\eeq
where we just employed our horizontal notation $\overline{\Gamma}^c_{ab}=q_f{}^c\Gamma^f_{de}q_a{}^dq_b{}^e$.
Furthermore, using \eqref{useid},
\beq
e_b{}^B\overline{D}_a e_B{}^c=e_b{}^B q_a{}^dq_f{}^c \pa_de_B{}^f+e_b{}^B q_a{}^d q_f{}^c \Gamma_{dg}^fe_B{}^g=e_b{}^B \overline{\pa}_a e_B{}^c+\overline{\Gamma}_{ab}^c\,,
\eeq
we eventually find
\beq\label{tensg}
\gamma^c_{ab}=e_b{}^B\overline{D}_a e_B{}^c\qquad e_B{}^c \overline{D}_{a}e_b{}^B=-\gamma^c_{ab}\,,
\eeq
which we stress are fully horizontal equations. 
This is the abstract-index version of \eqref{de}.\\

The interplay between the various derivatives discussed here is important. Each of the derivatives ($D_a$, $\overline{D}_a$, $\sD_A$) is useful in different contexts and computations, although the starting point, especially when the Carrollian structure is embedded, is the general connection $D_a$, which controls the temporal evolution as well as the horizontal transport. The relationship among these connections is straightforward, as we just reviewed.

A useful identity is that the Levi-Civita-Carroll divergence of a horizontal vector, $X = X^A e_A$, is given by
\begin{align}
\sD_A X^A = \frac{1}{\sqrt{q}} e_A \left[ \sqrt{q}X^A \right]\, ,
\end{align}
where $\sqrt{q}$ is the area element computed from the spatial metric $q_{AB}$ (see footnote \ref{detq}). This plays a role in the application of Stokes theorem to Carrollian structures, to which we now turn our attention.\\

\subsubsection{Volume Form and Stokes Theorem} \lb{sec:vol}
Given a Carrollian structure, the volume form can be chosen to be 
\begin{align}
\volN := k \wedge \volS\,, \qquad \text{where} \qquad \volS = \sqrt{q} e^1 \wedge e^2 \wedge ... \wedge e^{d}
\end{align}
is understood as the spatial volume form. The volume form obeys
\begin{align}
\volS = \iota_\ell \volN \qquad \text{and} \qquad \rd \volS = \theta \volN \,,
\end{align}
where we recalled the expansion $\theta$ from \eqref{expansion-scalar}. Furthermore, we can show the following formula for a function $f$ and a horizontal vector $X = X^Ae_A$, 
\begin{align}
\left( (\ell + \theta)[f] + (\sD_A + \ac_A)X^A \right)\volN = \rd \left( f\volS + X^A \vol_A \right)\,,
\end{align}
where we defined the contracted volume form $\vol_A := \iota_{e_A} \volN$.

This expression can be converted into abstract index notation. First, we compute\footnote{Here and in the following, whenever an equation requires multiple non-trivial steps, we will display them in square brackets on the LHS of the corresponding step.}
\begin{equation}
    \begin{aligned}
        (\sD_A + \ac_A)X^A &= (\sD_A + \ac_A) (e_a{}^A X^a)\\ 
\left[\sD_A X^a=e_A[X^a]\right] \ \ \ \ \ &= e_a{}^A e_A[X^a]+X^a \sD_A e_a{}^A +\ac_a X^a\\ 
\left[q_c{}^a \sD_A e_a{}^A=e_c{}^B \Gamma^A_{AB}\right] \ \ \ \ \ &= q_a{}^b\pa_bX^a+\Gamma^A_{AB} e_a{}^B X^a+\ac_a X^a\\ 
\left[\Gamma^A_{AB}e_c{}^B=e_a{}^A\Gamma^a_{bc}e_A{}^b=q_a{}^b\Gamma^a_{bc}\right] \ \ \ \ \ &= q_a{}^b \left( \pa_b X^a + \Gamma^a_{bc}X^c \right) +\ac_a X^a\\
&= (\overline{D}_a+\ac_a)X^a\,,
    \end{aligned}
\end{equation}
where $\ac_A e_a{}^A=\ac_a$, since $\ell^a\ac_a=0$. In this derivation, our previous result \eqref{de} comes in handy. Using moreover that 
\beq
X^A \nu_A=X^a e_a{}^A \iota_{e_A{}^b\pa_b}\nu_{\cN}=X^a q_a{}^b \iota_{b}\nu_{\cN}=X^a\iota_a \nu_{\cN}=X^a\nu_a\,,
\eeq
we then have
\begin{align}
\left( (\ell + \theta)[f] + (\overline{D}_a + \ac_a)X^a \right)\volN = \rd \left( f\volS + X^a \vol_a \right)\,.
\end{align}

The Stokes theorem therefore reads
\begin{align}
\int_\N \left( (\ell + \theta)[f] + (\sD_A + \ac_A)X^A \right)\volN =\int_{\pa \N} \left( f\volS + X^A \vol_A \right)\,, \lb{Stokes}
\end{align}
or, alternatively
\begin{align}
\int_\N \left( (\ell + \theta)[f] + (\overline{D}_a + \ac_a)X^a \right)\volN =\int_{\pa \N} \left( f\volS + X^a \vol_a \right)\,.
\end{align}\\

\subsection{Curvature}\label{RiemannCur}

In this section, we will construct the curvature tensors associated with the Carrollian connection $D_a$. We will do so for the standard Carrollian connection suitable for embeddings, introduced in subsection \ref{sec:CC}. 

Let us recall here that this Carrollian connection is determined by its action on the ruled Carrollian structure, that is (\ref{Dq}-\ref{Dk}),\footnote{The expansion tensor is, by definition, fully horizontal, $\ell^a \theta_{ab}=0$. Hence we will never use the notation $\overline{\theta}_{ab}$ to denote its horizontal projection. The overline on $\theta$ will always refer exclusively to the independent tensor $\btheta_{ab}$ introduced in \eqref{bthe}, which is also horizontal. This avoids any possible confusion between the intrinsic horizontality of $\theta_{ab}$ and the distinct tensor $\btheta_{ab}$.}
\beq
D_a q_{bc}&=&-k_b\theta_{ac}-k_c\theta_{ab}\label{Dqab}\\
D_a \ell^b &=&\theta_a{}^b+\omega_a\ell^b\label{Dla}\\
D_a k_b&=& -\omega_a k_b-k_a(\pi_b+\varphi_b)+\btheta_{ab}\,.\label{Dka}
\eeq
We furthermore recall that, in these expressions, the quantities 
\beq
\omega_a=\kappa k_a+\pi_a \quad  \text{and} \quad \btheta_{(ab)}
\eeq
are external inputs, undetermined given the ruled Carrollian structure. 

While so far we refrained from explicitly writing the connection symbols associated with $D_a$, it is useful and instructive to do it now, in order to construct the curvature tensors. Using the equations above, one has
\beq
\eqref{Dqab}:& \qquad \pa_a q_{bc}-2\Gamma_{a(b}^dq_{c)d}=-2k_{(b}\theta_{c)a}\\
\eqref{Dla}:& \qquad \pa_a \ell^b+\Gamma_{ac}^b \ell^c=\theta_a{}^b+\omega_a\ell^b\\
\eqref{Dka}:& \qquad \pa_a k_b-\Gamma_{ab}^c k_c= -\omega_a k_b-k_a(\pi_b+\varphi_b)+\btheta_{ab}\,.
\eeq
This implies
\beq
\Gamma^d_{a(b}q_{c)d}&=&\frac12(\pa_aq_{bc}+k_b\theta_{ac}+k_c\theta_{ab})\label{Gq}\\
\Gamma_{a\ell}^b&=&-\pa_a\ell^b+\theta_a{}^b+\omega_a \ell^b\\
\Gamma_{ab}^ck_c&=&(\pa_a+\omega_a)k_b+k_a(\pi_b+\varphi_b)-\btheta_{ab}\,.\label{gk}
\eeq

From these equations, we can derive the connection symbols using its projections
\beq
\Gamma^a_{bc}=\Gamma^d_{bc}k_d\ell^a+\Gamma^d_{bc}q_d{}^a\,.\label{gab}
\eeq
To compute the last term, we start from \eqref{Gq}, and take the cyclic permutation of indices $abc+bca-cab$, in the same manner one derives the usual Christoffel symbols. 
This gives
\beq
\Gamma_{ab}^d q_{cd}=\frac12(\pa_a q_{bc}+\pa_bq_{ac}-\pa_cq_{ab})+k_c\theta_{ab}\,,\label{ggq}
\eeq
which, multiplying with $q^{ce}$, becomes\footnote{Note, this indeed gives back \eqref{ggq} when contracted with $q_{ce}$: 
\begin{equation}
    \begin{aligned}
        \Gamma_{ab}^d q_{d}{}^c q_{ce}&=\frac{q_e{}^{d}}{2}(\pa_a q_{bd}+\pa_bq_{ad}-\pa_dq_{ab})\\
&=\frac{1}{2}(\pa_a q_{be}+\pa_bq_{ae}-\pa_eq_{ab})-\frac{k_e\ell^{d}}{2}(\pa_a q_{bd}+\pa_bq_{ad}-\pa_dq_{ab})\\
&=\frac{1}{2}(\pa_a q_{be}+\pa_bq_{ae}-\pa_eq_{ab})+k_e \theta_{ab}\,.
    \end{aligned}
\end{equation}}
\beq
\Gamma_{ab}^d q_{d}{}^e=\frac{q^{ed}}{2}(\pa_a q_{bd}+\pa_bq_{ad}-\pa_dq_{ab})\,.
\eeq
Putting this together with \eqref{gk} and using \eqref{gab}, we thus obtain
\beq\label{CC}
\Gamma^a_{bc}=\left((\pa_b+\omega_b)k_c+k_b(\pi_c+\varphi_c)-\btheta_{bc}\right)\ell^a+\frac{q^{ad}}{2}(\pa_b q_{cd}+\pa_cq_{bd}-\pa_dq_{bc})\,.
\eeq
This is an important result, it expresses the general Carrollian connection symbols in terms of the internal and external geometric data of the Carrollian structure.\\

From this, we can derive and confirm some identities
\beq
&\Gamma^a_{[bc]}=0 \qquad
\Gamma^a_{\ell\ell}=-(\ell-\kappa)[\ell^a] \qquad
\Gamma^b_{\ell(a}q_{c)b}=\frac12 \ell[q_{ac}]&\\
&\Gamma^b_{a\ell}k_b=\ell^b\pa_a k_b+\omega_a\qquad
\Gamma_{ab}^cq_c{}^b=\frac12 q^{bc}\pa_a q_{bc}\qquad 
\Gamma_{a\ell}^b q_b{}^c=\theta_a{}^c-q_{b}{}^{c}\pa_a \ell^b\,,&
\eeq
useful for the computation of the Riemann tensor below.
The latter is related to the connection symbols as in \eqref{Riem}, which is a general equation applying also to our Carrollian structure. We thus have
\beq
R^a{}_{bcd}=e_c [\Gamma^a_{db}]-e_d[\Gamma^a_{cb}]+\Gamma^e_{db}\Gamma^a_{ce}-\Gamma^e_{cb}\Gamma^a_{de}-C_{cd}{}^e\Gamma^a_{eb}\,.
\eeq
We will confine our attention to the abstract index notation in the coordinate basis, where $e_a=\pa_a$ and $C_{cd}{}^e=0$, such that
\beq
R^a{}_{bcd}=\pa_c \Gamma^a_{db}-\pa_d\Gamma^a_{cb}+\Gamma^e_{db}\Gamma^a_{ce}-\Gamma^e_{cb}\Gamma^a_{de}\,.
\eeq
Although a full, manifestly covariant decomposition of this tensor lies beyond the scope of this review, its essential features can already be appreciated through the projections and limiting cases derived below.

Using its definition, and the properties of our connection recapitulated in (\ref{Dqab}-\ref{Dka}), one has
\begin{equation}\label{Rl}
\begin{aligned}
R^e{}_{\ell ab} =& [D_a, D_b] \ell^e\\
=& D_a(\theta_b{}^e+\omega_b\ell^e)-D_b(\theta_a{}^e+\omega_a\ell^e)\\
=&2D_{[a}\theta_{b]}{}^e+2D_{[a} \omega_{b]} \ell^e+\omega_b (\theta_a{}^e+\omega_a\ell^e)-\omega_a (\theta_b{}^e+\omega_b\ell^e)\\
=&2(D_{[a}-\omega_{[a})\theta_{b]}{}^e+2 D_{[a}\omega_{b]}\ell^e\,.
\end{aligned}
\end{equation}
Similarly, we derive
\begin{equation}\label{Rk}
\begin{aligned}
R^a{}_{bcd}k_a =& -[D_c, D_d] k_b\\
=& 
D_d ( \btheta_{cb} - \omega_c k_b - k_c (\pi_b +\ac_b) ) - (c \leftrightarrow d ) \\
=& D_d \btheta_{cb} - k_b D_d \omega_c - \omega_c ( \btheta_{db} - \omega_d k_b - k_d (\pi_b +\ac_b) ) \\
& - (\pi_b +\ac_b)( \btheta_{dc} - \omega_d k_c - k_d (\pi_c +\ac_c) )  - k_c D_d (\pi_b+\ac_b)- (c \leftrightarrow d ) \\
\left[\eqref{btheta_[ab]}\right]\quad =& - 2 ( D_{[c} + \omega_{[c} )\btheta_{d]b} + 2 D_{[c} \omega_{d]} k_b - 2 k_{[c} ( D_{d]} +\pi_{d]} +\ac_{d]} ) (\pi_b+\ac_b) - (\pi_b+\ac_b) \vor_{cd}\,.
\end{aligned}
\end{equation}

Eventually, one can also gather the useful identity
\begin{equation}\label{Rq}
\begin{aligned}
R^e{}_{bcd}q_{ae}+R^e{}_{acd}q_{be} =& -[D_c, D_d] q_{ab}\\
=& -D_c(-k_a \theta_{bd}-k_b\theta_{ad})+D_d(-k_a \theta_{bc}-k_b\theta_{ac}) \\
= & 2k_a ( D_{[c} - \omega_{[c})\theta_{d]b}+ 2k_b ( D_{[c} - \omega_{[c})\theta_{d]a} - 2\theta_{a[c}\btheta_{d]b} - 2\theta_{b[c}\btheta_{d]a} \\
 & + 2(\pi_a+\ac_a) \theta_{b[c} k_{d]}+ 2(\pi_b+\ac_b) \theta_{a[c} k_{d]}\,.
\end{aligned}
\end{equation}

\subsubsection{Riemann-Carroll Tensor}\label{curva}
We have already defined the Levi-Civita-Carroll covariant derivative $\sD_A$ which acts on horizontal tensors according to \eqref{hder} and is compatible with the spatial metric $q_{AB}$. We can use this covariant derivative to define an analog of the Riemann curvature tensor. The Riemann-Carroll tensor $\sR^A{}_{BCD}$ is defined via the commutator of the Levi-Civita-Carroll covariant derivative as follows,
\begin{align}
[\sD_C, \sD_D] X^A = \sR^A{}_{BCD} X^B + \vor_{CD} \ell[X^A]\,. \lb{Rie-Car}
\end{align}
The vorticity term $\vor_{AB}$ appears following the fact that the Carrollian frames are not holonomic and obey \eqref{C-comm}. By expressing the LHS in terms of the Christoffel-Carroll symbols $\gamma^A_{BC}$ given in \eqref{Chris-Car}, the Riemann-Carroll tensor is given in components by\footnote{Note that $C_{AB}^C$ is identically zero, and thus here we simply applied \eqref{Riem}.}
    \begin{align}
\sR^A{}_{BCD} = e_C [ \gamma^A_{BD} ] -  e_D[ \gamma^A_{BC}]+ \gamma^A_{CE}\gamma^E_{BD} - \gamma^A_{DE}\gamma^E_{BC}\,. \lb{RC}
\end{align}

This tensor shares some standard properties of the conventional Riemann tensor, including the antisymmetry of the last two indices and the algebraic Bianchi identity,
\begin{align}
\sR^A{}_{BCD} = -\sR^A{}_{BDC}\,, \qquad \text{and} \qquad \sR^A{}_{[BCD]} = 0\,. \lb{RC-property}
\end{align}
It is nevertheless important to note that the Riemann-Carroll tensor, with the first index lowered $\sR_{ABCD} = q_{AE} \sR^E{}_{BCD}$, is not antisymmetric in its first two indices. This fact follows from 
\begin{equation}
[\sD_C, \sD_D]q_{AB} = -\sR_{ABCD} - \sR_{BACD} + \vor_{CD}\ell[q_{AB}]\,.
\end{equation}
Since $[\sD_C, \sD_D]q_{AB} = 0$ from the metric compatibility of $\sD_A$, we arrive at 
\begin{align}
\sR_{(AB)CD} = \vor_{CD} \theta_{AB}\,. 
\end{align}
where we recalled \eqref{adtheta}, that is, $\theta_{AB} = \frac12 \ell[q_{AB}]$.\\

Interestingly, due to this property, the Ricci-Carroll tensor, $\sR_{AB} := \sR^C{}_{ACB}$, is not entirely symmetric, and its antisymmetric components are given by the vorticity,
\begin{align}
\sR_{[AB]} = \tfrac{1}{2}\theta \vor_{AB}\,. \lb{Ric-anti}
\end{align}
To prove this, one uses the contracted version of \eqref{RC}, together with $\gamma^A_{AB}=e_B[\ln \sqrt{q}]$ and \eqref{thC}.

We would like to write \eqref{RC} in an abstract index notation by defining the horizontal tensor $\sR^a{}_{bcd} := e_A{}^a \sR^A{}_{BCD} e_b{}^B e_c{}^C e_d{}^D$. We have 
\begin{equation}
\begin{aligned}
\sR^a{}_{bcd} =e_A{}^a e_b{}^B e_d{}^D \overline{\pa}_c  \gamma^A_{BD}+ \gamma^a_{ce}\gamma^e_{bd} - (c \leftrightarrow d)\,. \label{RCbar-0}
\end{aligned}
\end{equation}
To manipulate the first term, we recall \eqref{gGbar}, which can be alternatively expressed as\footnote{We emphasize that $\overline{\pa}_a e_b{}^A=q_a{}^c q_b{}^d \pa_c e_d{}^A\neq q_a{}^c  \pa_c e_b{}^A$. One should be careful with this.}
\begin{align}
 e_b{}^A \overline{\pa}_a e_A{}^c  =  \gamma^c_{ab} - \overline{\Gamma}^c_{ab} = -  e_A{}^c \overline{\pa}_a e_b{}^A\,.\label{dbg}
\end{align}
With the help of the Leibniz rule, we show
\begin{equation}
    \begin{aligned}
       \overline{\pa}_c  \gamma^A_{BD}  e_A{}^a e_b{}^B e_d{}^D &= \overline{\pa}_c \left( \gamma^e_{fg} e_e{}^A e_B{}^f e_D{}^g \right)e_A{}^a e_b{}^B e_d{}^D \\
&= \overline{\pa}_c  \gamma^a_{bd} + \gamma^e_{bd} e_A{}^a \overline{\pa}_c e_e{}^A  + \gamma^a_{ed} e_b{}^B\overline{\pa}_c e_B{}^e + \gamma^a_{be} e_d{}^D\overline{\pa}_c e_D{}^e \\
\left[\eqref{dbg}\right]\quad &= \overline{\pa}_c  \gamma^a_{bd} - \gamma^e_{bd} \left(\gamma^a_{ce} - \overline{\Gamma}^a_{ce} \right) + \gamma^a_{ed} \left(\gamma^e_{cb} - \overline{\Gamma}^e_{cb} \right) + \gamma^a_{be} \left(\gamma^e_{cd} - \overline{\Gamma}^e_{cd} \right) \\
\left[\overline{\Gamma}^c_{ab}=q_f{}^c\Gamma^f_{de}q_a{}^dq_b{}^e\right]\quad&= q_c{}^h( \pa_h  \gamma^e_{fg}  + \Gamma^e_{hi}\gamma^i_{fg}-\Gamma^i_{hf}\gamma^e_{ig} - \Gamma^i_{hg}\gamma^e_{if}) q_e{}^a q_b{}^f q_d{}^g \\
& \ \ \ - \gamma^e_{bd}\gamma^a_{ce} + \gamma^a_{ed}\gamma^e_{cb} + \gamma^a_{be}\gamma^e_{cd}  \\
\left[\eqref{ol} \right]\quad &= \overline{D}_c \gamma^a_{bd}- \gamma^e_{bd}\gamma^a_{ce} + \gamma^a_{ed} \gamma^e_{cb} + \gamma^a_{be}\gamma^e_{cd}\,. \label{forcomment}
    \end{aligned}
\end{equation}
We note that while $\Gamma^a_{bc}$ are connection symbols on the Carrollian manifold, $\gamma^a_{bc}$ is a tensor (see \eqref{tensg}) and thus can be acted upon with the covariant derivative $D_a$. 

Finally, substituting the above result in \eqref{RCbar-0}, we arrive at the final result
\begin{align}\lb{RCbar}
\sR^a{}_{bcd} = \overline{D}_c \gamma^a_{bd} - \overline{D}_d \gamma^a_{bc} +\gamma^a_{ed}\gamma^e_{cb}-\gamma^a_{ec}\gamma^e_{bd}\,,
\end{align}
expressing the Riemann-Carroll tensor with abstract index notation. One notes that, by design, this tensor is readily horizontal, and given that it contains the lift of the Levi-Civita-Carroll connection, it does not capture curvature involving the null temporal direction.\\

\subsubsection{Horizontal Curvature}

The Riemann-Carroll tensor just derived can be compared to the horizontal projector of the curvature tensor $R^a{}_{bcd}$. 

To do so, we start by noticing that
\beq
-[D_c,D_d]e_b{}^A=R^e{}_{bcd}e_e{}^A\,,
\eeq
contracted with $e_A{}^a$, gives
\beq\label{RDDe}
R^e{}_{bcd}q_e{}^a=-e_A{}^a [D_c,D_d]e_b{}^A\,.
\eeq
Projecting all the lower indices with $q_a{}^b$, we obtain the horizontal curvature tensor
\beq
\overline{R}^a{}_{bcd}=q_b{}^fq_c{}^gq_d{}^h R^e{}_{fgh} q_e{}^a=-e_A{}^a q_b{}^f q_c{}^g q_d{}^h 
[D_c,D_d]e_b{}^A\,.
\eeq

We wish to compute this tensor, and thus we need to evaluate $D_ae_b{}^A$. It is simpler to start with  $e_c{}^A D_a e_A{}^b$, whose decomposition (see \eqref{deco}) gives
\beq
e_c{}^A D_a e_A{}^b=q_a{}^d e_c{}^A D_d e_A{}^b +k_a e_c{}^A \ell^eD_e e_A{}^b\,.\label{is1}
\eeq
The second term can be evaluate as follows
\beq
e_c{}^A \ell^eD_e e_A{}^b=\ell[e_A{}^b]e_c{}^A+\ell^e\Gamma_{ed}^be_A{}^d e_c{}^A= [\ell,e_A]^be_c{}^A+q_c{}^aD_a\ell^b\,,\label{s1}
\eeq
where in the second equality we have added and subtracted $e_A{}^a\pa_a \ell^b e_c{}^A$ and used $e_c{}^A e_A{}^d =q_c{}^d$. For the $[\ell,e_A]$ commutator, using \eqref{C-comm}, we get
\beq
[\ell,e_A]^be_c{}^A=\varphi_A \ell^b e_c{}^A=\varphi_c \ell^b\,,
\eeq
which inserted back in \eqref{s1} gives
\beq
e_c{}^A \ell^eD_e e_A{}^b=\varphi_c \ell^b+q_c{}^aD_a\ell^b=\ell^b(\varphi_c+\pi_c)+\theta_c{}^b\,,
\eeq
where we used \eqref{Dla} and $q_a{}^b \omega_b=\pi_b$.
Therefore, we can process \eqref{is1} to get
\begin{equation}
    \begin{aligned}
        e_c{}^A D_a e_A{}^b&=q_a{}^d e_c{}^A D_d e_A{}^b +k_a (\ell^b(\varphi_c+\pi_c)+\theta_c{}^b)\\
\left[\delta_e^b=q_e{}^b+k_e\ell^b\right]\quad &= q_a{}^d e_c{}^A  q_e{}^b D_d e_A{}^e +q_a{}^d e_c{}^A k_e \ell^b D_d e_A{}^e +k_a (\ell^b(\varphi_c+\pi_c)+\theta_c{}^b)\\
\left[\eqref{tensg}\right]\quad &= \gamma^b_{ac}-q_a{}^d D_d k_e \ell^b q_c{}^e+k_a (\ell^b(\varphi_c+\pi_c)+\theta_c{}^b)\\
&=\gamma^b_{ac}-\btheta_{ac} \ell^b +k_a (\ell^b(\varphi_c+\pi_c)+\theta_c{}^b)\,,
    \end{aligned}
\end{equation}
where we used $e_c{}^A k_e D_d e_A{}^e=-e_c{}^A  D_d k_e e_A{}^e=-q_c{}^e D_d k_e$.

Since we want $D_a e_c{}^A$, we can process the LHS and derive
\begin{equation}
    \begin{aligned}
        e_c{}^A D_a e_A{}^b&=D_a q_c{}^b-e_A{}^b D_a e_c{}^A\\
\left[q_c{}^b=\delta_c^b-k_c\ell^b\right]\quad &=-k_c D_a\ell^b-D_a k_c \ell^b-e_A{}^b D_a e_c{}^A\\
\left[(\ref{Dla}-\ref{Dka})\right]\quad &=-k_c (\omega_a\ell^b+\theta_a{}^b)-\ell^b(\btheta_{ac}-\omega_ak_c-k_a(\varphi_c+\pi_c))-e_A{}^b D_a e_c{}^A\,.
    \end{aligned}
\end{equation}
Eventually, this equation and the previous one combine to give 
\beq
e_A{}^b D_a e_c{}^A&=&-k_c (\omega_a\ell^b+\theta_a{}^b)-\ell^b(\btheta_{ac}-\omega_ak_c-k_a(\varphi_c+\pi_c))-e_c{}^A D_a e_A{}^b\\
&=&-k_c (\omega_a\ell^b+\theta_a{}^b)-\ell^b(\btheta_{ac}-\omega_ak_c-k_a(\varphi_c+\pi_c))\\
&&-\gamma^b_{ac}+\btheta_{ac} \ell^b -k_a (\ell^b(\varphi_c+\pi_c)+\theta_c{}^b)\\
&=&-\gamma^b_{ac}-k_c \theta_a{}^b-k_a\theta_c{}^b\,,
\eeq
which, upon multiplying by $e_b{}^B$, gives our sought-for result
\beq\label{DeT}
D_ae_c{}^B=-\left(k_c\theta_a{}^b+k_a\theta_c{}^b+\gamma^b_{ac}\right) e_b{}^B\,.
\eeq

This is exactly what we needed to then evaluate 
\begin{equation}
\begin{aligned}
\overline{R}^a{}_{bcd}&=-e_A{}^aq_c{}^g q_d{}^h q_b{}^f [D_g,D_h]e_f{}^A\\
\left[q_b{}^f k_f=0 \right]\quad&=2e_A{}^aq_{[c}{}^g q_{d]}{}^h q_b{}^f[\theta_h{}^i e_i{}^A D_g k_f+\theta_f{}^i e_i{}^A D_g k_h+D_g(\gamma^i_{hf}e_i{}^A)] \\
\left[\eqref{Dka} \ , \ \eqref{btheta_[ab]}\right]\quad &= \overline{D}_c \gamma^a_{bd} - \overline{D}_d \gamma^a_{bc} +\gamma^a_{ed}\gamma^e_{cb}-\gamma^a_{ec}\gamma^e_{bd}+\theta_d{}^a\btheta_{cb}-\theta_c{}^a\btheta_{db}-\theta_b{}^a\varpi_{cd}\,.
\end{aligned}
\end{equation}
This is an important result, and, comparing it with \eqref{RCbar}, we can finally derive
\beq\label{RR}
\overline{R}^a{}_{bcd}=\sR^a{}_{bcd}+\theta_d{}^a\btheta_{cb}-\theta_c{}^a\btheta_{db}-\theta_b{}^a\varpi_{cd}\,.
\eeq
This is reminiscent of the Gauss equation for $\S$ -- see e.g. \eqref{HGauss} -- with $\theta_a{}^b$ and $\btheta_{ab}$ extrinsic curvatures, and the term involving $\varpi_{ab}$ controlling integrability of the horizontal distribution. It further shows that the Riemann-Carroll tensor captures the intrinsic spatial geometry and is insensitive to the null evolution and curvature, whereas the horizontal projection of the Riemann tensor still carries a notion of parallel transport in null time.\\

\subsubsection{Simplified Frameworks and Flatness}

\paragraph{Carrollian Cartesian Structure} Although not diffeomorphism-covariant, it is instructive to express these tensors in the analogue of the Cartesian coordinates for the ruled Carrollian structure. The strongest condition one can intrinsically impose is
\beq
\pa_a\ell^b=0 \qquad \pa_a q_{bc}=0\qquad \pa_a k_b=0\,,
\eeq
and thus $\varphi_a=0=\varpi_{ab}$. The connection symbols are then given by
\beq
\Gamma^a_{bc}=\left(\omega_bk_c+k_b\pi_c-\btheta_{bc}\right)\ell^a=\left(\kappa\k_bk_c+2k_{(b}\pi_{c)}-\btheta_{bc}\right)\ell^a \,.
\eeq
A straightforward computation yields
\begin{equation}
\begin{aligned}
R^a{}_{bcd} &= \pa_c \Gamma^a_{db}-\pa_d\Gamma^a_{cb}+\Gamma^e_{db}\Gamma^a_{ce}-\Gamma^e_{cb}\Gamma^a_{de} \\
& = 2\ell^a \left( -k_{[c}(\pa_{d]}-\omega_{d]}) \omega_b + k_b (\pa_{[c}-\omega_{[c}) \pi_{d]} - (\pa_{[c}+\omega_{[c})\btheta_{d]b} \right)\,.
\end{aligned}
\end{equation}
One remarks that this curvature is non-vanishing even in Cartesian coordinates, due to the extrinsic terms $\omega_a$ and $\btheta_{(ab)}$ in the Carrollian connection -- see discussion around \eqref{omtbar}.

\paragraph{Intrinsic Curvature} The opposite situation arises when the connection is entirely dictated by intrinsic data. This is reached imposing
\beq
\omega_a=0\qquad \btheta_{(ab)}=0 \,.
\eeq
The connection symbols are then given by
\beq\label{intcur}
\Gamma^a_{bc}=\left(\pa_bk_c+k_b\varphi_c-\btheta_{[bc]}\right)\ell^a+\frac{q^{ad}}{2}(\pa_b q_{cd}+\pa_cq_{bd}-\pa_dq_{bc}) \,.
\eeq
In this case, the curvature tensor becomes entirely a function of the intrinsic data, and expresses how the non-constancy of $k_a$, $\ell^a$, and $q_{ab}$ bend the Carrollian geometry. This situation is more akin to the usual pseudo-Riemannian one, in which the curvature tensor is by construction only a function of the intrinsic geometric data. We will not display the Riemann tensor in this case, as it does not acquire a particularly illuminating form.\\  

\paragraph{Covariant Flatness} We can now address under which conditions the Riemann tensor can be covariantly set to zero, thereby providing a definition of flat ruled Carrollian structures.  Suppose that all the extrinsic data ($\omega_a,\btheta_{(ab)}$) are zero. Then, the Riemann tensor depends on the ruled Carrollian data $k_a$, $\ell^a$, and $q_{ab}$, as one can directly infer from the connection symbols in \eqref{intcur}. A trivial Ehresmann connection $k$ (that is, $\rd k=0$) immediately leads to $\varphi_a=0=\varpi_{ab}$ (see \eqref{d-k}). The latter condition in turn implies  $\btheta_{[ab]}=0$, such that $\btheta_{ab}=0$. If we further require that $\theta_{ab}=0$, that is, $q_{ab}$ is time independent, then we see that all the contractions of the Riemann considered in (\ref{Rl}-\ref{Rq}) are identically zero.

Then, applying \eqref{deco} on all the indices, and using (\ref{Rl}-\ref{Rq}), we compute
\beq\label{RFLAT}
R^a{}_{bcd}=\overline{R}^a{}_{bcd}+q_e{}^aq_b{}^fq_c{}^mk_dR^e{}_{fmh}\ell^h-q_e{}^aq_b{}^fk_cq_d{}^mR^e{}_{fmh}\ell^h\,.
\eeq

Then, inserting $\theta_a{}^b=0$ in \eqref{DeT}
\beq
D_ae_c{}^B=-\left(k_c\theta_a{}^b+k_a\theta_c{}^b+\gamma^b_{ac}\right) e_b{}^B=-\gamma^b_{ac} e_b{}^B\,,
\eeq
and using \eqref{RDDe}, we can evaluate  the last two terms and gather
\begin{equation}
    \begin{aligned}
q_e{}^aq_b{}^fq_c{}^mR^e{}_{fmh}\ell^h&= -q_b{}^fq_c{}^m\ell^h e_A{}^a[D_m,D_h]e_f{}^A\\
&=q_b{}^fq_c{}^m\ell^he_A{}^a\left(D_m(\gamma^g_{hf}e_g{}^A)-D_h(\gamma^g_{mf}e_g{}^A)\right)\\
\left[e_g{}^Ae_A{}^a=q_g{}^a\,, \ell^h\gamma^g_{hf}=0\right] \ \ \  &= q_b{}^fq_c{}^m\ell^h\left(D_m\gamma^g_{hf}-D_h\gamma^g_{mf}\right)q_g{}^a\\
\left[\ell^hD_m\gamma^g_{hf}=-\gamma^g_{hf}D_m\ell^h=0\right] \ \ \ &=-q_g{}^aq_b{}^fq_c{}^m\ell^hD_h\gamma^g_{mf}\\
\left[D_h q_g{}^a=0\,, q_g{}^a\gamma^g_{mf}=\gamma^a_{mf}\right] \ \ &= -\ell^hD_h\gamma^a_{bc}\,.
    \end{aligned}
\end{equation}
Therefore, \eqref{RFLAT} becomes
\beq
R^a{}_{bcd}=\overline{R}^a{}_{bcd}-\ell^h D_h \gamma^a_{bc} k_d+\ell^h D_h \gamma^a_{bd} k_c\,.
\eeq

A flat ruled Carrollian structure can thus be obtained by separately requiring 
\beq
\overline{R}^a{}_{bcd}=0\qquad \text{and} \qquad D_\ell \gamma^a_{bc}=0\,.
\eeq
Note that the former condition, thanks to \eqref{RR}, can be recast as
\beq
\overline{R}^a{}_{bcd}=\sR^a{}_{bcd}=0\,.
\eeq
Thus, covariant flatness is reached by requiring that the degenerate metric on the base $q_{AB}$ is simultaneously flat in the usual Euclidean sense ($\sR^a{}_{bcd}=0$), but also, and importantly, it is also covariantly conserved along the temporal direction ($D_\ell \gamma^a_{bc}$). Note that this last equation is a tensorial covariant statement, as explained below \eqref{forcomment}. 

As we saw, flatness naturally depends on both intrinsic and extrinsic features -- a reflection of the fact that, unlike in the pseudo-Riemannian case, the ambient Levi-Civita connection does not induce a unique metric-compatible and torsionless connection on a null hypersurface. These aspects will be further explored in the next section. \\

This concludes our overview of the essential aspects of Carrollian geometry. We have outlined the construction of Carrollian structures, their compatible connections, and the corresponding curvature tensors, providing a coherent framework that parallels the pseudo-Riemannian case while revealing its profound departures. The degeneracy of the metric and the ensuing breakdown of the Levi-Civita theorem emerge as defining features, reshaping the very notion of geometry on null manifolds. So far, this review has offered a foundational perspective on the geometry of null manifolds, developed from an intrinsic standpoint, without embedding it in an ambient space. We now turn to the complementary viewpoint, where such structures are realized as hypersurfaces embedded in an ambient spacetime.

\newpage

\section{Embedding of Null Hypersurfaces and Rigging}\label{rigg}

In the previous sections, we introduced and explored Carrollian structures in depth, starting from their definition and progressing through the construction of general Carrollian connections, as well as the associated curvature tensors -- all defined intrinsically on the Carrollian manifold. As anticipated, our primary motivation is their application to  null hypersurfaces in general relativity. The purpose of this section is to embed the Carrollian structure into an ambient spacetime as a null hypersurface and to demonstrate how the intrinsic Carrollian geometry and connection introduced earlier naturally emerge from the extrinsic geometry of this embedding.

We now consider a Carrollian manifold $\N$ as a codimension-1 null hypersurface embedded in a $(d+2)$-dimensional Lorentzian spacetime. The intrinsic geometry of $\N$ discussed in section \ref{sec:modern}, as well as the Carrollian connection in section \ref{sec:connection}, can be induced from the spacetime geometry using the Mars-Senovilla rigging technique \cite{Mars:1993mj}. This method provides a rigorous, coordinate-independent geometric decomposition of spacetime along a hypersurface -- which may be null, timelike, or spacelike -- extending the Arnowitt-Deser-Misner (ADM) or Gauss-Codazzi formalism to general hypersurfaces. It also offers a consistent framework for handling null hypersurfaces, where the conventional ADM decomposition fails, that is, becomes singular. \\

The $(d+2)$-dimensional Lorentzian spacetime $\M$ is endowed with a Lorentzian metric $g = g_{\mu\nu} \rd x^\mu \otimes \rd x^\nu$ with  inverse $g^{-1} = g^{\mu\nu}\pa_\mu \otimes \pa_\nu$, and a Levi-Civita (that is, torsionless and metric-compatible) connection $\nabla$ -- as discussed in section \ref{421}. The rigging technique equips the $(d+1)$-dimensional hypersurface $\N$ with a rigging structure $(\N, k, n)$, consisting of a normal 1-form $n = n_\mu \rd x^\mu$ and a nowhere-vanishing rigging vector $k = k^\mu \pa_\mu$ that is transverse to $\N$ and tangent-bundle dual to the normal form, satisfying $\iota_k n = k^\mu n_\mu = 1$. In general, the norms of $n$ and $k$ are non-zero, and we define them respectively as
\begin{align}
2\rho_n := g^{-1}(n,n)=n_\mu g^{\mu\nu}n_\nu\,, \qquad \text{and}  \qquad 2\rho_k := g(k,k)=k^\mu g_{\mu\nu}k^\nu\,. 
\end{align}
Later, as we are interested in the case where $\N$ is null, we will impose $\rho_n \Neq 0$.  

The rigging structure, and thus $n_\mu$ and $k^\mu$, provides a notion of transversality to the hypersurface. A vector $X \in T\M$ and a 1-form $\omega \in \Omega^1(\M)$ are tangent to $\N$ when they satisfy $\iota_X n =X^\mu n_\mu \Neq 0$ and $\iota_k \omega = k^\mu \omega_\mu \Neq 0$, respectively. We highlight how this construction and the differential geometry behind is very similar in spirit to the way we constructed the projector and the Ehresmann connection in previous sections: the notion of transversality is perfectly described in terms of tangent and dual co-tangent bundles. While for spacelike and timelike hypersurfaces this can be made to coincide with the raising and lowering of indices using the metric, for null hypersurfaces one has to distinguish between the two procedures -- metric dual versus tangent-bundle dual. With these data, we define the rigged projector, which is a rank-$(1,1)$ tensor, as
\begin{align}
\Pi := \mathbb{I} - n \otimes k\,, \qquad \text{or in components} \qquad \Pi_\mu{}^\nu = \delta_\mu^\nu - n_\mu k^\nu.  \lb{projector}
\end{align}
It satisfies 
\begin{align}
\Pi_\mu{}^\alpha\Pi_\alpha{}^\nu=\Pi_\mu{}^\nu \qquad k^\nu \Pi_\nu{}^\mu = 0 \qquad  \Pi_\mu{}^\nu n_\nu = 0\,. 
\end{align}

The rigged projector is designed to project a tensor on $\M$ to a tensor on $\N$. For instance, given a vector $X \in T\M$, the vector $\tilde{X} := \Pi(X, \cdot) = X^\nu \Pi_\nu{}^\mu \pa_\mu$ is tangent to $\N$ and obeys $\iota_{\tilde{X}} n =0$. In a similar manner, given a 1-form $\omega \in \Omega^1(\M)$, the 1-form $\tilde{\omega} := \Pi(\cdot, \omega) = \rd x^\mu \Pi_\mu{}^\nu \omega_\nu$ also lies in $\Omega^1(\N)$ and obeys $\iota_k \tilde{\omega}= 0$. 

Introducing the metric duals $n^\mu=g^{\mu\nu}n_\nu$ and $k_\mu=g_{\mu\nu}k^\nu$, the rigged projector can be used to define the following vector and 1-form on $\N$
\begin{align}
\tilde{n} &:=  n^\nu \Pi_\nu{}^\mu  \pa_\mu = (n^\mu - 2\rho_n k^\mu) \pa_\mu\\
\tilde{k} &:= \frac{1}{1-4\rho_n \rho_k}  \Pi_\mu{}^\nu k_\nu  \rd x^\mu = \frac{1}{1-4\rho_n \rho_k} \left( k_\mu-2\rho_k n_\mu \right) \rd x^\mu\,. \lb{k-bar}
\end{align}

It immediately follows from the definition that $\iota_{\tilde{n}} n =0$ and $\iota_k \tilde{k} = 0$. Moreover, thanks to the prefactor in $\tilde k$, one can verify that $\tilde{n}$ and $\tilde{k}$ are tangent-bundle dual, that is $\iota_{\tilde{n}} \tilde{k} = 1$. Their norms are
\begin{align}
g (\tilde{n}, \tilde{n}) = -2\rho_n (1-4\rho_n\rho_k) \qquad \text{and} \qquad g^{-1} (\tilde{k},\tilde{k}) = - \frac{2\rho_k}{1-4\rho_n\rho_k}\,.\lb{norm-bar}
\end{align}\\

\subsection{Null Rigging and Induced Carrollian Structure} \lb{sec:null-rigg}

\paragraph{Non-null Hypersurface} Different rigging vectors define different rigging structures. For instance, when considering timelike or spacelike surfaces, one can choose that the rigging vector $k$ is proportional to the metric dual of $n$,
\begin{align}
k^\mu = \frac{1}{2\rho_n} n^\mu \qquad \text{and} \qquad 2\rho_k = \frac{1}{2\rho_n}\,, \qquad \text{for the normal rigging structure}. \nonumber
\end{align}
The induced metric on the surface can be obtained from the projection of the bulk metric,
\begin{align}
\gamma_{\mu\nu} = \Pi_\mu{}^\alpha \Pi_\nu{}^\beta g_{\alpha \beta} = \left( \delta_\mu^\alpha - \frac{1}{2\rho_n}n_\mu n^\alpha \right)\left( \delta_\nu^\beta - \frac{1}{2\rho_n}n_\nu n^\beta \right)g_{\alpha \beta} = g_{\mu\nu} - \frac{1}{2\rho_n} n_\mu n_\nu\,.  \label{hmn}
\end{align}
This choice of rigging is obviously singular when $\rho_n \to 0$ and the surface becomes null, giving rise to all singularities encountered when considering the null limit of various geometric quantities. This is the reason why for null hypersurfaces we cannot identify the metric dual of $n$ with the rigging vector $k$.

\paragraph{Null Hypersurface} Given a normal 1-form $n$, the condition $\iota_k n = 1$ does not uniquely determine the rigging vector $k$, as one can always make a shift, $k \to k + \lambda$, by a vector $\lambda$ satisfying $\iota_\lambda n = 0$. This new rigging vector will, in general, have a different norm from $k$, depending on the choice of $\lambda$. In other words, by choosing an appropriate vector $\lambda$, we can adjust the norm $\rho_k$ of the rigging vector to any value we desire. For our purposes, we will work with a null rigging, where $\rho_k = 0$.\footnote{We will use the same condition in the next section, to describe the so-called sCarrollian structure. When both the normal and the rigging are null, plus when introducing null dyads for the space (and thus exclusively in $4$ spacetime dimensions), this reduces to the Newman-Penrose tetrads. We will discuss this further in section \ref{7}. \label{F32}} It immediately follows that $\tilde{k}_\mu = k_\mu$, and that $k_\mu$ is automatically tangent to the null hypersurface $\N$. Hence, we shall denote $\tilde{k}$ simply as $k$, and we will also denote the tangent vector by $\tilde{n}^\mu = \ell^\mu$. Note that, although $\ell^\mu$ and $n^\mu$ are, by definition, equal on $\N$, they have different bulk extensions, unless $\rho_n$ is zero everywhere in the bulk. For simplicity, we will simply write $\rho_n = \rho$, and we recall that $\rho\Neq 0$ defines the locus of points of the isolated null hypersurface $\cN$. We also have the relations
\begin{align}
\ell^\mu =n^\nu\Pi_\nu{}^\mu= n^\mu - 2\rho k^\mu \qquad \text{and} \qquad 2\rho = n_\mu n^\mu = - \ell_\mu \ell^\mu\,.  \label{lvsn}
\end{align}
The rigged metric is given by
\begin{equation}
\begin{aligned}
h_{\mu\nu} = \Pi_\mu{}^\alpha \Pi_\nu{}^\beta g_{\alpha \beta} &=\left( \delta_\mu^\alpha - n_\mu k^\alpha \right)\left( \delta_\nu^\beta - n_\nu k^\beta \right)g_{\alpha \beta} = g_{\mu\nu} - 2n_{(\mu} k_{\nu)}\,,\label{rm}
\end{aligned}
\end{equation}
and, when evaluated on $\rho=0$, we will use the notation $q_{\mu\nu}\Neq h_{\mu\nu}$.

We also define the horizontal projector $q_\mu{}^\nu$ satisfying
\begin{align}
\Pi_\mu{}^\nu = k_\mu \ell^\nu + q_\mu{}^\nu \qquad \ell^\mu q_\mu{}^\nu =0 \qquad q_\mu{}^\nu k_\nu = 0  \qquad q_\mu{}^\alpha q_\alpha{}^\nu = q_\mu{}^\nu\,.
\end{align}
Therefore, we have two projectors: $\Pi_\mu{}^\nu$ projects to the null hypersurface, while $q_\mu{}^\nu$ projects to the horizontal sub-bundle thereof. In the right frame and simplifying, one has $\delta=\text{diag}(1,1,1,\dots,1)$, $\Pi=\text{diag}(0,1,1,\dots,1)$ and $q=\text{diag}(0,0,1,\dots,1)$.\\

While it is possible to perform all computations directly in the ambient spacetime with abstract coordinates $x^\mu$, it is also crucial to understand how to translate between tensors on $\M$ and intrinsic tensors on the Carrollian manifold $\N$, with coordinates $x^a$. This will allow us to link with the intrinsic analysis on $\N$ developed in previous sections.  We will do so by properly introducing the   mixed indices tensors $\Pi_a{}^\mu$ and $\Pi_\mu{}^a$, which provide a way to deduce tensors on $\N$ from their ambient avatars. In other words, these tensors translate between the ambient and intrinsic descriptions: $\Pi_a{}^\mu$ implements the pushforward of tangent vectors, while $\Pi_\mu{}^a$ encodes the pullback of spacetime data onto the hypersurface. This will also highlight another advantage of the rigging technique. 

Let us first review the standard differential geometry of embeddings. The Carrollian manifold $\N$ is embedded as a null hypersurface in the spacetime $\M$ via the inclusion map $i: \N \hookrightarrow \M$, which identifies points in $\N$ with points in $\M$. The pushforward $i_*: T\N \to T\M$ maps a vector $\tilde{X} = \tilde{X}^a \pa_a \in T\N$ to a vector $X = X^\mu \pa_\mu \in T\M$ as\footnote{If, in coordinates, the inclusion map is $x^\mu(x^b)$, then one explicitly has $e_a{}^\mu=\frac{\pa x^\mu(x^b)}{\pa x^a}$.}
\begin{align}
X^\mu \pa_\mu &= i_* (\tilde{X}^a \pa_a)  = \tilde{X}^a i_*(\pa_a) = \tilde{X}^a e_a{}^\mu \pa_\mu\,, \quad \text{where we defined} \quad e_a{}^\mu \pa_\mu := i_*(\pa_a)\,.  
\end{align}
In addition, the pullback $i^*: T^*\M \to T^*\N$ maps a 1-form $\omega = \omega_\mu \rd x^\mu \in T^*\M$ to a 1-form $\tilde{\omega} = \tilde{\omega}_a \rd x^a \in T^*\N$ according to the definition
\begin{align}
X^\mu \omega_\mu = \tilde{X}^a \tilde{\omega}_a\,, \qquad \text{which infers} \qquad \tilde{\omega}_a = e_a{}^\mu \omega_\mu\,. 
\end{align}
Alternatively, we can write
\begin{align}
\omega_a \rd x^a = i^* (\omega_\mu \rd x^\mu) = \omega_\mu i^*(\rd x^\mu) = e_a{}^\mu \omega_\mu \rd x^a\,,
\end{align}
which implies
\begin{align}
\rd x^a e_a{}^\mu  = i^* (\rd x^\mu)\,.
\end{align}

Next, we want to define the dual basis $e_\mu{}^a$ such that $e_a{}^\mu e_\mu{}^b = \delta_a^b$. If this basis exists, it can provide a notion of projection of a vector $X^a = X^\mu e_\mu{}^a$ and a lift of a 1-form $\omega_\mu = e_\mu{}^a \omega_a$.

However, without additional geometric structure, $e_\mu{}^a$ is not unique, with the freedom corresponding to the choice of how one projects a $T\M$ vector to the subspace $T\N$. Put differently, the pushforward $i_*: T\N \to T\M$ is injective but not surjective, and thereby does not by itself specify how to extend tangent or cotangent data off the hypersurface. To find $e_\mu{}^a$, one must specify a complement to $T\N$ inside $T\M$, so that every spacetime vector can be uniquely decomposed into tangential and transverse components. Different choices of complements lead to different basis $e_\mu{}^a$.  

For spacelike and timelike hypersurfaces, one can choose $e_\mu{}^a = h^{ab} e_b{}^\nu g_{\mu\nu}$, where $h^{ab}$ is the inverse of the induced rigged metric $h_{ab} = e_a{}^\mu e_b{}^\nu g_{\mu\nu}$ on the hypersurfaces. Then one has
\beq
e_a{}^\mu e_\mu{}^b = e_a{}^\mu h^{bc} e_c{}^\nu g_{\mu\nu}=h^{bc}h_{ac}=\delta_a^b\,.
\eeq
However, this fails when working with null hypersurfaces, as the induced metric is not invertible. The null rigging structure comes again to the rescue: it provides exactly the additional geometric structure required to define the tensor $e_\mu{}^a$. This is also another merit of the rigging technique, especially when dealing with the geometry of a null hypersurface.

Algebraically, the equation $e_a{}^\mu e_\mu{}^b = \delta_a^b$ imposes $(d+1)^2$ constraints for $(d+2)(d+1)$ unknowns in $e_\mu{}^a$. Given the null rigging structure, we impose the condition $k^\mu e_\mu{}^a =0$, which provides additional $(d+1)$ constraints. Therefore, with the null rigging, we have $(d+1)^2+(d+1)=(d+2)(d+1)$ constraints, which  completely fix the tensor $e_\mu{}^a$. Stated differently, this means that both the basis $e_a = e_a{}^\mu \pa_\mu$ of $T\N$ and the basis  $e^a = \rd x^\mu e_\mu{}^a$ of $T^*\N$ can be respectively completed with $k$ and $n$ such that $(e_a, k)$ and $(e^a, n)$ form orthogonal bases for $T\M$ and $T^*\M$, with thus $e_a{}^\mu e_\mu{}^b = \delta_a^b$, and $k^\mu e_\mu{}^a=0=e_a{}^\mu n_\mu$. Indeed, the rigging vector $k$ and the normal $n$ provide the transverse direction to $\N$, as we have already explained.\\

With this structure, any vector $X\in T\M$ can be decomposed as 
\begin{align}
X^\mu &= (X^\nu e_\nu{}^a) e_a{}^\mu + (X^\nu n_\nu) k^\mu  = X^\nu \left( e_\nu{}^a e_a{}^\mu + n_\nu k^\mu  \right)\,. 
\end{align}
We thus obtain the completeness relation
\begin{align}
\delta_\nu^\mu =  e_\nu{}^a e_a{}^\mu + n_\nu k^\mu\,.  
\end{align}
Comparing to the definition of the rigged projector \eqref{projector}, we conclude that
\begin{align}
e_\nu{}^a e_a{}^\mu = \Pi_\nu{}^\mu\,. 
\end{align}

Finally, let us define the  mixed indices tensors 
\begin{align}
\Pi_a{}^\mu := e_a{}^\nu \Pi_\nu{}^\mu = e_a{}^\mu \qquad
\Pi_\mu{}^a := \Pi_\mu{}^\nu e_\nu{}^a = e_\mu{}^a\,.
\end{align}
They obey the conditions 
\begin{align}
\Pi_a{}^\mu \Pi_\mu{}^b = \delta_a^b = q_a{}^b + k_a \ell^b \qquad \text{and} \qquad \Pi_\mu{}^a \Pi_a{}^\nu = \Pi_\mu{}^\nu\,. \label{piapia}
\end{align}
Since the maps $\Pi_a{}^\mu$ and $\Pi_\mu{}^a$ will appear often in what follows, we will refer to them as "soldering".\footnote{Strictly speaking, a soldering map is defined when the dimensions of the two manifolds are equal. Nonetheless, we will use the word soldering here in a generalized sense.} Our choice $k^\mu e_\mu{}^a=0$ is what ultimately led to \eqref{piapia}. We prefer to think of $\Pi_a{}^\mu$ as an a-priori distinct entity from $e_a{}^\mu$, since the latter coincides with the former only under specific assumptions -- see also the discussion below \eqref{spia}. 

Then, the intrinsic tensors on $\cN$ are given by
\beq
\ell^a \Neq n^\mu \Pi_\mu{}^a \qquad k_a \Neq \Pi_a{}^\mu k_\mu\qquad q_{ab}\Neq \Pi_a{}^\mu\Pi_b{}^\nu g_{\mu\nu}\,,\label{rcs}
\eeq
whereas one has
\beq\label{pkpn}
\Pi_a{}^\mu n_\mu =0 \qquad k^\mu \Pi_\mu{}^a=0\,.
\eeq
This gives a unique way of distinguishing between intrinsic and extrinsic vectors and forms on $\N$: the triple $\ell^a,k_a,q_{ab}$ forms exactly the basic ingredients of the intrinsic ruled Carrollian structure $\RCarr$ on $\cN$, while the normal one form $n_\mu$ and its tangent bundle dual null rigging $k^\mu$ define how the null hypersurface is plunged into the bulk. We emphasize that \eqref{rcs} establishes a clear relationship between the null rigging $k$ in the ambient space and the ruling $k$ on the Carrollian manifold.

While our presentation so far and in the following will not confine to adapted coordinates, we note that if we choose $x^\mu = (x_N, x^a)$, with $n\approx \rd x_N$, the tangent basis to the embedding becomes trivial, $e_a{}^\mu = \delta^\mu_a$. This corresponds to the so-called trivial embedding, so that coordinates on $\N$ are identified with coordinates on $\M$. In this case only, one could then define the map $\Pi_a{}^b$. While this is clearly a convenient choice, it is important to appreciate that in general, one should distinguish between coordinates on $\M$ and coordinates on $\N$. In the following, we will always utilize the indices $a,b,c,\dots$ exclusively for tensors in $\N$, and employ the soldering $\Pi_a{}^\mu$ and $\Pi_\mu{}^a$ to relate quantities in $\N$ to quantities in $\M$.\\

\subsection{Rigged Connection} \lb{sec:rigg-connection}

Having introduced the null rigging structure and how it induces the ruled Carrollian structure on $\N$,  we next discuss how the rigged connections exactly gives rise to the standard Carrollian connection discussed in section \ref{sec:CC}. First, let us recall that, a tensor $T_a{}^b$ on $\N$ lifts to a tensor on $\M$ as $T_\mu{}^\nu=\Pi_\mu{}^aT_a{}^b\Pi_b{}^\mu$, which by design satisfies $k^\mu T_\mu{}^\nu =0$ and $T_\mu{}^\nu n_\nu =0$, thanks to \eqref{pkpn}. The rigged connection is defined in the following way. For a tensor field $T_a{}^b$ on $\N$, it is given by the bulk Levi-Civita covariant derivative projected onto $\N$ using the soldering, 
\begin{align}
D_a T_b{}^c = \Pi_a{}^\mu \Pi_b{}^\nu (\nabla_\mu T_\nu{}^\rho)\Pi_\rho{}^c\,.\label{rc}
\end{align}
One can show that the rigged projector is covariantly-conserved once projected onto $\N$
\begin{align}
\Pi_a{}^\mu \Pi_b{}^\nu (\nabla_\mu \Pi_\nu{}^\rho)\Pi_\rho{}^c = -\Pi_a{}^\mu \Pi_b{}^\nu (k^\rho \nabla_\mu n_\nu + n_\nu \nabla_\mu k^\rho)\Pi_\rho{}^c = 0\,,
\end{align}
which followed from $\Pi_\mu{}^\nu n_\nu =0$ and $k^\mu \Pi_\mu{}^\nu =0$. \\

Let us examine the properties of the rigged connection. The torsionless condition is given by the property that $T\cN$ is an integrable distribution in $T{\cal M}$. In other words, the rigged connection is torsionless if and only if $n$ satisfies Frobenius theorem. Indeed, repeatedly using \eqref{rc} and the definition of $\Pi_\mu{}^\nu$, one has
\begin{equation}
\begin{aligned}
T_{ab}{}^c D_c F& = -[D_a, D_b] F \\
& = -D_a (\Pi_b{}^\rho \nabla_\rho F) + D_b (\Pi_a{}^\rho \nabla_\rho F) \\
& = -\Pi_a{}^\mu \Pi_b{}^\nu \nabla_\mu (\Pi_\nu{}^\rho \nabla_\rho F) + \Pi_a{}^\mu \Pi_b{}^\nu \nabla_\nu (\Pi_\mu{}^\rho \nabla_\rho F) \\
& = -\Pi_a{}^\mu \Pi_b{}^\nu \left( \nabla_\mu \Pi_\nu{}^\rho  - \nabla_\nu \Pi_\mu{}^\rho  \right) \nabla_\rho F - \Pi_a{}^\mu \Pi_b{}^\nu [\nabla_\mu, \nabla_\nu] F\\
\left[ T_{\mu\nu}{}^\rho=0 \right]\quad &= -\Pi_a{}^\mu \Pi_b{}^\nu \left( \nabla_\mu \Pi_\nu{}^\rho  - \nabla_\nu \Pi_\mu{}^\rho  \right) \nabla_\rho F\\
\left[ \Pi_a{}^\mu n_\mu=0 \right]\quad &= \Pi_a{}^\mu \Pi_b{}^\nu \left( \nabla_\mu n_\nu - \nabla_\nu n_\mu  \right) k[F]\,,
\end{aligned}
\end{equation}
where $T_{\mu\nu}{}^\rho=0$ follows from the bulk Levi-Civita connection being torsion-free. Now we impose the integrability of the distribution, that is, we require $n \wedge \rd n = 0$,\footnote{This is Frobenius theorem: a distribution is integrable if and only if $n \wedge \rd n = 0$.} which in turn implies $\nabla_{[\mu} n_{\nu]}=\alpha_{[\mu}n_{\nu]}$ for a given 1-form $\alpha_\mu$. Inserting this in the previous equation gives
\beq
T_{ab}{}^c D_c F= \Pi_a{}^\mu \Pi_b{}^\nu \left( \nabla_\mu n_\nu - \nabla_\nu n_\mu  \right) k[F]=2\Pi_a{}^\mu \Pi_b{}^\nu \alpha_{[\mu}n_{\nu]}k[F]=0\,,\label{tf}
\eeq
where we simply used that $\Pi_a{}^\mu n_\mu=0$. As we will confine our attention to distributions satisfying Frobenius theorem, the rigged connection is torsionless. This is an important result, which justifies from the embedding of the null hypersurface the intrinsic torsionless condition we imposed in section \ref{sec:CC}.

In the computations that follow, we will use that
\beq
&\Pi_a{}^\mu = e_a{}^\nu \Pi_\nu{}^\mu=  e_a{}^\nu (k_\nu\ell^\mu+q_\nu{}^\mu)=k_a\ell^\mu+q_a{}^\mu\,,\label{piamu}&\\
&\Pi_\mu{}^a =  \Pi_\mu{}^\nu e_\nu{}^a=(k_\mu\ell^\nu+q_\mu{}^\nu) e_\nu{}^a=k_\mu\ell^a+q_\mu{}^a\,.&
\eeq
Here, we defined the mixed-indices tensors $e_a{}^\mu q_\mu{}^\nu=q_a{}^\nu$ and $ q_\mu{}^\nu e_\nu{}^a=q_\mu{}^a$. Much like the soldering provides a way to project tensors from $\M$ to $\N$, the mixed-indices tensors $q_a{}^\mu$ and $q_\mu{}^a$ give us a way to project from $\M$ directly to the horizontal sub-bundle of $T\N$ and $T^*\N$, because, indeed, $\ell^a q_a{}^\mu=\ell^a\Pi_a{}^\mu-\ell^\mu=n^\nu \Pi_\nu{}^a \Pi_a{}^\mu-\ell^\mu=n^\nu \Pi_\nu{}^\mu-\ell^\mu=0$.\\

Let us consider how the rigged connection acts on the Carrollian data. For the Carrollian vector, we have 
\begin{equation}
\begin{aligned}
D_a \ell^b &= \Pi_a{}^\mu (\nabla_\mu \ell^\nu) \Pi_\nu{}^b  \\
&= \Pi_a{}^\mu \nabla_\mu \ell^\nu q_\nu{}^b + \Pi_a{}^\mu \nabla_\mu \ell^\nu k_\nu \ell^b \\
& = q_a{}^\mu \nabla_\mu \ell^\nu q_\nu{}^b + \Pi_a{}^\mu \nabla_\mu \ell^\nu k_\nu\ell^b +k_a\ell^\mu \nabla_\mu \ell^\nu q_\nu{}^b\,. \label{Dl}
\end{aligned}
\end{equation}
The last term vanishes on the null hypersurface due to the fact that $\nabla_\ell \ell^\mu \propto \ell^\mu$. To derive this, we use the integrability condition $\nabla_{[\mu} n_{\nu]}=\alpha_{[\mu}n_{\nu]}$ to write
\begin{equation}
    \begin{aligned}
n^\nu \nabla_\nu n^\beta&=g^{\mu\beta}n^\nu (\nabla_\mu n_\nu+\alpha_\nu n_\mu-\alpha_\mu n_\nu)\\
&=g^{\mu\beta}(\nabla_\mu \rho+n^\nu \alpha_\nu n_\mu)\\
&=(k[\rho]+n^\nu \alpha_\nu)n^\beta\,. \lb{nabla-n-derive}
\end{aligned}
\end{equation}
This derivation is performed  assuming that $\rho\Neq 0$ and therefore $n^\gamma n_\gamma=0$. Note, however, that it is important to keep track of derivatives of $\rho$ in the normal direction, as those are not vanishing. Hence, we used $\pa_\mu\rho = \Pi_\mu{}^\nu \pa_\nu \rho+n_\mu k[\rho]\Neq n_\mu k[\rho]$ in the last line. Using \eqref{lvsn}, one has that $\ell^\mu\Neq n^\mu$, and thus one can further demonstrate
\begin{equation}
    \begin{aligned}
\ell^\nu \nabla_\nu \ell^\beta&=n^\nu \nabla_\nu (\Pi_\gamma{}^\beta n^\gamma)\\
&=-n^\nu(k^\beta \nabla_\nu n_\gamma + n_\gamma \nabla_\nu k^\beta)n^\gamma+n^\nu\Pi_\gamma{}^\beta \nabla_\nu n^\gamma\\
&=-n_\gamma n^\gamma (k[\rho]+n^\nu \alpha_\nu) k^\beta+(k[\rho]+n^\nu \alpha_\nu)\ell^\beta\\
&=(k[\rho]+\ell^\nu \alpha_\nu)\ell^\beta\,.
\end{aligned}
\end{equation}
We therefore conclude that 
\begin{align}\label{lnl}
\ell^\nu \nabla_\nu \ell^\mu \Neq \left( k[\rho]+\ell^\nu \alpha_\nu  \right) \ell^\mu\,.
\end{align}
This can be inserted in \eqref{Dl}, and using $\ell^\mu q_\mu{}^a=\ell^\mu q_\mu{}^\nu e_\nu{}^a=0$ we get
\begin{equation}
\begin{aligned}
D_a \ell^b = q_a{}^\mu \nabla_\mu \ell^\nu q_\nu{}^b + \Pi_a{}^\mu \nabla_\mu \ell^\nu k_\nu\ell^b \,.\label{Dl2}
\end{aligned}
\end{equation}
First, we remark that the expansion tensor on $\cN$ is determined from the bulk as
\begin{align}
\theta_a{}^b = q_a{}^\mu \nabla_\mu \ell^\nu q_\nu{}^b\,.
\end{align}
Then, we define from the bulk 
\begin{align}
\omega_a = \Pi_a{}^\mu \nabla_\mu \ell^\nu k_\nu\,,\label{omabulk}
\end{align}
with components $\omega_a = \pi_a + \kappa k_a$ given by  
\begin{align}
\pi_a = q_a{}^\mu \nabla_\mu \ell^\nu k_\nu \qquad \text{and} \qquad \kappa = \ell^\mu \nabla_\mu \ell^\nu k_\nu\,.  
\end{align}
Eventually, from \eqref{Dl2}, one obtains
\beq
D_a\ell^b=\theta_a{}^b+\omega_a \ell^b\,.
\eeq
This is exactly the intrinsic definition and structure of the Carrollian connection that we provided in \eqref{Dell}, demonstrating its bulk origin. \\

Our result \eqref{lnl} proves that on $\N$, the Carrollian vector is the null geodesic generator with inaffinity $\kappa$ given by 
\begin{align}
\kappa = k[\rho]+\ell^\nu \alpha_\nu\,. 
\end{align}
This shows that $\kappa$ depends on extrinsic data, in particular on the norm of $n$ away from $\cN$ ($k[\rho]$), which is independent from the intrinsic data $\ell^a$, $k_a$, and $q_{ab}$. This is the reason why it parametrizes the Carrollian connection. This feature was anticipated from an intrinsic standpoint in and around eq. \eqref{omtbar}. We here proved this fact for $\kappa$, and we will now show it for $\btheta_{(ab)}$.\\

In a similar manner, we now compute the action of the rigged connection on the Ehresmann connection $k_a$. To do so, we extend eq. \eqref{d-k} in the proximity of the null hypersurface in the bulk, such that 
\beq
\pa_{[\mu}k_{\nu]}=\varphi_{[\mu}k_{\nu]}-\frac12 \varpi_{\mu\nu}\,.\label{dkb}
\eeq
Using this, we then obtain
\begin{equation}
\begin{aligned}\label{ppk}
D_a k_b &= \Pi_a{}^\mu \Pi_b{}^\nu \nabla_\mu k_\nu \\
\left[\eqref{piamu}\right]\quad & = q_a{}^\mu q_b{}^\nu \nabla_\mu k_\nu + (\Pi_a{}^\mu \ell^\nu \nabla_\mu k_\nu)k_b + k_a (\ell^\mu q_b{}^\nu \nabla_\mu k_\nu) \\
\left[ \ell^\nu k_\nu=1\right]\quad & = q_a{}^\mu q_b{}^\nu \nabla_\mu k_\nu -(\Pi_a{}^\mu  \nabla_\mu \ell^\nu k_\nu)k_b + k_a (2\ell^\mu q_b{}^\nu \nabla_{[\mu} k_{\nu]}+ \ell^\mu q_b{}^\nu \nabla_\nu k_\mu) \\
& = q_a{}^\mu q_b{}^\nu \nabla_\mu k_\nu -(\Pi_a{}^\mu  \nabla_\mu \ell^\nu k_\nu)k_b + k_a (\ell^\mu q_b{}^\nu ( 2\ac_{[\mu} k_{\nu]}- \vor_{\mu\nu})-  q_b{}^\nu \nabla_\nu \ell^\mu k_\mu) \\
\left[\eqref{omabulk}\right]\quad & = \btheta_{ab}- \omega_a k_b - k_a (\pi_b+\ac_b)\,,
\end{aligned}
\end{equation}
where we introduced
\begin{align}
\btheta_{ab} = q_a{}^\mu q_b{}^\nu \nabla_\mu \k_\nu\,.
\end{align}
Here again we note that, while the skew-symmetric part of $\btheta_{ab}$ is constrained due to \eqref{dkb}, its symmetric part parametrize the Carrollian connection with extrinsic data, and encodes our freedom in choosing $k_a$. \\

Finally, recalling \eqref{rm} and the fact that $q_{\mu\nu}\Neq h_{\mu\nu}$, we compute on $\cN$\footnote{Since we are relating an expression on $\N$ to a bulk tensor acted upon by derivatives, on the RHS of the first line, we need to insert the bulk induced metric $h_{\mu\nu}$, which evaluates to $q_{ab}$ only when projected -- see the discussion in \eqref{hqhq} below.\label{ff}}
\begin{equation}
\lb{Dq-derive}
\begin{aligned}
D_a q_{bc} 
&= \Pi_a{}^\alpha \Pi_b{}^\beta \Pi_c{}^\gamma \nabla_\alpha h_{\beta \gamma} \\
&= \Pi_a{}^\alpha \Pi_b{}^\beta \Pi_c{}^\gamma \nabla_\alpha \left(\Pi_\beta{}^\mu \Pi_\gamma{}^\nu g_{\mu \nu} \right)\\
& = 2\Pi_a{}^\alpha \nabla_\alpha \Pi_\beta{}^\mu \Pi_{(b}{}^\beta \Pi_{c)}{}^\nu g_{\mu\nu} \\
\left[\Pi_a{}^\mu n_\mu =0\right]\quad & = -2 \Pi_a{}^\alpha \nabla_\alpha n_\beta \Pi_{(b}{}^\beta \Pi_{c)}{}^\nu k^\mu g_{\mu\nu} \\
\left[\eqref{rcs}\right]\quad & =  -2 \Pi_a{}^\alpha \nabla_\alpha n_\beta \Pi_{(b}{}^\beta k_{c)}\,.
\end{aligned}
\end{equation}
Since on the null surface we have $n^\mu \Neq \ell^\mu$ and $\ell^\mu \ell_\mu = 0$, we can show that
\begin{equation}
\begin{aligned}\label{ppn}
\Pi_a{}^\alpha \nabla_\alpha n_\beta \Pi_{b}{}^\beta &= \Pi_a{}^\alpha \nabla_\alpha n^\gamma g_{\beta \gamma} \Pi_{b}{}^\beta \\
\left[\eqref{projector}\right]\quad &= \Pi_a{}^\alpha \nabla_\alpha n^\gamma (\Pi_\gamma{}^\delta + n_\gamma k^\delta) g_{\beta \delta} \Pi_{b}{}^\beta \\
&\Neq D_a\ell^c  \Pi_c{}^\gamma \Pi_\gamma{}^\delta g_{\beta \delta} \Pi_{b}{}^\beta +  \Pi_a{}^\alpha \nabla_\alpha\rho k_b \\
&= D_a\ell^c  q_{cb} + \pa_a \rho k_b \\
\left[\pa_a \rho\Neq 0\right]\quad & = \theta_{ab}\,.
\end{aligned}
\end{equation}
Altogether, we obtain
\begin{align}
D_a q_{bc} = - (\theta_{ab} k_{c} + \theta_{ac} k_b)\,,
\end{align}
which is exactly the intrinsic equation \eqref{Dq}.\\

Therefore, we derived (\ref{Dq}-\ref{Dk}) from the ambient space and its Levi-Civita connection, through the null rigging procedure. This demonstrates that the induced connection on $\cN$ cannot be metric compatible, whereas it trivially has no torsion. As already touched upon, while other intrinsic connections are available and certainly interesting in various approaches to Carrollian physics, we are here promoting this specific connection as the canonical one arising from embedding Carroll structures in pseudo-Riemannian manifolds.\\ 

\subsection{Gauss, Codazzi-Mainardi, and Null Brown-York Stress Tensor}

In this final subsection, we first revisit the Gauss and Codazzi-Mainardi equations for non-null hypersurfaces. We then extend them to the null case, showing how the intrinsic Riemann tensor associated with the Carrollian connection in section \ref{RiemannCur} relates to the rigged projection of the bulk Riemann tensor. Finally, we show that, from purely geometric arguments, one can construct the null Brown-York tensor, whose conservation laws, via the Codazzi-Mainardi relation, directly follow from the bulk vacuum Einstein equations. \\

\subsubsection{Non-null Hypersurface}\label{531}

If $n_\mu$ is the normal 1-form to a non-null hypersurface, the induced metric and canonical projector have already been discussed in \eqref{hmn}, and they are given by
\beq
\gamma_{\mu\nu}=g_{\mu\nu}-\epsilon n_\mu n_\nu\qquad \gamma_\mu{}^\nu=\delta_\mu^\nu-\epsilon n_\mu n^\nu\,,
\eeq
where $\epsilon=\frac1{2\rho}$ controls the nature of the surface and the normalization of the normal. If $\rho >0$, then the surface is timelike, while $\rho <0$ implies that the surface is spacelike. In both cases, we do not need to introduce the rigging vector. The ambient metric and the normal 1-form provide all the geometric data to canonically induce the intrinsic geometry on the hypersurface. This will however prevent us to take a smooth null limit, as we discuss in the next section. There, we will also explain that we should distinguish between the induced metric $\gamma_{\mu\nu}$ in \eqref{hmn} and the rigged metric $h_{\mu\nu}$ in \eqref{rm}. The latter requires extra structure, while the former is the starting point of this subsection.\\

As usual, we will denote the coordinates intrinsic to the hypersurface as $x^a$. The hypersurface is embedded in $\M$, with the vielbein for the embedding $e_a{}^\mu$ satisfying
\beq
\gamma_\nu{}^\mu=e_\nu{}^a \e_a{}^\mu \qquad \delta_a^b=e_a{}^\mu e_\mu{}^b\,,
\eeq
and thus expressing the fact that, since the dimensionality of the spaces is different, $e_\mu{}^a$ is a partial inverse of $e_a{}^\mu$. In keeping with the notation  above, we will call the vielbein for the embedding the "soldering", and denote them
\beq
e_a{}^\mu=\gamma_a{}^\mu\qquad e_\mu{}^a=\gamma_\mu{}^a\,.
\eeq
One should be careful: these quantities are maps between different spaces, and thus should not be  confused with the projector $\gamma_\mu{}^\nu$. The reason why we denote them $\gamma_a{}^\mu$ and $\gamma_\mu{}^a$ is because one typically works in adapted coordinates to the embedding, in which case one would have $n_a=0=n^a$, and thus $\gamma_a{}^\mu=\delta_a^\mu$ and $\gamma_\mu{}^a=\delta_\mu^a$, which are exactly the $a\mu$-components of the projector $\gamma_\mu{}^\nu$. The latter, in this specific setting, is simply $\gamma=\text{diag}(0,1,\dots,1)$.\\

Given the soldering $\gamma_a{}^\mu$, one defines the induced connection via
\beq
D_aV^b=\gamma_a{}^\mu \nabla_\mu V^\nu \gamma_\nu{}^b\,.\label{dind}
\eeq
The extrinsic curvature, also called second fundamental form, is defined as
\beq
K_a{}^b=\gamma_a{}^\mu \nabla_\mu n^\nu \gamma_\nu{}^b\,.\label{ext}
\eeq
This is customarily introduced with both indices down. Here, we introduce it with mixed indices, as this parallels the null case discussed below.

Then, consider a vector $V^a$ tangent to the hypersurface. One can use the defining equation for the Riemann tensor (see \eqref{rdef}) of the induced connection to evaluate
\begin{equation}
\begin{aligned}
[D_c,D_d] V^a&= R^a{}_{bcd}V^b\\
\left[\eqref{dind}\right]\quad &= D_c(\gamma_d{}^\mu \gamma_\nu{}^a \nabla_\mu V^\nu)-(c\leftrightarrow d)\\
\left[\eqref{dind}\right]\quad &= 2\gamma_d{}^\mu \gamma_\nu{}^a \gamma_c{}^\rho \nabla_{[\rho}\nabla_{\mu]}V^\nu +\gamma_c{}^\rho \gamma_d{}^\sigma \nabla_\rho \gamma_\sigma{}^\mu \gamma_\nu{}^a \nabla_\mu V^\nu - (c\leftrightarrow d)\\
& \ \ +\gamma_c{}^\rho \gamma_\beta{}^a \nabla_\rho \gamma_\nu{}^\beta \gamma_d{}^\mu \nabla_\mu V^\nu - (c\leftrightarrow d)\\
\left[\eqref{ext}\right]\quad&=\gamma_d{}^\mu \gamma_\nu{}^a \gamma_c{}^\rho R^\nu{}_{\gamma\rho\mu}V^b \gamma_b{}^\gamma -\epsilon K_{[cd]} n^\mu \gamma_\nu{}^a\nabla_\mu V^\nu+\epsilon V^b K_c{}^a K_{db}-(c\leftrightarrow d)\\
&=V^b\left[ \gamma_\nu{}^a \gamma_b{}^\gamma \gamma_c{}^\rho \gamma_d{}^\mu R^\nu{}_{\gamma\rho\mu}-2\epsilon K_{b[c}K_{d]}{}^a\right]\,,
\end{aligned}
\end{equation}
where we used that $V^\mu=V^b \gamma_b{}^\mu$ and $\gamma_\mu{}^\nu=\delta_\mu^\nu-\epsilon n_\mu n^\nu$.

Since this is true for any vector field, we can relate the intrinsic Riemann tensor to the projection of the bulk Riemann tensor. This is the famous Gauss equation,
\beq
R^a{}_{bcd}=\gamma_\nu{}^a \gamma_b{}^\gamma \gamma_c{}^\rho \gamma_d{}^\mu R^\nu{}_{\gamma\rho\mu}-2\epsilon K_{b[c}K_{d]}{}^a\,.
\eeq\\

The Codazzi-Mainardi equation describes the relationship between the covariant derivative of the extrinsic curvature and the Riemann tensor. It is given by
\beq
D_c K_d{}^a-D_d K_c{}^a= [D_c,D_d]n^a= R^a{}_{bcd} n^b= \gamma_\nu{}^a \gamma_c{}^\rho \gamma_d{}^\mu R^\nu{}_{\gamma\rho\mu}n^\gamma\,,
\eeq
where we used that $n^a K_a{}^b=0$.

The Gauss and Codazzi-Mainardi equations (often referred to as Gauss-Codazzi equations) express how the induced metric -- and thus its curvature tensors -- is linked to the second fundamental form of a submanifold. The objective of the next subsection is to reproduce these results in the case of a degenerate induced metric, and thus for a null hypersurface.\\

\subsubsection{Null Hypersurface} \lb{sec:NGC}

We start with the Gauss equation for the null surface $\N$. In the same spirit as the non-null cases, we consider 
\begin{equation}
\lb{Gauss-derive}
\begin{aligned}
R^a{}_{bcd} V^b 
&= [D_c, D_d] V^a   \\
\left[\eqref{rc}\right]\quad &= D_c \left( \Pi_d{}^\delta \nabla_\delta V^\alpha \Pi_\alpha{}^a \right) - (c \leftrightarrow d) \\ 
\left[\eqref{rc}\right]\quad &=  \Pi_c{}^\gamma \Pi_d{}^\mu \Pi_\nu{}^a \nabla_\gamma \left( \Pi_\mu{}^\delta \nabla_\delta V^\alpha \Pi_\alpha{}^\nu \right)- (c \leftrightarrow d) \\
&= \Pi_\alpha{}^a \Pi_c{}^\gamma \Pi_d{}^\delta [\nabla_\gamma, \nabla_\delta] V^\alpha +  2\Pi_{[c}{}^\gamma \Pi_{d]}{}^\mu \nabla_\gamma  \Pi_\mu{}^\delta \nabla_\delta V^\alpha \Pi_\alpha{}^a \\
& \ \ +  2\Pi_{[c}{}^\gamma \Pi_{d]}{}^\delta \nabla_\delta V^\alpha \nabla_\gamma \Pi_\alpha{}^\nu \Pi_\nu{}^a \\
\left[\eqref{projector}\right]\quad &= V^b \Pi_\alpha{}^a \Pi_b{}^\beta \Pi_c{}^\gamma \Pi_d{}^\delta R^\alpha{}_{\beta \gamma \delta}  - 2 \Pi_c{}^\gamma \Pi_d{}^\mu \nabla_{[\gamma} n_{\mu]} k^\delta \nabla_\delta V^\alpha \Pi_\alpha{}^a \\
& \ \ - 2\Pi_{[c}{}^\gamma \Pi_{d]}{}^\delta \nabla_\delta V^\alpha n_\alpha \nabla_\gamma k^\nu \Pi_\nu{}^a\,,
\end{aligned}
\end{equation}
where we used that $V^\beta=V^b\Pi_b{}^\beta$, $\nabla_\gamma \Pi_\mu{}^\delta=-\nabla_\gamma n_\mu k^\delta-n_\mu \nabla_\gamma k^\delta$, and $\Pi_d{}^\mu n_\mu=0=k^\nu\Pi_\nu{}^a$.
The second term vanishes due to our requirement of integrability, i.e., Frobenius theorem. The third term can be evaluated using the Leibniz rule and recalling that $V^\alpha n_\alpha =0$,
\begin{equation}
\begin{aligned}
- 2\Pi_{[c}{}^\gamma \Pi_{d]}{}^\delta \nabla_\delta V^\beta n_\beta \nabla_\gamma k^\alpha \Pi_\alpha{}^a
& = 2\Pi_{[c}{}^\gamma \Pi_{d]}{}^\delta \nabla_\delta n_\beta V^\beta \nabla_\gamma k^\alpha \Pi_\alpha{}^a \\
[V^\beta=V^b\Pi_b{}^\beta] \qquad & = V^b ( \Pi_d{}^\delta \nabla_\delta n_\beta \Pi_b{}^\beta) ( \Pi_c{}^\gamma \nabla_\gamma k^\alpha \Pi_\alpha{}^a )  - (c \leftrightarrow d) \\
[\eqref{ppn}]\qquad & = V^b \theta_{db}( \Pi_c{}^\gamma \nabla_\gamma k_\mu g^{\mu \alpha} \Pi_\alpha{}^a )  - (c \leftrightarrow d) \\
[\delta_\mu^\nu=\Pi_\mu{}^\nu+n_\mu k^\nu]\qquad & = V^b \theta_{db}\left( \Pi_c{}^\gamma \nabla_\gamma k_\nu (\Pi_\mu{}^\nu + n_\mu k^\nu) g^{\mu \alpha} \Pi_\alpha{}^a \right)  - (c \leftrightarrow d) \\
[\Pi_\mu{}^\nu = \Pi_\mu{}^e \Pi_e{}^\nu]\qquad & = V^b \theta_{db}\left( \Pi_c{}^\gamma \nabla_\gamma k_\nu \Pi_e{}^\nu \Pi_\mu{}^e g^{\mu \alpha} \Pi_\alpha{}^a \right)  \\
& \ \ \  +\frac12 V^b \theta_{db}\left( \Pi_c{}^\gamma \nabla_\gamma (k_\nu k^\nu) n_\mu g^{\mu \alpha} \Pi_\alpha{}^a \right) - (c \leftrightarrow d) \\
[\eqref{ppk}, \ \ k_\nu k^\nu=0]\qquad  & = V^b \theta_{db}D_c k_e \Pi_\mu{}^e g^{\mu \alpha} \Pi_\alpha{}^a  - (c \leftrightarrow d)\,.\label{term3}
\end{aligned}
\end{equation}
Now, using the various properties of the rigged projector and the soldering, such as \eqref{piapia}, we evaluate
\begin{equation}
\begin{aligned}
q_a{}^c&=\delta_a^c-k_a\ell^c\\
&=\Pi_a{}^\beta \Pi_\beta{}^c-\Pi_a{}^\mu   k_\mu n^{\beta} \Pi_\beta{}^c\\
&=\Pi_a{}^\mu g_{\mu\nu} (\delta_\alpha^\nu-n_\alpha k^\nu) g^{\alpha \beta} \Pi_\beta{}^c\\
&=\Pi_a{}^\mu g_{\mu\nu} \Pi_\alpha{}^\nu g^{\alpha \beta} \Pi_\beta{}^c\\
&=\Pi_a{}^\mu g_{\mu\nu} \Pi_b{}^\nu g^{\alpha \beta}\Pi_\alpha{}^b \Pi_\beta{}^c\\
&=q_{ab}g^{\alpha \beta}\Pi_\alpha{}^b \Pi_\beta{}^c\,.
\end{aligned}
\end{equation}
Since $q_a{}^c=q_{ab}q^{bc}$, we conclude that
\begin{align}
g^{\alpha \beta}\Pi_\alpha{}^a \Pi_\beta{}^b = q^{ab}\,.
\end{align}
This can be used in our derivation \eqref{term3} to finally gather
\begin{equation}
\begin{aligned}
- 2\Pi_{[c}{}^\gamma \Pi_{d]}{}^\delta \nabla_\delta V^\beta n_\beta \nabla_\gamma k^\alpha \Pi_\alpha{}^a &= V^b \theta_{db}D_c k_e \Pi_\mu{}^e g^{\mu \alpha} \Pi_\alpha{}^a  - (c \leftrightarrow d)\\
&= V^b \theta_{db}D_c k_e q^{ea}  - (c \leftrightarrow d)\\
\left[\eqref{ppk}\right]\quad &= V^b \theta_{db}\left(\btheta_c{}^a - k_c (\pi^a + \ac^a) \right)  - (c \leftrightarrow d)\,,
\end{aligned}
\end{equation}
where we used that, since $\pi_a$ and $\varphi_a$ are horizontal, $(\pi_e+\varphi_e)q^{ea}=\pi^a+\varphi^a$.

Putting things together, we eventually arrive at the Gauss equation
\begin{align}
R^a{}_{bcd} = \Pi_\alpha{}^a \Pi_b{}^\beta \Pi_c{}^\gamma \Pi_d{}^\delta R^\alpha{}_{\beta \gamma \delta} + 2\theta_{b[c}\left( k_{d]}(\pi^a +\ac^a)  -\btheta_{d]}{}^a\right) \, . \lb{NGauss}
\end{align}
To the best of our knowledge, this is the first time that this equation has been written for a null hypersurface in this form, and thus it represents a novel result.\\

From the Gauss equation, we can derive the Codazzi-Mainardi equation for the null hypersurface. To do so, let us first define the null Weingarten tensor
\begin{align}
W_a{}^b := D_a \ell^b = \theta_a{}^b + \omega_a \ell^b\,, \qquad \text{and the trace} \qquad W := W_a{}^a\,. \lb{null-Wein}
\end{align}

Before continuing, we break to demonstrate an important feature of the Weingarten tensor, which is that one can impose that it is shift invariant, deriving thereby the associated transformation of $\omega_a$. Using \eqref{shift}, \eqref{newshift}, and the fact that $\theta_{ab} q^{bc}=\theta_a{}^c$, one has
\beq
\delta_{\zeta}\theta_a{}^b=\theta_{ac}\delta_{\zeta}q^{cb}=-\theta_{ac}(\ell^c\zeta^b+\ell^b\zeta^c)=-\zeta_c\theta_{a}{}^{c}\ell^b\,,
\eeq
where we used that both $\zeta_a$ and $\theta_{ab}$ are orthogonal to $\ell^a$. Therefore, one has
\beq
\delta_{\zeta}W_a{}^b=0 \qquad \Leftrightarrow \qquad \delta_{\zeta}\omega_a=\zeta_b \theta_a{}^b\,,
\eeq
which provides how $\omega_a$ transforms on $\N$ under shifts of the Ehresmann connection, which are the intrinsic counterpart of changing the ruling. Therefore, this condition ensures that the Weingarten map is independent of the choice of rigging vector.

Then, consider 
\begin{equation}
\lb{CM-derive}
\begin{aligned}
D_c W_d{}^a - D_d W_c{}^a & = [D_c,D_d] \ell^a  \\
& =  R^a{}_{bcd} \ell^b\\
\left[\eqref{NGauss}\right]\quad & = \left( \Pi_\alpha{}^a \Pi_b{}^\beta \Pi_c{}^\gamma \Pi_d{}^\delta R^\alpha{}_{\beta \gamma \delta} - 2  \theta_{b[c}\btheta_{d]}{}^a + 2 (\pi^a +\ac^a) \theta_{b[c}k_{d]}  \right)\ell^b\,.
\end{aligned}
\end{equation}
Using that $ \theta_{bc} \ell^b =0$, we arrive at the Codazzi-Mainardi equation
\begin{equation}
\begin{aligned}
D_c W_d{}^a - D_d W_c{}^a  = \Pi_\alpha{}^a \Pi_c{}^\gamma \Pi_d{}^\delta R^\alpha{}_{\beta \gamma \delta} \ell^\beta \, .\lb{NCM}
\end{aligned}
\end{equation}

Furthermore, by contracting the $a$ and $c$ indices, we can derive
\begin{equation}
\begin{aligned}
D_c W_d{}^c - D_d W &=  \Pi_\alpha{}^\gamma \Pi_d{}^\delta R^\alpha{}_{\beta \gamma \delta} \ell^\beta \\
& =  \Pi_d{}^\delta (\delta_\alpha^\gamma - n_\alpha k^\gamma) R^\alpha{}_{\beta \gamma \delta} \ell^\beta \\
& = \Pi_d{}^\delta R_{\beta\delta} \ell^\beta - \Pi_d{}^\delta R_{\alpha \beta \gamma \delta} n^\alpha \ell^\beta k^\gamma\,.
\end{aligned}
\end{equation}
Since $n^\alpha \Neq \ell^\alpha$, the last term vanishes on the null hypersurface due to the antisymmetry of the Riemann tensor, namely $R_{(\alpha \beta) \gamma \delta} = 0$. Introducing the Einstein tensor $G_{\alpha\beta}$, we obtain
\begin{align}\label{cl}
D_b W_a{}^b - D_a W =  \Pi_a{}^\alpha R_{\alpha \beta} \ell^\beta =  \Pi_a{}^\alpha G_{\alpha \beta} \ell^\beta\,.
\end{align} 
The last equality holds on the null hypersurface because $\Pi_a{}^\alpha G_{\alpha \beta} \ell^\beta=\Pi_a{}^\alpha R_{\alpha \beta} \ell^\beta-\frac12 \Pi_a{}^\alpha R g_{\alpha \beta} \ell^\beta$, and the last term vanishes since $\Pi_a{}^\alpha  g_{\alpha \beta} \ell^\beta = \Pi_a{}^\alpha \ell_\alpha \Neq  \Pi_a{}^\alpha n_\alpha = 0$. 

Equation \eqref{cl} provides an intrinsic tensor on the null hypersurface which is covariantly conserved on-shell of the projected Einstein equations. Indeed, defining
\beq\label{nBY}
T_a{}^b=\frac1{8\pi G}\left(W_a{}^b-\delta_a{}^b W\right)\,,
\eeq
one has 
\beq
D_b T_a{}^b = \frac1{8\pi G} \Pi_a{}^\alpha G_{\alpha \beta} \ell^\beta \ \hat{=} \ 0\,,
\eeq
on-shell of the bulk equations of motion and in the absence of matter. Once split into temporal and spatial components on the Carrollian manifold (projecting with $\ell^a$ and $q_c{}^a$, respectively), the conservation of the null Brown-York stress tensor becomes exactly \eqref{EE}, that is, the Raychaudhuri and Damour equations. We conclude remarking that the null Brown-York stress tensor can be defined intrinsically on the null manifold without referring to an ambient space, and thus without linking it to a rigging construction. On the other end, the important feature of this tensor is exactly equation \eqref{cl}, which links its divergence to the bulk Einstein equations. This result, however, requires the notion of a rigging projector and thus of a rigging vector.

The tensor $T_a{}^b$ has been named the null Brown-York stress tensor because of the property we just derived, and because it reproduces from the gravitational covariant phase space the charges on a null hypersurface, exactly like its non-null counterpart.\footnote{Incidentally, this is the reason why there is a prefactor of $8\pi G$ in \eqref{nBY}.}\\ 

This concludes our overview of the first advanced topic, which is the plunging of the intrinsic Carrollian structure in an ambient pseudo-Riemannian manifold. We have seen how the null rigging procedure naturally generalizes the usual formalism for spacelike and timelike to encompass also null hypersurfaces. Moreover, we have derived the intrinsic Carrollian connection as stemming from the rigged connection induced from the bulk Levi-Civita one. This cleanly shows how the standard connection is torsionless but not metric-compatible, as its non-metric connection symbols capture important information about the extrinsic geometry of the embedded surface. Eventually, we derived the Gauss and Codazzi-Mainardi equations, relating the intrinsic Riemann tensor to its ambient counterpart. This allowed us to introduce the null Brown-York stress tensor, and derive that its conservation is nothing but the Einstein equations projected to the null hypersurface. \\

\newpage
%%%%%%%%
\section{sCarrollian Structure} \lb{sec:sCarr}

In the previous section, we examined how the Carrollian structure of a null hypersurface can be embedded in a higher-dimensional ambient spacetime. We showed that both the intrinsic ruled Carrollian structure and the Carrollian connection -- introduced respectively in Sections \ref{sec:modern} and \ref{sec:Carr-connection} -- can be induced from the bulk data encoded in the null rigging structure. This framework allowed us to derive the Gauss and Codazzi-Mainardi equations for the null hypersurface, relating the bulk curvature tensors to their intrinsic Carrollian counterparts. We concluded with the introduction of the null Brown-York tensor, whose conservation laws reproduce the Einstein equations in the bulk. In the present section, we extend this analysis beyond the null case, to hypersurfaces of arbitrary causal character.

A central motivation for this extension is to deepen the link between the physics of null hypersurfaces and that of the surrounding bulk. While it is often sufficient to study a single null hypersurface -- or boundary -- in isolation, there are many situations where one must also understand how physical degrees of freedom propagate into the bulk, in directions transverse to the null hypersurface. A natural way to approach this problem is to regard the null hypersurface $\mathcal N$ as part of a continuous family of hypersurfaces that may be timelike, spacelike, or null. Concretely, one can view the spacetime $\mathcal M$ -- or a suitable region thereof -- as foliated by the level sets of a smooth function $r(x)$,
whose zero level, $r(x)=0$, defines the null hypersurface $\mathcal N = \mathcal H_{r=0}$. The remaining level sets $\mathcal H_r$ then form a one-parameter family of hypersurfaces that approach $\mathcal N$ as $r \to 0$. For small but finite $r$, the hypersurfaces $\mathcal H_r$ can be chosen to be timelike. In this case, they are commonly referred to as stretched horizons. We will denote a single such surface by $\mathcal H$, although we stress that our formalism here applies equally-well to spacelike hypersurfaces.\\

In the conventional ADM formulation, the geometry induced on a stretched horizon differs substantially from the ruled Carrollian structure intrinsic to $\mathcal N$. This mismatch makes the null limit subtle and can lead to singular behavior, primarily because the standard ADM construction relies on the inverse induced metric, which ceases to exist in the null case where the induced metric becomes degenerate. Here, once again, the rigging technique proves invaluable. The null rigging structure introduced for $\mathcal N$ naturally extends to non-null hypersurfaces, providing a unified geometric treatment of hypersurfaces of any causal type within the same formalism.

The main message we wish to convey is that a natural extension of the ruled Carrollian structure can be induced on a stretched horizon. We refer to this extension as the stretched Carrollian or sCarrollian structure. In this framework, the ruled Carrollian structure arises as the null limit of the sCarrollian one when approaching $\mathcal N$, and all associated geometric quantities remain regular throughout this limiting process.

A sCarrollian structure provides a more intricate description of a timelike hypersurface, where a non-degenerate metric can be defined and inverted without difficulty. Yet, as already emphasized, its true advantage lies in its ability to smoothly intertwine with the null case and to provide a uniform language for all hypersurfaces, regardless of causal character. What does the sCarrollian procedure accomplish on a non-null hypersurface? Essentially, it identifies a preferred vector field $\ell$ together with an auxiliary 1-form $k$, which remains null in the ambient spacetime. When the hypersurface is non-null, the existence of an invertible metric makes $k$ redundant, since it carries no additional intrinsic information. Yet this is precisely the datum required to describe null hypersurfaces, and its inclusion on non-null ones becomes essential for maintaining a democratic treatment of all hypersurfaces within a single coherent geometric framework.\\

\subsection{Stretched Carrollian Structure}

We now turn to the construction of the stretched Carrollian structure. There are two equivalent ways to proceed. One approach is to endow the stretched horizon $\H$ directly with a stretched Carrollian structure, denoted $\SCarr$, and then show that $\SCarr$ can be induced from the bulk when $\H$ is embedded in the ambient pseudo-Riemannian manifold $\M$. This follows the same reasoning adopted earlier for the ruled Carrollian structure $\RCarr$ on the null hypersurface $\N$. Alternatively, we may start from the embedding picture and derive $\SCarr$ as the geometric structure induced on $\H$; this second viewpoint is more natural, since the necessary framework of the rigging structure has already been developed in section~\ref{rigg}.

Because the construction once again relies on the null rigging structure, the presentation will closely parallel that of section~\ref{rigg}. The only essential difference is that we now allow the stretching function $2\rho = n_\mu n^\mu$, which determines the causal nature of the hypersurface and vanishes on $\N$, to take nonzero values on $\H$. Consequently, $\rho$ must be tracked explicitly in all computations that follow.\\

Let us recap the differential geometry of the embedding and the null rigging structure presented in section \ref{sec:null-rigg}. The stretched horizon $\H$ is embedded in the ambient pseudo-Riemannian spacetime $\M$ via the inclusion map $i: \H \hookrightarrow \M$. It defines the pushforward $i_*: T\H \to T\M$ and the pullback $i^*: T^*\M \to T^*\H$, characterized by the embedding vielbein $e_a{}^\mu = i_* (\pa_a) = i^* (\rd x^\mu)$. To define the orthogonal dual $e_\mu{}^a$ such that $e_a{}^\mu e_\mu{}^b = \delta_a^b$, an additional geometric structure is required. We then introduce the null rigging\footnote{The nullity of the rigging structure refers to $k^
\mu k_\mu=0$, which is unrelated to the nullity of the hypersurface we are describing. A null rigging structure can describe hypersurfaces of any causal character.} structure $(\H, k, n)$ such that
\begin{align}
k^\mu n_\mu =1 \qquad n_\mu n^\mu = 2\rho \qquad k^\mu k_\mu =0 \qquad e_a{}^\mu n_\mu = 0 \qquad \text{and} \qquad k^\mu e_\mu{}^a =0 \, . 
\end{align}
Similar to the null case, we will furthermore impose that $\H$ is an integrable submanifold of $\M$, whose normal 1-form $n$ satisfies the Frobenius integrability condition $n \wedge \rd n = 0$. 

Introducing the rigged projector 
\begin{align}
\Pi_\mu{}^\nu = \delta_\mu^\nu - n_\mu k^\nu \quad \Rightarrow \quad \Pi_\mu{}^\nu n_\nu=0=k^\mu \Pi_\mu{}^\nu\,,
\end{align}
we define the following vector and its tangent-bundle dual 1-form, tangent to $T\H$ and $T^*\H$, respectively, as follows:
\begin{align}
\ell^\mu = n^\nu \Pi_\nu{}^\mu = n^\mu - 2\rho k^\mu\,, \qquad \text{and} \qquad k_\mu = \Pi_\mu{}^\nu k_\nu \, .\lb{ell-n}
\end{align}
As we have shown in the previous section, and will demonstrate again later, the vector $\ell^\mu$ and the 1-form $k_\mu$ are the lifts of the Carrollian vector $\ell^a$ and the Ehresmann connection $k_a$, respectively. We also remind the reader that $\ell^\mu$ and $k_\mu$ are tangent-bundle dual, meaning that $\ell^\mu k_\mu =1$. The norm squared of the Carrollian vector is $\ell^\mu \ell_\mu = - n^\mu n_\mu = -2\rho$, and one has $\ell^\mu n_\mu=0$. We emphasize that the vertical vector $\ell^\mu$ does not coincide with the normal vector $n^\mu$ unless $\rho = 0$, which holds only when evaluated on the null hypersurface $\N$. The relation $\ell^\mu = n^\mu - 2\rho k^\mu$ for $\rho \neq 0$ is central to the distinction between $\H$ and $\N$. We also define the horizontal projector $q_\mu{}^\nu$ as
\begin{align}
\Pi_\mu{}^\nu = q_\mu{}^\nu + k_\mu \ell^\nu \,,
\end{align}
satisfying $\ell^\mu q_\mu{}^\nu = 0 = q_\mu{}^\nu k_\nu$ and $q_\mu{}^\rho q_\rho{}^\nu = q_\mu{}^\nu$. This leads to the completeness relation,
\begin{align}
\delta_\mu^\nu = \Pi_\mu{}^\nu + n_\mu k^\nu = q_\mu{}^\nu + k_\mu \ell^\nu + n_\mu k^\nu\,.
\end{align}

Next, by defining the rigged metric $h_{\mu\nu} = \Pi_\mu{}^\alpha \Pi_\nu{}^\beta g_{\alpha \beta}$, we obtain the following decomposition of the spacetime metric:
\begin{align}\label{hqhq}
g_{\mu\nu} = h_{\mu\nu} + 2n_{(\mu} k_{\nu)}\,, \qquad \text{and} \qquad h_{\mu\nu} = q_{\mu\nu} - 2\rho k_\mu k_\nu\,, 
\end{align}
where we recall that $q_{\mu\nu} := q_\mu{}^\alpha q_\nu{}^\beta g_{\alpha\beta}$. We refrain from calling $h_{\mu\nu}$ the induced metric. Indeed, one should contrast this bulk tensor with $\gamma_{\mu\nu}$ introduced in \eqref{hmn}. These tensors agree on the hypersurface $\H$, consistently with the fact that the induced metric is unique on $\H$, but their bulk extensions differ. In other words, these metrics agree when inserting vector fields on $\H$ only. While the rigged metric $h_{\mu\nu}$ requires knowledge of $g_{\mu\nu}$, $n_\mu$, and the rigging vector $k^\mu$, the canonical induced metric $\gamma_{\mu\nu}$ depends only on $g_{\mu\nu}$ and $n_\mu$. The former is not necessary for non-null hypersurfaces, and in fact we derived in section \ref{531} all the relevant equations without introducing a rigging vector. However, as we emphasized at various stages in this section, on a null hypersurface only the rigged metric is well-defined, and thus we need to import this tool to non-null hypersurfaces in the spirit of providing a unified smoothly-connected framework. The fact that the rigged projector $\Pi_\mu{}^\nu$ has a different bulk extension compared with the canonical projector $\gamma_\mu{}^\nu$ will have important repercussions in the study of extrinsic curvature tensors. 

We emphasize that $q_{\mu\nu}$ is degenerate, whereas $h_{\mu\nu}$ is not, and thus as long as $\rho\neq 0$, the rigged metric $h_{\mu\nu}$ on $\H$ is invertible. The non-invertibility of $q_{\mu\nu}$ stems from
\beq
\ell_\nu=\ell^\mu g_{\mu\nu}=\ell^\mu q_{\mu\nu} -2\rho k_\nu+n_\nu=\ell^\mu q_{\mu\nu}+\ell_\nu \quad \Rightarrow \quad  \ell^\mu q_{\mu\nu}=0\,,
\eeq
from which one simply deduces
\beq\label{lh}
\ell^\mu h_{\mu\nu}=-2\rho k_\nu \,,
\eeq
which can also be directly checked computing $\ell^\mu \Pi_\mu{}^\alpha \Pi_\nu{}^\beta g_{\alpha\beta}$.
Equation \eqref{lh} encodes all the subtleties in taking $\rho\to 0$, expressing how the metric becomes degenerate, thus coinciding with $q_{\mu\nu}$ in the limit.\\

To translate between the indices on $\M$ and the internal indices on $\H$, we define the soldering as
\begin{align}
\Pi_a{}^\mu := e_a{}^\nu \Pi_\nu{}^\mu\,, \qquad \text{and} \qquad  \Pi_\mu{}^a :=  \Pi_\mu{}^\nu e_\nu{}^a\,.\label{spia}
\end{align}
These objects satisfy the following conditions,
\begin{align}
\Pi_a{}^\mu \Pi_\mu{}^b  = \delta_a^b\,, \qquad \text{and} \qquad  \Pi_\mu{}^a \Pi_a{}^\nu = \Pi_\mu{}^\nu\,. 
\end{align}

It is important to remark that we have $\Pi_\mu{}^a = e_\mu{}^a$ and $\Pi_a{}^\mu = e_a{}^\mu$. At first sight, this may appear to be a superficial change of notation, but in fact it reflects a substantive choice we made. It holds because we defined the embedding vielbein and its dual to be tangent to $T\H$ and $T^*\H$, respectively, satisfying the conditions $e_a{}^\mu n_\mu = 0 = k^\mu e_\mu{}^a$. If instead we adopted a different definition, for instance $e_\mu{}^a = h^{ab} e_b{}^\nu g_{\mu\nu}$ (which can be introduced in the case of a non-null hypersurface), then the condition $k^\mu e_\mu{}^a = 0$ would fail, and consequently $\Pi_\mu{}^a \neq e_\mu{}^a$. We stress once more that we shall use the rigging structure to define $e_\mu{}^a$, as this ensures that our construction applies uniformly to any hypersurface, independent of its causal nature.\\

We can now define the following intrinsic tensors on the stretched horizon $\H$:
\begin{align}
\ell^a = n^\mu \Pi_\mu{}^a, \qquad k_a = \Pi_a{}^\mu k_\mu\,, \qquad \text{and} \qquad h_{ab} = \Pi_a{}^\mu \Pi_b{}^\nu g_{\mu\nu} = q_{ab} - 2\rho k_a k_b\,. \lb{sCarr}
\end{align}
The vector $\ell^a$ is the Carrollian vector, $k_a$ is the Ehresmann connection, and $h_{ab}$ is the rigged metric given by the pullback of the spacetime metric onto $\H$. As we described above, the rigged metric coincides with the unique induced metric once  restricted to $\H$. The stretching $\rho$ is related to the norm squared of the Carrollian vector, that is $2\rho = -h_{ab} \ell^a \ell^b$. All of these are completely intrinsic to $\H$. The collection $\SCarr = (\H, h_{ab}, \ell^a, k_b, \rho)$ defines the stretched Carrollian structure -- called the sCarrollian structure for brevity. 

Following \eqref{lh}, from the sCarrollian structure one has the relation 
\begin{align}
h_{ab} \ell^b + 2\rho k_a = 0\,,  \lb{h-ell}
\end{align}
which demonstrates that on a sCarrollian structure the Ehresmann connection is not an extra datum to be determined, as it is already encoded in $\ell^a, h_{ab}$ and $\rho$. Therefore, strictly-speaking, $k_a$ should not be an independent entry in the definition of $\SCarr$. Nonetheless, for $\rho\to 0$, $h_{ab}\to q_{ab}$ and $h_{ab}\ell^b\to 0$, so $k_a$ in this case must be independently introduced as extra data (see \eqref{kaq}), which expresses the ruling of the hypersurface.

From the inverse bulk metric $g^{\alpha \beta}$, using the soldering we can define
\begin{align}
q^{ab} := g^{\alpha \beta} \Pi_\alpha{}^a \Pi_\beta{}^b \,. \lb{qI}
\end{align}
Interestingly, while the pullback of the metric $g_{\alpha \beta}$ on $\M$ yields the metric $h_{ab}$ on $\H$, the procedure \eqref{qI} does not yield the inverse of $h_{ab}$, as one can verify that $q^{ac}h_{cb} \neq \delta_b^a$ as follows:
\begin{equation}
\begin{aligned}\label{qhq}
q^{ac}h_{cb} &= g^{\alpha \gamma} \Pi_\alpha{}^a \Pi_\gamma{}^c \Pi_c{}^\delta \Pi_b{}^\beta g_{\delta\beta} \\
\left[\Pi_\gamma{}^c\Pi_c{}^\delta = \Pi_\gamma{}^\delta = \delta_\gamma^\delta - n_\gamma k^\delta \right] \quad
& = g^{\alpha \gamma} \Pi_\alpha{}^a (\delta_\gamma^\delta - n_\gamma k^\delta) \Pi_b{}^\beta g_{\delta\beta} \\
\left[\eqref{sCarr} \right] \quad
& = \Pi_b{}^\beta \Pi_\beta{}^a - k_b \ell^a \\
\left[\Pi_b{}^\beta\Pi_\beta{}^a = \delta_b^a = q_b{}^a + k_b \ell^a \right] \quad
& = q_b{}^a\,.
\end{aligned}
\end{equation}
One can also check that $q^{ab} k_b = k^\alpha \Pi_\alpha{}^a =0$. Since we have the decomposition \eqref{sCarr} of the metric, $h_{ab} = q_{ab} - 2\rho k_a k_b$, the above computation thus infers $q^{ac}q_{cb} = q_b{}^a$, thereby justifying the above definition of $q^{ab}$ from the bulk metric. This is another manifestation of the fact that the rigged projector is "oblique", i.e., it does not coincide with the "orthogonal" projector $\gamma_\mu{}^\nu$. \\

As we already stressed at various stages so far, unlike the degenerate metric $q_{ab}$ in the case of the null hypersurface $\N$, the metric $h_{ab}$ on the stretched horizon $\H$ is perfectly invertible. Its inverse is given by
\begin{align}
h^{ab} = q^{ab} - \frac{1}{2\rho} \ell^a \ell^b\,, \qquad \text{satisfying} \qquad h^{ac}h_{cb} = q_b{}^a +k_b \ell^a = \delta_b^a\,. 
\end{align}
From this result, it is readily clear that the inverse $h^{ab}$ is undefined when $\rho = 0$. 
Similarly, one notes that while $k_\mu$ is null in the bulk, its projection to the surface is not null, and instead its norm on $\H$ satisfies $k_a h^{ab}k_b=-\frac1{2\rho}$. This is another indication that such a 1-form must be introduced as extra datum on a null hypersurface, as its norm would be divergent in the limit. Hence, to smoothly connect the geometry of the stretched horizon $\H$ to that of the null hypersurface $\N$, we refrain from using the inverse metric $h^{ab}$, and instead rely only on $h_{ab}, q^{ab},\ell^a, k_b$, which admit a smooth null limit.\\

%%%%%%%%%%%%%%%%%%%%%
\subsection{sCarrollian Connection}

As discussed in section \ref{sec:rigg-connection}, the Carrollian connection is induced on $\N$ from the bulk Levi-Civita connection. Here, we parallel that analysis and define the sCarrollian connection using the soldering,
\begin{align}
D_a T_b{}^c = \Pi_a{}^\alpha \Pi_b{}^\beta (\nabla_\alpha T_\beta{}^\gamma) \Pi_\gamma{}^c\,,
\end{align}
where $T_b{}^c$ is a generic tensor on the stretched horizon $\H$, and $T_\beta{}^\gamma = \Pi_\beta{}^b T_b{}^c \Pi_c{}^\gamma$ denotes its lift to the spacetime $\M$.

The sCarrollian connection is torsionless, as one can explicitly check following the same derivation as in section \ref{sec:rigg-connection}. How does this connection act on the sCarrollian structure? We can start by evaluating $D_a k_b$. The derivation \eqref{ppk} does not rely on the null condition $\rho = 0$, and thus it holds for both the null hypersurface $\N$ and the non-null hypersurface $\H$.  We therefore have that 
\begin{align}
D_a k_b = \Pi_a{}^\alpha \Pi_b{}^\beta \nabla_\alpha k_\beta = \btheta_{ab} - \omega_a k_b - k_a (\pi_b + \ac_b)\,,\label{dks}
\end{align}
where we recall the definitions
\begin{align}
\btheta_{ab} := q_a{}^\alpha q_b{}^\beta \nabla_\alpha k_\beta, \qquad \text{and} \qquad \omega_a := \Pi_a{}^\alpha \nabla_\alpha \ell^\beta k_\beta = \pi_a + \kappa k_a\,. \label{B}
\end{align}
We also define the lifts of the Carrollian acceleration and the Carrollian vorticity as in \eqref{dkb}, $\pa_{[\alpha} k_{\beta]} = \ac_{[\alpha} k_{\beta]} - \frac{1}{2} \vor_{\alpha\beta}$. \\

Next, we evaluate $D_a \ell^b$, following the same line of derivation as in \eqref{Dl},
\begin{equation}
\begin{aligned}
D_a \ell^b &= \Pi_a{}^\alpha \nabla_\alpha \ell^\beta \Pi_\beta{}^b \\
& = q_a{}^\alpha \nabla_\alpha \ell^\beta q_\beta{}^b + \Pi_a{}^\alpha \nabla_\alpha \ell^\beta k_\beta \ell^b + k_a \ell^\alpha \nabla_\alpha \ell^\beta q_\beta{}^b \\
& = \theta_a{}^b +\omega_a \ell^b + k_a A^b \,,\label{dls}
\end{aligned}
\end{equation}
where, besides $\omega_a$, we define the connection coefficients 
\begin{align}
\theta_a{}^b := q_a{}^\alpha \nabla_\alpha \ell^\beta q_\beta{}^b\,, \qquad \text{and} \qquad A^b := \ell^\alpha \nabla_\alpha \ell^\beta q_\beta{}^b\,.\label{A}
\end{align}
The last term encodes the acceleration of the Carrollian vector in the horizontal directions, satisfying $A^a k_a=0$. It is a new feature of the sCarrollian connection, which vanishes on the null hypersurface $\N$, as we proved in \eqref{lnl}. On the non-null hypersurface $\H$, however, the horizontal acceleration $A^a$ does not necessarily vanish, providing a new term in \eqref{dls} compared to \eqref{Dl2}. 

Considering $A_a = q_{ab} A^b$, we can show that 
\begin{equation}
\begin{aligned}
A_a & = q_a{}^\alpha \ell^\beta \nabla_\beta \ell_\alpha \\
\left[\ell_\alpha = n_\alpha -2\rho k_\alpha, \ q_a{}^\alpha k_\alpha =0   \right] \quad
&= q_a{}^\alpha \ell^\beta \left(\nabla_\beta n_\alpha - 2\rho \nabla_\beta k_\alpha \right)\\
\left[\nabla_{[\alpha} n_{\beta]} = \alpha_{[\alpha} n_{\beta]}  \right] \quad
&= q_a{}^\alpha \ell^\beta \left(\nabla_\alpha n_\beta - \alpha_\alpha n_\beta +\alpha_\beta n_\alpha - 2\rho \nabla_\beta k_\alpha \right) \\
\left[\ell^\beta n_\beta =0, \ q_a{}^\alpha n_\alpha =0  \right] \quad
&= q_a{}^\alpha \ell^\beta \left(\nabla_\alpha n_\beta - 2\rho \nabla_\beta k_\alpha \right) \\
\left[n_\beta = \ell_\beta +2\rho k_\beta   \right]  \quad
&= q_a{}^\alpha \ell^\beta \left(\nabla_\alpha \ell_\beta + 2k_\beta \nabla_\alpha\rho  + 4\rho \partial_{[\alpha} k_{\beta]} \right)\,.
\end{aligned}
\end{equation}
Note that $q_a{}^\alpha n_\alpha=0$ because $q_a{}^\alpha n_\alpha=\Pi_a{}^\beta q_\beta{}^\alpha n_\alpha=\Pi_a{}^\beta(\Pi_\beta{}^\alpha-k_\beta\ell^\alpha)n_\alpha=0$, since $\Pi_\beta{}^\alpha n_\alpha=0=\ell^\alpha n_\alpha$.

Using
\beq
q_a{}^\alpha \ell^\beta(\pa_{[\alpha} k_{\beta]})= q_a{}^\alpha \ell^\beta(\ac_{[\alpha} k_{\beta]} - \frac{1}{2} \vor_{\alpha\beta})=\frac{\ac_a}{2}\,,
\eeq
and $\ell^\beta \nabla_\alpha \ell_\beta = \tfrac{1}{2} \nabla_\alpha (\ell^\beta \ell_\beta) = -\nabla_\alpha \rho$, we arrive to the final result
\beq
A_a = \overline{\pa}_a \rho + 2\rho \ac_a \,,
\eeq
where we recall the definition of the horizontal derivative $\overline{\pa}_a \rho = q_a{}^\alpha \nabla_\alpha \rho$. This final result demonstrates that 
\beq
\lim_{\rho \to 0}A^a=0\,,\label{Anl}
\eeq
which in turn explains why the action of the connection on $\ell$ on $\H$ (see \eqref{dls}) smoothly transitions to \eqref{Dl2} in the null limit.\\

In general, the tensor $\theta_{ab} = \theta_a{}^c q_{cb} = q_a{}^\alpha q_b{}^\beta \nabla_\alpha \ell_\beta$ is not symmetric, except when evaluated on the null hypersurface $\N$. To evaluate its antisymmetric part, we use the relation $\ell_\beta = n_\beta - 2\rho k_\beta$ given in \eqref{ell-n} to show that 
\begin{equation}
\lb{theta[ab]}
\begin{aligned}
\theta_{[ab]} &= q_a{}^\alpha q_b{}^\beta \nabla_{[\alpha} \ell_{\beta]} \\
\left[q_b{}^\beta k_\beta = 0\right]  \quad
& = q_a{}^\alpha q_b{}^\beta \nabla_{[\alpha} n_{\beta]} - 2\rho q_a{}^\alpha q_b{}^\beta \nabla_{[\alpha} k_{\beta]} \\
\left[\nabla_{[\alpha} n_{\beta]} = \alpha_{[\alpha} n_{\beta]}, \ q_a{}^\alpha n_\alpha =0 \right]  \quad
& = - 2\rho q_a{}^\alpha q_b{}^\beta \pa_{[\alpha} k_{\beta]} \\
\left[\eqref{dkb}\right] \quad
& = \rho \vor_{ab}\,.
\end{aligned}
\end{equation}
The fact that $\theta_{[ab]}$ is related to the Carrollian vorticity is reminiscent of $\btheta_{[ab]} = - \tfrac{1}{2}\vor_{ab}$, and indeed the combination $\theta_{ab} + 2\rho \btheta_{ab}$ is always symmetric. As claimed, we explicitly see that $\theta_{[ab]}$ vanishes when evaluated on the null hypersurface $\N$, where $\rho = 0$. \\

We have studied how the sCarrollian connection acts on $\ell^a$ and $k_a$. We now focus on its action on the metric, $D_a h_{bc}$. Before proceeding, we recall that the 1-form $n_\mu$ and vector $k^\mu$ are extrinsic quantities, annihilated by the rigged projector. These geometric data instruct us about the geometric structure of the ambient space in the proximity of the surface. We thus define the extrinsic curvature tensors on $\H$
\beq
& N_{ab} : = \Pi_a{}^\alpha \nabla_\alpha n_\beta \Pi_b{}^\beta = \Pi_a{}^\alpha \nabla_\alpha \ell_\beta \Pi_b{}^\beta + 2 \Pi_a{}^\alpha \nabla_\alpha(\rho k_\beta) \Pi_b{}^\beta \,, & \lb{news}\\
& S_a{}^b := \Pi_a{}^\alpha \nabla_\alpha k^\beta \Pi_\beta{}^b\,, & \label{shape}
\eeq
where we recall that $n_\beta = \ell_\beta + 2\rho k_\beta$ from \eqref{ell-n}. We remark that both these tensors do not coincide with the extrinsic curvature defined in \eqref{ext} since they are defined using the rigged projector. The tensor $S_a{}^b$ is related to $K_a{}^b$ only when the rigged projector is chosen to be non-null, and proportional to $n^\mu$ itself -- a setup not admitting a smooth null limit, and thus disregarded in what follows. We have chosen to name the tensor in \eqref{news} with "N" to recall that it is associated to the normal $n_\mu$, while the quantity in \eqref{shape} is named "S" because it reminds us of the shape operator in deformation theory.\\

We want to relate these tensors to 
\beq
D_a k_b =\Pi_a{}^\alpha\nabla_\alpha k_\beta \Pi_b{}^\beta\,,\label{dkkk}
\eeq
and to the non-null analogue of the Weingarten tensor defined in \eqref{null-Wein},
\begin{align}
W_a{}^b := D_a \ell^b=\Pi_a{}^\alpha\nabla_\alpha \ell^\beta \Pi_\beta{}^b\,. \lb{W-Wb}
\end{align}
To do so, we evaluate
\begin{equation}
\lb{Kb}
\begin{aligned}
S_a{}^b & = \Pi_a{}^\alpha \nabla_\alpha k^\beta \Pi_\beta{}^b \\
& = \Pi_a{}^\alpha \nabla_\alpha k_\gamma g^{\beta\gamma} \Pi_\beta{}^b \\
\left[\delta^\gamma_\mu = \Pi_\mu{}^\gamma + n_\mu k^\gamma= \Pi_\mu{}^c \Pi_c{}^\gamma+n_\mu k^\gamma   \right] \quad
& = \Pi_a{}^\alpha \nabla_\alpha k_\gamma (\Pi_\mu{}^c \Pi_c{}^\gamma + n_\mu k^\gamma) g^{\mu\beta} \Pi_\beta{}^b \\
\left[k^\gamma \nabla_\alpha k_\gamma =0\right]\quad & = \Pi_a{}^\alpha \Pi_c{}^\gamma \nabla_\alpha k_\gamma \Pi_\mu{}^c g^{\mu\beta} \Pi_\beta{}^b\\
\left[\eqref{qI}\right]\quad & =\Pi_a{}^\alpha \Pi_c{}^\gamma \nabla_\alpha k_\gamma q^{cb}\\
& = D_a k_c q^{cb}\,.
\end{aligned}
\end{equation}
Next, we compute
\begin{equation}
\begin{aligned}
\Pi_a{}^\alpha \nabla_\alpha \ell_\beta \Pi_b{}^\beta & = \Pi_a{}^\alpha \nabla_\alpha \ell^\gamma g_{\beta\gamma} \Pi_b{}^\beta\\
\left[\delta^\gamma_\mu = \Pi_\mu{}^\gamma + n_\mu k^\gamma= \Pi_\mu{}^c \Pi_c{}^\gamma+n_\mu k^\gamma   \right]\quad & = \Pi_a{}^\alpha \nabla_\alpha \ell^\mu (\Pi_\mu{}^c \Pi_c{}^\gamma+n_\mu k^\gamma) g_{\beta\gamma} \Pi_b{}^\beta\\
\left[\eqref{W-Wb} \ ,  \ \eqref{sCarr} \ , \ k^\gamma g_{\gamma\beta}\Pi_b{}^\beta=k_b\right]\quad & = W_a{}^c h_{cb}+\Pi_a{}^\alpha \nabla_\alpha \ell^\mu n_\mu k_b\\
\left[ \ell^\mu n_\mu =0\right]\quad & = W_a{}^c h_{cb}-\Pi_a{}^\alpha \ell^\mu \nabla_\alpha n_\mu k_b\\
\left[ n_\mu=\ell_\mu+2\rho k_\mu \ , \ \ell^\mu \nabla_\alpha \ell_\mu=- \nabla_\alpha\rho\right]\quad & = W_a{}^c h_{cb}+\Pi_a{}^\alpha(\nabla_\alpha\rho-2\ell^\mu \nabla_\alpha (\rho k_\mu)) k_b\\
\left[l^\mu k_\mu=1\right]\quad & = W_a{}^c h_{cb}+\Pi_a{}^\alpha(-\nabla_\alpha\rho-2 \rho \ell^\mu \nabla_\alpha k_\mu) k_b\\
\left[\ell^\mu=\Pi_c{}^\mu \ell^c\right]\quad & = W_a{}^c h_{cb}-D_a\rho k_b-2 \rho \Pi_a{}^\alpha \ell^c\Pi_c{}^\mu\nabla_\alpha k_\mu k_b\\
\left[\eqref{dkkk}\right]\quad & = W_a{}^c h_{cb}-D_a\rho k_b-2 \rho D_a k_c  \ell^c k_b\,.
\end{aligned}
\end{equation}
This informs us that \eqref{news} is
\begin{equation}
\begin{aligned}
N_{ab} & = \Pi_a{}^\alpha \nabla_\alpha \ell_\beta \Pi_b{}^\beta + 2 \Pi_a{}^\alpha \nabla_\alpha(\rho k_\beta) \Pi_b{}^\beta\\
\left[\eqref{dkkk} \ , \ \Pi_a{}^\alpha \nabla_\alpha \rho=D_a\rho\right]\quad & = W_a{}^c h_{cb}-D_a\rho k_b-2 \rho D_a k_c  \ell^c k_b + 2\rho D_a k_b+2 D_a\rho k_b \\
\left[q_b{}^c = \delta_b^c - k_b\ell^c  \right] \quad & =W_a{}^c h_{cb}+D_a\rho k_b + 2\rho D_a k_c q_b{}^c \\
\left[\eqref{qhq} \right]\quad & =(W_a{}^d+2\rho D_a k_c q^{cd}) h_{db}+D_a\rho k_b \\
\left[\eqref{Kb}\right]\quad &= (W_a{}^d+2\rho S_a{}^d) h_{db}+D_a\rho k_b\,,   \lb{N}
\end{aligned}
\end{equation}
which expresses how these various tensors are related.\\

We now evaluate $D_a h_{bc}$. In \eqref{Dq-derive}, we have demonstrated that
\begin{equation}
\begin{aligned}
\Pi_a{}^\alpha \Pi_b{}^\beta \Pi_c{}^\gamma \nabla_\alpha \left(\Pi_\beta{}^\mu \Pi_\gamma{}^\nu g_{\mu \nu} \right) & = 2\Pi_a{}^\alpha \nabla_\alpha \Pi_\beta{}^\mu \Pi_{(b}{}^\beta \Pi_{c)}{}^\nu g_{\mu\nu} \\
& = -2 \Pi_a{}^\alpha \nabla_\alpha n_\beta \Pi_{(b}{}^\beta \Pi_{c)}{}^\nu k^\mu g_{\mu\nu} \\
& =  -2 \Pi_a{}^\alpha \nabla_\alpha n_\beta \Pi_{(b}{}^\beta k_{c)}\,,
\end{aligned}
\end{equation}
without ever imposing $\rho=0$. Therefore, this result carries over to this section, except that now $\Pi_\beta{}^\mu \Pi_\gamma{}^\nu g_{\mu \nu}=h_{\beta\gamma}$, leading to $\Pi_a{}^\alpha \Pi_b{}^\beta \Pi_c{}^\gamma \nabla_\alpha \left(\Pi_\beta{}^\mu \Pi_\gamma{}^\nu g_{\mu \nu} \right)=D_a h_{bc}$.\footnote{The difference with the null case in \eqref{Dq-derive} is that, there,  $h_{ab}\Neq q_{ab}$, see \eqref{sCarr} and the discussion in footnote \ref{ff}.} Thus, given \eqref{news}, we obtain the compact final result 
\beq
D_a h_{bc}  =  -2 \Pi_a{}^\alpha \nabla_\alpha n_\beta \Pi_{(b}{}^\beta k_{c)}= -2 N_{a(b} k_{c)}\,.\label{dhs}
\eeq
This shows that the sCarrollian connection $D_a$ is not compatible with the induced metric $h_{ab}$, just as the Carrollian connection is not compatible with the metric $q_{ab}$. Since the stretched horizon $\H$ is not a null hypersurface, there exists a Levi-Civita metric-compatible connection. However, such a connection depends on the use of the inverse induced metric $h^{ab}$, which, as before, we avoid using. It is not hard to verify that, starting from \eqref{N}, in the $\rho\to 0$ limit, $N_{ab} \Neq \theta_{ab}$, and therefore $D_a h_{bc} \Neq D_a q_{bc}$ in a smooth manner. Yet again, we see that the price to pay in order to treat hypersurfaces of all causal nature democratically is to give up metric-compatibility in the connection. This is expected, as we showed from the intrinsic perspective in section \ref{sec:Carr-connection} that we cannot impose metric compatibility and torsionless for a general Carrollian geometry, and thus the null limit of a timelike hypersurface becomes intractable if we describe it with the metric compatible connection.\\

Let us also evaluate $D_a q_{bc}$ on the stretched horizon $\H$. Using the relation \eqref{ell-n}, that is $q_{ab} = h_{ab} + 2\rho k_a k_b$, we have that, 
\begin{equation}
\begin{aligned}
D_a q_{bc} &= D_a h_{bc} + 2 D_a \rho k_b k_c + 4\rho D_{a}k_{(b}k_{c)} \\
\left[\eqref{dhs}\right]\quad &= - 2\left( N_{a(b} -2\rho D_{a}k_{(b}-D_a \rho k_{(b}\right)k_{c)}\\
\left[\eqref{N}\right]\quad &= - 2\left( (W_a{}^d+2\rho S_a{}^d) h_{d(b} -2\rho D_{a}k_{(b}\right)k_{c)}\\ 
\left[\eqref{Kb}\right]\quad & = - 2\left(W_a{}^d h_{d(b}+2\rho D_a k_e q_{(b}{}^{e} -2\rho D_{a}k_{(b}\right)k_{c)}\\
& = - 2\left(W_a{}^d h_{d(b}-2\rho D_a k_d k_{(b}\ell^{d}\right)k_{c)}\, .
\end{aligned}
\end{equation}
We can then process these two contributions. The first one gives
\begin{equation}
\begin{aligned}
W_a{}^c h_{cb}  & = D_a \ell^c h_{cb} \\
\left[ \eqref{dls}\right] \quad & = (\theta_a{}^c +\omega_a \ell^c + k_a A^c)h_{cb} \\
\left[\eqref{h-ell} \ , \ \theta_a{}^c k_c=0=A^c k_c\right]\quad & = \theta_{ab} -2\rho \omega_a k_b + k_a A_b \,,
\end{aligned}
\end{equation}
whereas the second contribution can be evaluated directly from \eqref{dks}, $-2\rho D_a k_d \ell^d k_{b}= 2\rho  \omega_a k_{b}$, and cancels the second term in the first contribution. Therefore, putting things together, the final result is
\beq
D_a q_{bc}= -2\left(\theta_{a(b} + k_a A_{(b}\right)k_{c)} \label{dqs}\,.
\eeq
This equation beautifully encodes how the connection acts on the degenerate metric, while also demonstrating the smoothness of the null limit. Indeed, as shown in \eqref{Anl} , one has $A_a \Neq 0$, and thus \eqref{dqs} reduces to \eqref{Dq} on $\N$. For completeness, we additionally evaluate $D_a q^{bc}$ as follows
\begin{equation}
\lb{DqI}
\begin{aligned}
D_a q^{bc} &= D_a \left( g^{\beta \gamma} \Pi_\beta{}^b \Pi_\gamma{}^c \right) \\
&= \Pi_a{}^\alpha \nabla_\alpha\left( g^{\beta \gamma} \Pi_\beta{}^\mu \Pi_\gamma{}^\nu \right) \Pi_\mu{}^b \Pi_\nu{}^c \\
\left[\Pi_\beta{}^\mu = \delta_\beta^\mu -n_\beta k^\mu, \ k^\mu\Pi_\mu{}^b =0 \right] \quad
&= -2 \Pi_a{}^\alpha n^\nu \nabla_\alpha k^\mu  \Pi_{\mu}{}^{(b} \Pi_{\nu}{}^{c)} \\
\left[\eqref{shape}, \ \ell^c = n^\nu\Pi_\nu{}^c \right] \quad
&= - 2S_a{}^{(b} \ell^{c)}\, .
\end{aligned}
\end{equation}\\

We conclude this subsection with a brief recap of the various tensors we just introduced on the stretched horizon $\H$. The tensors $N_{ab}$ and $S_a{}^b$ are the extrinsic curvatures of the hypersurface in the rigging formalism, related as in \eqref{N}. We found that their decompositions in terms of the Carrollian data is
\beq
& N_{ab} = \theta_{ab}+2\rho \btheta_{ab}+k_a(A_b-2\rho (\pi_b+\ac_b))+k_b(D_a-2 \omega_a)\rho \label{Ncs} & \\
& S_a{}^b  = \btheta_{ac}q^{cb}-k_a (\pi^b+\varphi^b) \lb{Wb-decom}\,, &
\eeq
while we also recall
\beq
W_a{}^b = \Pi_a{}^\alpha \nabla_\alpha \ell^\beta \Pi_\beta{}^b  = \theta_a{}^b + \omega_a \ell^b + k_a A^b\,. \lb{W-decom}
\eeq\\

%%%%%%%%%%%%%%%%%%%%%
\subsection{Gauss, Codazzi-Mainardi, and sCarrollian Stress Tensor}

In this final section, we discuss the Gauss equation, the Codazzi-Mainardi equation, and the sCarrollian stress tensor, following closely the discussion in section \ref{sec:NGC}.

We begin with the Gauss equation, which relates the intrinsic curvature tensor on the stretched horizon $\H$ to its bulk counterpart. We can directly import the derivation we did in the null case, since \eqref{Gauss-derive} was computed without ever imposing the null condition $\rho=0$,
\begin{equation}
\begin{aligned}\label{gs}
R^a{}_{bcd} V^b = [D_c, D_d]V^a = V^b \Pi_\alpha{}^a \Pi_b{}^\beta \Pi_c{}^\gamma \Pi_d{}^\delta R^\alpha{}_{\beta \gamma \delta} - 2\Pi_{[c}{}^\gamma \Pi_{d]}{}^\delta \nabla_\delta V^\alpha n_\alpha \nabla_\gamma k^\nu \Pi_\nu{}^a\,,
\end{aligned}
\end{equation}
for an arbitrary vector $V^b$ on $\H$. We remind the reader that this equation holds when $\H$ is an integrable submanifold of the spacetime $\M$, satisfying Frobenius theorem. 

Using $V^\alpha n_\alpha=0$ and $V^\alpha=\Pi_b{}^\alpha V^b$, the second term becomes
\beq
- 2\Pi_{[c}{}^\gamma \Pi_{d]}{}^\delta \nabla_\delta V^\alpha n_\alpha \nabla_\gamma k^\nu \Pi_\nu{}^a=2\Pi_{[c}{}^\gamma \Pi_{d]}{}^\delta V^b \Pi_b{}^\alpha \nabla_\delta n_\alpha \nabla_\gamma k^\nu \Pi_\nu{}^a\,.
\eeq
This expression exactly features the extrinsic tensor \eqref{news} and \eqref{shape}. Therefore, the Gauss equation for the stretched horizon using the rigged projector acquires the evocative form
\begin{align}
R^a{}_{bcd} = \Pi_\alpha{}^a \Pi_b{}^\beta \Pi_c{}^\gamma \Pi_d{}^\delta R^\alpha{}_{\beta \gamma \delta} - 2 S_{[d}{}^a N_{c]b} \,. \lb{HGauss}
\end{align}
As we discussed below \eqref{dhs}, $N_{ab}\Neq \theta_{ab}$, which can also be derived simply from \eqref{Ncs}. Moreover, \eqref{Wb-decom} is unaffected by the null limit. Thus, we obtain
\begin{equation}
\begin{aligned}
R^a{}_{bcd} & = \Pi_\alpha{}^a \Pi_b{}^\beta \Pi_c{}^\gamma \Pi_d{}^\delta R^\alpha{}_{\beta \gamma \delta} - 2 S_{[d}{}^a N_{c]b}\\
& \Neq \Pi_\alpha{}^a \Pi_b{}^\beta \Pi_c{}^\gamma \Pi_d{}^\delta R^\alpha{}_{\beta \gamma \delta} + 2\theta_{b[c}\left( k_{d]}(\pi^a +\ac^a)  -\btheta_{d]}{}^a\right)\,, 
\end{aligned}
\end{equation}
where we used that $N_{ab}$ is symmetric on $\N$. The null limit exactly reproduces \eqref{NGauss} -- a non-trivial confirmation. Both \eqref{HGauss} and its null limit are novel results of this review. \\

We now proceed to the Codazzi-Mainardi equation. First, following the same approach as in \eqref{CM-derive}, we consider
\begin{equation}
\begin{aligned}
D_c W_d{}^a - D_d W_c{}^a & = [D_c,D_d] \ell^a  \\
& =  R^a{}_{bcd} \ell^b\\
\left[\eqref{HGauss}  \right] \quad
& = \left( \Pi_\alpha{}^a \Pi_b{}^\beta \Pi_c{}^\gamma \Pi_d{}^\delta R^\alpha{}_{\beta \gamma \delta} - 2 S_{[d}{}^a N_{c]b} \right)\ell^b\,.
\end{aligned}
\end{equation}
Recalling the decomposition \eqref{Ncs}, one has
\begin{align}
N_{cb}\ell^b = \left(D_c - 2 \omega_c\right)\rho\,.
\end{align}
We finally arrive at 
\begin{equation}
\lb{D-W}
\begin{aligned}
D_c W_d{}^a - D_d W_c{}^a  = \Pi_\alpha{}^a \Pi_c{}^\gamma \Pi_d{}^\delta R^\alpha{}_{\beta \gamma \delta} \ell^\beta- 2 S_{[d}{}^a \left(D_{c]} - 2 \omega_{c]}\right)\rho\,.
\end{aligned}
\end{equation}
Indeed, the last term vanishes when $\rho =0$, and again we confirm that in the null limit one recovers \eqref{NCM}. This is the Codazzi-Mainardi equation for $\ell^\mu$. \\

It is instructive to derive the Codazzi-Mainardi equation for $n^\mu$. To do so, we perform a similar computation for $D_d k_b$, that is 
\begin{equation}
\lb{D-Wb}
\begin{aligned}
D_c D_d k_b - D_d D_c k_b 
& =  - R^e{}_{bcd} k_e \\
\left[\eqref{HGauss}\right] \quad
& = -\left( \Pi_\alpha{}^e \Pi_b{}^\beta \Pi_c{}^\gamma \Pi_d{}^\delta R^\alpha{}_{\beta \gamma \delta} - 2S_{[d}{}^e N_{c]b} \right)k_e \\
& = \Pi_b{}^\alpha \Pi_c{}^\gamma \Pi_d{}^\delta R_{\alpha \beta \gamma \delta} k^\beta\,,
\end{aligned}
\end{equation}
where we used that, from \eqref{Wb-decom}, $S_a{}^b k_b=0$.

Then, recalling the relation \eqref{Kb}, $S_d{}^a = D_d k_b q^{ba}$ and the formula \eqref{DqI}, we have that
\begin{equation}
\begin{aligned}
D_c S_d{}^a - D_d S_c{}^a &= q^{ab}D_c D_{d} k_{b} -2 D_{d} k_{b} S_c{}^{(a} \ell^{b)} - (c \leftrightarrow d)\,.  
\end{aligned}
\end{equation}
To proceed, we remark $D_{[d} k_b S_{c]}{}^b=D_{[d} k_b D_{c]} k_a q^{ab}=0$.
Using then $D_d k_b \ell^b = - \omega_d$ (directly derived from \eqref{dks}), and \eqref{D-Wb}, we gather
\begin{equation}
\begin{aligned}
D_c S_d{}^a - D_d S_c{}^a &= q^{ab}\Pi_b{}^\alpha \Pi_c{}^\gamma \Pi_d{}^\delta R_{\alpha \beta \gamma \delta} k^\beta + S_c{}^a \omega_d - S_d{}^a \omega_c \,.
\end{aligned}
\end{equation}
The first term in the RHS can be further manipulated using \eqref{qI},
\begin{equation}
\begin{aligned}
q^{ab}\Pi_b{}^\alpha R_{\alpha \beta \gamma \delta} k^\beta & = g^{\mu\nu}\Pi_\mu{}^a \Pi_\nu{}^b\Pi_b{}^\alpha R_{\alpha \beta \gamma \delta} k^\beta \\
\left[\Pi_\nu{}^b\Pi_b{}^\alpha = \Pi_\nu{}^\alpha = \delta_\nu^\alpha - n_\nu k^\alpha   \right] \quad
&= \Pi_\mu{}^a (g^{\mu\alpha} - n^\mu k^\alpha) R_{\alpha \beta \gamma \delta} k^\beta \\
\left[k^\alpha k^\beta R_{\alpha \beta \gamma \delta} =0 \right] \quad
&= \Pi_\alpha{}^a R^\alpha{}_{\beta \gamma \delta} k^\beta \,.
\end{aligned}
\end{equation}
This eventually leads us to
\begin{equation}
\lb{D-Kb}
\begin{aligned}
(D_c+\omega_c) S_d{}^a - (D_d +\omega_d)S_c{}^a &= \Pi_\alpha{}^a \Pi_c{}^\gamma \Pi_d{}^\delta R^\alpha{}_{\beta \gamma \delta} k^\beta \,.
\end{aligned}
\end{equation}\\

Then, using that $n^\mu=\ell^\mu+2\rho k^\mu$, we can combine \eqref{D-W} and \eqref{D-Kb} and derive the Codazzi-Mainardi equation for $n^\mu$
\begin{equation}
\begin{aligned}
\Pi_\alpha{}^a \Pi_c{}^\gamma \Pi_d{}^\delta R^\alpha{}_{\beta \gamma \delta} n^\beta &= 2\rho D_c S_d{}^a -2\rho D_d S_c{}^a+S_{d}{}^a D_{c}\rho - S_{c}{}^a D_{d}\rho +D_c W_d{}^a - D_d W_c{}^a\\
& = D_c(W_d{}^a+2\rho S_d{}^a) - D_d(W_c{}^a+2\rho S_c{}^a)-S_{d}{}^a D_{c}\rho + S_{c}{}^a D_{d}\rho\,.\label{dsR}
\end{aligned}
\end{equation}

By contracting the $a$ and $c$ indices, using $R^{(\alpha \beta)}{}_{\gamma \delta} = 0$, the LHS becomes
\begin{align}
\Pi_\alpha{}^a \Pi_c{}^\gamma \Pi_d{}^\delta R^{\alpha\beta}{}_{\gamma \delta} n_\beta = \Pi_\alpha{}^\gamma \Pi_d{}^\delta R^{\alpha\beta}{}_{\gamma \delta} n_\beta = \left(\delta_\alpha^\gamma - n_\alpha k^\gamma \right)\Pi_d{}^\delta R^{\alpha\beta}{}_{\gamma \delta} n_\beta = \Pi_d{}^\delta R_\delta{}^\beta n_\beta \, .
\end{align}
Furthermore, $\Pi_d{}^\delta R_\delta{}^\beta n_\beta = \Pi_d{}^\delta (G_\delta{}^\beta + \frac{1}{2}R \delta_\delta^\beta) n_\beta = \Pi_d{}^\delta G_\delta{}^\beta n_\beta$, since $\Pi_d{}^\delta n_\delta =0$. Therefore,
\begin{align}
\Pi_a{}^\alpha G_\alpha{}^\beta n_\beta  = D_b \left(W_a{}^b-W \delta_a{}^b+2\rho (S_a{}^b -S\delta_a^b)\right)-\left( S_a{}^b - S \delta_a^b \right) D_b \rho \,,
\end{align}
where we defined the traces $W := W_a{}^a$ and $S = S_a{}^a$. It is evident that this equation gives exactly \eqref{cl} when evaluated on the null hypersurface $\N$.\\

Importantly, on a non-null hypersurface this projection of Einstein equations seems to lead to a non-conserved tensor, due to the presence of the stretching function $\rho$. Nonetheless, one can define  the sCarrollian stress tensor on the stretched horizon $\H$ as
\begin{align}\label{sBY}
T_a{}^b := \frac{1}{8\pi G}\left(W_a{}^b-W \delta_a{}^b+2\rho (S_a{}^b -S\delta_a^b) \right)\,.
\end{align}
We then arrive at
\begin{align}
\Pi_a{}^\alpha G_\alpha{}^\beta n_\beta  = 8\pi G D_b T_a{}^b-\left( S_a{}^b - S \delta_a^b \right) D_b \rho \,. \lb{DT-H}
\end{align}
This encodes Einstein equations as (non-)conservation laws for the sCarrollian stress tensor. This leads to a conserved stress tensor whenever one can impose $D_a\rho=0$ -- see below --, or $S_a{}^b-S \delta_a{}^b=0$. In the latter case, one has that the tensor $T_a{}^b=\frac1{8\pi G}(W_a{}^b-W\delta_a{}^b)$, which is as in \eqref{nBY}, is exactly conserved on shell, both for a stretched and a null horizon.

Importantly, although the purpose of this section was to study the geometry of the stretched horizon, namely a non-null hypersurface located near a null one, this assumption has not been explicitly used in our derivation. We only considered the limit $\rho = 0$ as a consistency check with the findings of the previous section. The results obtained here therefore apply to any hypersurface $\H$ embedded in the pseudo-Riemannian spacetime $\M$, independently of whether a null hypersurface $\N$ is present.\\

If a null hypersurface $\N$ is indeed present, one can then construct a conserved sCarrollian stress tensor by consistently setting $D_a\rho=0$. Let us derive this result. As explained at the beginning of this section, the stretched horizon $\H_r$ and the null hypersurface $\N$ can be viewed as leaves of a foliation defined by a function $r(x) = \text{constant}$, with $r(x)=0$ corresponding to $\N$. Since $r(x)$ is constant on both $\H$ and $\N$, we have
\beq
D_a r(x)=0\,.
\eeq
Now the task is to identify $\rho$ with $r$. The strategy is to exploit the scaling freedom of the sCarrollian structure. Recall that the ruled Carrollian structure is defined up to internal transformations, as discussed in section \ref{sec:C-symm}. Among these is the rescaling of the Carrollian vector and the Ehresmann connection. This notion of symmetry also applies to the sCarrollian structure. However, the rescaling $\ell^a \to \Phi \ell^a$ for a function $\Phi(x)$ simultaneously rescales the stretching as $\rho \to \Phi^2 \rho$. Since $\rho \Neq 0 \Neq r$, one can set the scale of the Carrollian vector such that $\rho = r$. By construction, this ensures $D_a \rho = D_a r = 0$, thereby eliminating the last term in \eqref{DT-H}. The associated sCarrollian stress tensor is thus conserved.\\

Let us conclude this review demonstrating in detail how this procedure works. On the stretched horizon, we define two sCarrollian structures, $\SCarr = \left(\H, h_{ab}, \ell^a, k_b, \rho \right)$ and $\widehat{\SCarr} = \left(\H, h_{ab}, \hat{\ell}^a, \hat{k}_b, \hat{\rho} \right)$, where $\hat{\rho} = r$ is identified with the foliation function. The two sCarrollian structures are related by the rescaling transformations, 
\begin{align}
\ell^a = \Phi \hat{\ell}^a\,, \qquad k_a = \frac{1}{\Phi} \hat{k}_a\,, \qquad \text{and} \qquad \rho = \Phi^2 \hat{\rho} = \Phi^2 r\,.  
\end{align}
From the bulk perspective, these transformations correspond to the rescalings
\begin{align}
n_\mu = \Phi \hat{n}_\mu\,, \qquad \text{and} \qquad k^\mu = \frac{1}{\Phi} \hat{k}^\mu \,.
\end{align}
The rescaling transformations preserve the spacetime metric $g_{\mu\nu}$ and the rigged projector $\Pi_\mu{}^\nu$, and therefore also preserve the soldering $\Pi_a{}^\mu$ and $\Pi_\mu{}^a$, the induced metrics $h_{ab}$ and $q^{ab}$, as well as the sCarrollian connection $D_a$. On the other hand, one has 
\begin{align}
W_a{}^b &= D_a\ell^b = D_a(\Phi \hat\ell^b)=\Phi \hat{W}_a{}^b + D_a \Phi \hat{\ell}^b\\
S_a{}^b &= \Pi_a{}^\alpha \nabla_\alpha k^\beta \Pi_\beta{}^b = \Pi_a{}^\alpha \nabla_\alpha (\Phi^{-1} \hat{k}^\beta) \Pi_\beta{}^b = \Phi^{-1} \hat{S}_a{}^b \, .
\end{align}
The sCarrollian stress tensor can thus be expressed as 
\begin{align}
T_a{}^b = \Phi \hat{T}_a{}^b + \frac{1}{8\pi G} \left( \hat{\ell}^b D_a \Phi  + \hat{\ell}^cD_c \Phi \delta_a^b \right) \, ,
\end{align}
where we used the sCarrollian stress tensor as in \eqref{sBY}, but for the rescaled tensors,
\begin{align}
\hat{T}_a{}^b  = \frac{1}{8\pi G}\left( \hat{W}_a{}^b - \hat{W} \delta_a^b  + 2\hat{\rho}\left(\hat{S}_a{}^b - \hat{S} \delta_a^b\right) \right)\,.
\end{align}
With this, we show that (recall $D_a\hat\ell^b=\hat W_a{}^b$)
\begin{equation}
\begin{aligned}
D_b T_a{}^b & = \Phi D_b \hat{T}_a{}^b + \hat{T}_a{}^b D_b \Phi + \frac{1}{8\pi G} \left( \hat{W} D_a \Phi - \hat{W}_a{}^bD_b \Phi + \hat{\ell}^b [D_a,D_b]\Phi \right) \\
&= \Phi D_b \hat{T}_a{}^b  + \frac{2r}{8\pi G} \left( \hat{S}_a{}^b - \hat{S}\delta_a^b \right)D_b \Phi \, ,
\end{aligned}
\end{equation}
where we used the torsionless condition of the sCarrollian connection, $[D_a,D_b]\Phi = 0$, and we recall that $\hat{\rho} = r$ in this setup. 

We then show how the last term in \eqref{DT-H} transforms,
\begin{equation}
\begin{aligned}
\left( S_a{}^b - S \delta_a^b \right) D_b \rho &=  \Phi^{-1}\left( \hat{S}_a{}^b - \hat{S} \delta_a^b \right) D_b (\Phi^2 r) = 2r \left( \hat{S}_a{}^b - \hat{S} \delta_a^b \right) D_b \Phi \, ,
\end{aligned}
\end{equation}
where we recall $D_a r =0$. We have all the elements to eventually demonstrate that \eqref{DT-H}, after the rescaling, becomes the conservation laws of the improved sCarrollian stress tensor
\begin{align}
\frac{1}{8\pi G}\Pi_a{}^\alpha G_\alpha{}^\beta \hat{n}_\beta = D_b \hat{T}_a{}^b \ \hat{=} \ 0 \, ,
\end{align}
in the absence of matter fields. This demonstrates that one can construct a conserved stress tensor directly from the Einstein equations on purely geometric grounds, in a form that applies uniformly to hypersurfaces of any causal character. In particular, its null limit is well defined and reproduces the null Brown-York stress tensor previously discussed.\\

This concludes our analysis of the sCarrollian structure and its extension to hypersurfaces of arbitrary causal character. Following the logic of the previous section, we began from the rigging construction to induce the sCarrollian structure and its associated connection. We then examined the various extrinsic curvatures and their role in the Gauss and Codazzi-Mainardi equations. Finally, we derived an improved sCarrollian stress tensor whose conservation laws precisely reproduce the Einstein equations on the hypersurface.

\newpage

\section{Historical Context and Applications}\label{7}

At this stage of the presentation, having developed Carrollian geometry intrinsically and shown how it is induced from a bulk Lorentzian spacetime, it is useful to step back and place this construction in its historical and conceptual context. Historically, the geometry of null hypersurfaces was developed in the reverse order, starting from their embedding in spacetime and only later being reinterpreted in Carrollian terms. The purpose of this section is to recall this lineage and to situate the present approach within the broader literature, without interrupting the pedagogical flow of the preceding sections.\\

\subsection{Asymptotically Flat Gravity}

Historically, the Carrollian geometric structure first appeared in the study of isolated systems in General Relativity. In that setting, the central problem was to characterize the gravitational field far away from sources, where spacetime should approach Minkowski space \cite{Bondi, Sachs:1961zz, Penrose:1965am}. The key conceptual step, due to Penrose \cite{Penrose:1962ij} and developed systematically by Tamburino and Winicour \cite{Tamburino:1966zz}, Geroch \cite{Geroch:1977big}, and Ashtekar and Hansen \cite{Ashtekar:1978zz}, is the conformal completion of the physical spacetime $(\M, g_{\mu\nu})$: one introduces an unphysical metric
\begin{equation}
\hat{g}_{\mu\nu}=\Omega^2 g_{\mu\nu},
\end{equation}
on an auxiliary manifold $\hat{\M}$ with boundary $\cal I$, such that, on $\cal I$,
\begin{equation}
\Omega=0,
\qquad
\nabla_\mu\Omega\neq 0,
\qquad
g^{\mu\nu}\nabla_\mu\Omega \nabla_\nu\Omega = 0\,.
\end{equation}
By construction, the boundary $\cal I$ is therefore a null hypersurface. This construction replaces the vague operation of "going to infinity" by local differential geometry at a finite boundary, and provides the natural starting point for the asymptotic analysis of radiation, conserved quantities, and symmetries.

A second fundamental idea in this early literature is to detach the boundary from the bulk spacetime and study it as an abstract manifold endowed with fields. As reviewed by Geroch \cite{Geroch:1977big}, one separates the asymptotic data into universal (or geometrical) fields and physical fields. The universal fields define the asymptotic geometry and the asymptotic symmetry group, while the physical fields encode the asymptotic content of the bulk gravitational content. In particular, at null infinity one already encounters several of the structures that would later become standard in Carrollian geometry: a degenerate metric on a null manifold, a distinguished null direction generating the boundary time, a quotient description in terms of the space of null generators, and an infinite-dimensional enhancement of the translation sector by supertranslations. In modern language, $\cal I$ is a null manifold endowed with a Carrollian structure.

More concretely, the null generator is selected by the conformal factor through
\begin{equation}
\ell^\mu \propto g^{\mu\nu}\nabla_\nu\Omega\big|_{\cal I},
\end{equation}
while the pullback of the unphysical metric to $\cal I$ is degenerate and annihilated by $\ell^a$, where $x^a$ are boundary coordinates. Quotienting $\cal I$ by the integral curves of $\ell^a$ produces the base space ${\cal B}$, whose conformal geometry organizes much of the asymptotic structure. In the Minkowskian case, ${\cal B}\cong {\cal S}^2$, and supertranslations are naturally described by functions on ${\cal B}$. This is precisely the origin of the BMS enhancement \cite{Bondi, Sachs:1961zz}. In this sense, the asymptotic literature had already isolated many of the genuinely non-Riemannian features of null geometry.\footnote{In four spacetime dimensions, the asymptotic analysis at null infinity is also closely related to the Newman-Penrose formalism \cite{Newman:1961qr}, which provides a null-tetrad description of the radiative data. This is consistent with our earlier remark that, when both the normal and the rigging are null and one supplements the construction with a complex dyad on the spatial base, the resulting structure reduces to a Newman-Penrose tetrad (see Footnote \ref{F32}).}\\

At the same time, it is important to appreciate that this asymptotic construction does not lead to the most general intrinsic Carrollian geometry. First, $\cal I$ is a very special null manifold, selected by asymptotic flatness and by an ambient conformal completion. This imposes constraints on the topology and intrinsic geometry of this null manifold. Second, some of the most important intrinsic data of a generic null manifold are no longer free at $\cal I$. A particularly sharp example is the second fundamental tensor (see \eqref{Kmh})
\begin{equation}
\theta_{ab}=\frac12 \mathcal L_{\ell} q_{ab},
\end{equation}
which, in a general Carrollian manifold, is part of the intrinsic kinematics: its trace gives the expansion $\theta$, while its traceless part gives the shear $\sigma_{ab}$.\footnote{This tensor should not be confused with the asymptotic -- radial -- shear $C_{AB}$, as they are completely unrelated.} For a generic null hypersurface, $\theta_{ab}$ is genuine geometric data. At null infinity, however, the conformal completion and the asymptotic Einstein equations place this tensor in a much more rigid setting. Indeed, since the degenerate boundary metric $q_{ab}$ is an initial value datum in gravity, it can be set to any desired value. Furthermore, its shear $\sigma_{ab}$ must vanish by virtue of the leading order asymptotic Einstein equations. Thus, the boundary metric is conformally time-independent on $\cal I$, placing a restricting condition on its intrinsic geometry. Moreover, often the boundary metric is chosen to be altogether time independent, that is, it is Lie-dragged by the generator, $\mathcal L_{\ell} q_{ab}=0$. The nontrivial radiative information characterizing the bulk is not encoded in a free intrinsic $\theta_{ab}$, but in the asymptotic shear $C_{AB}$, featuring in the equivalence class of boundary connections \cite{Ashtekar:1981hw}, as we discussed in section \ref{422}. This is one precise sense in which $\cal I$ should be viewed as a distinguished and constrained realization of Carrollian geometry, rather than as providing its very definition.

From this viewpoint, the role of Henneaux's work \cite{Henneaux:1979vn} becomes conceptually clear. It should not be interpreted as the beginning of null geometry altogether, since many of its basic ingredients had already appeared in the asymptotic analysis of gravity (see references above). What was achieved in \cite{Henneaux:1979vn} is a formulation of zero-signature geometry directly and intrinsically, without starting from null infinity or from any ambient asymptotic construction. In that framework, one begins with a degenerate metric together with the additional data needed to identify the null direction, and one constructs intrinsically the corresponding geometric objects, including the second fundamental tensor, as we reviewed in Section \ref{zerosig}. This shifts the focus from the constrained asymptotic realization to free and intrinsic generic zero-signature spacetimes. Having a more general and intrinsic formulation of Carrollian geometry is crucial for some of its prominent applications (such as generic null hypersurface, condensed matter, and BKL regimes), as we review in section \ref{73} below.\\

This also explains the rationale behind the structure of the present review. We deliberately did not follow the historical chronology in the main development. Instead, we first introduced Carrollian geometry intrinsically, in its most general form, and only afterwards turned to its realization through embedded null hypersurfaces. In our view, this order is both conceptually cleaner and pedagogically preferable: only once the intrinsic data have been identified does it become clear which features are universal to null geometry and which arise from embedding it into an ambient spacetime. The earlier asymptotic works are unquestionably foundational and deserve explicit acknowledgment, but they do not replace the need for a general intrinsic treatment of Carrollian manifolds.\\

\subsection{Complementary Approaches}

The intrinsic viewpoint developed in this review is not the only one available in the literature. 
Carrollian geometry and Carrollian symmetry have also been formulated and used through a number of complementary frameworks, often motivated by different mathematical questions or by different physical applications. We list here several important directions and indicate their main features.\\

A first class of approaches is based on homogeneous spaces \cite{Figueroa-OFarrill:2018ilb}, kinematical groups \cite{Bacry:1968zf}, and algebraic classifications \cite{Figueroa-OFarrill:2017sfs}. In this perspective, Carrollian spacetimes are understood as homogeneous spaces associated with the Carroll group and its extensions, in close analogy with the description of Minkowski, de Sitter, Galilean, or Newton-Cartan geometries through cosets and kinematical Lie algebras. This route is particularly useful for classifying possible Carrollian backgrounds, for understanding contractions of relativistic symmetries, and for organizing invariant tensors and representation-theoretic data. It also makes manifest that Carrollian geometry can be approached directly from symmetry principles, without first introducing a null hypersurface embedded in an ambient spacetime.\\

A second important line of work starts from gauging the Carroll algebra \cite{Hartong:2015xda, Figueroa-OFarrill:2022mcy}. Here the basic idea is to treat the Carroll algebra as a local spacetime symmetry, in the same spirit in which one gauges the Poincar\'e or Bargmann algebra. This naturally leads to a first-order geometric description in terms of Carrollian vielbeins, spin connections, and gauge fields for local boosts and rotations. Such a framework is well suited for coupling matter to Carrollian backgrounds, for formulating Carrollian gravity theories, and for making contact with Cartan-geometric and gauge-theoretic methods. In several instances, this also provides a natural setting in which connections, the analogue of torsion, curvature, and compatibility conditions can be discussed \cite{Figueroa-OFarrill:2020gpr}. Closely related to this is a broader Cartan-geometric or first-order viewpoint on Carrollian structures. In these formulations, the emphasis is not placed on a degenerate metric alone, but rather on a collection of soldering forms, connections, and curvature constraints defining the geometry locally. This language is particularly natural in Chern-Simons-like constructions, in Lie algebra expansions, and in lower-dimensional gravity models. It is also useful whenever one wants to distinguish carefully between kinematical data, gauge redundancies, and dynamical fields. We will review some of its applications in the next subsection.\\

Another important perspective is the one based on Carroll/Newton-Cartan duality and on more general non-Lorentzian geometric correspondences. In such approaches, Carrollian and Newton-Cartan structures are viewed as dual or complementary degenerations of relativistic geometry, often exchanging the role of temporal and spatial directions \cite{Duval:2014uoa}. This viewpoint has proven useful in clarifying the relation between Galilean and Carrollian limits, in understanding ambient constructions, and in formulating generalized non-Lorentzian connections and geodesic structures \cite{Bekaert:2015xua}. It also makes clear that Carrollian geometry belongs to a larger family of degenerate geometric frameworks rather than constituting an isolated case.\\

A further route is provided by Carrollian geometries obtained from a small-$c$ limit of relativistic geometry. In this approach one starts from a Lorentzian metric, its inverse, and sometimes its Levi-Civita connection, and performs a systematic expansion around $c \to 0$ \cite{Dautcourt:1997hb}. This produces Carrollian data order by order, often together with natural candidates for Carrollian connections, stress tensors, and effective dynamics \cite{Hartong:2015xda}. As reviewed below, such expansions can be carried out at the level of matter theories, hydrodynamics \cite{Ciambelli:2018wre}, or gravity itself \cite{Hansen:2021fxi}, and are particularly useful when one wishes to keep track of which Carrollian objects are intrinsic and which descend from a parent relativistic theory. This point of view has also clarified the distinction between strict Carrollian limits and more general Carrollian regimes.\\

One may also mention the splitting formalism and related frame-based approaches, in which the geometry is organized through a distinguished decomposition between longitudinal and transverse directions \cite{Ciambelli:2018wre, Ciambelli:2018ojf}. In the Carrollian context, this is especially useful in hydrodynamics and in the analysis of null or near-null systems, where one wants to separate temporal evolution along the Carrollian direction from spatial structures living on the base manifold \cite{Ciambelli:2018xat, Petkou:2022bmz}. Such formulations often provide an efficient bridge between relativistic fluids, Galilean fluids, and Carrollian fluids, and help organize frame choices, constitutive relations, and covariance properties.\\

For broader overviews of these directions, as well as many additional references, we defer the reader to the recent reviews \cite{Bergshoeff:2022eog, deBoer:2023fnj, Bagchi:2025vri} and the references therein.\\

\subsection{Selected Applications of Carrollian Physics}\label{73}

Beyond its role in the geometry of null hypersurfaces, Carrollian physics has by now appeared in a rather broad range of physical settings. We list here some prominent examples.\\

A first and by now central application is flat-space holography. The observation that conformal Carrollian symmetries at null infinity reproduce the BMS symmetries of asymptotically flat gravity \cite{Duval:2014uva} has led to the proposal that gravity in asymptotically flat spacetime admits a dual description in terms of a Carrollian conformal field theory living on the codimension-one asymptotic null boundary \cite{Bagchi:2016bcd}. In this context, Carrollian geometry provides the natural boundary structure \cite{Ciambelli:2018wre}, Carrollian CFT data organize soft theorems, Ward identities, and celestial limits \cite{Donnay:2022aba}, and the flat limit of AdS/CFT leads to Carrollian amplitudes \cite{Alday:2024yyj}, which can be studied also intrinsically \cite{Bagchi:2022emh}.

Closely related, but conceptually also broader, is the development of Carrollian conformal field theory (see, e.g., \cite{Bagchi:2019xfx, Chen:2021xkw}). This includes the study of primaries, correlation functions, Ward identities, operator product expansions, and modular properties. Such field theories can arise from contractions of relativistic CFTs, from deformations, or from intrinsically defined symmetry principles, and they now constitute a subject in their own right, independently of holography.\\

Carrollian structures also play an important role in the study of null hypersurfaces, black-hole horizons, and cosmological horizons. In this setting, Carrollian geometry describes the intrinsic kinematics of the null surface \cite{Ciambelli:2019lap}, while the Einstein equations projected onto the hypersurface can often be rewritten as Carrollian conservation laws \cite{Donnay:2019jiz}. This has led to fruitful connections with the membrane paradigm, horizon fluids, and near-horizon symmetry analyses \cite{Penna:2018gfx}. In particular, it is this specific application that is the origin of the developments we reviewed in sections \ref{rigg} and \ref{sec:sCarr}, and it is the one most tied to the intrinsic geometric description of Carrollian physics. \\

A further important application is Carrollian hydrodynamics. Carrollian fluids can be obtained either as $c\to 0$ limits of relativistic fluids \cite{Ciambelli:2018xat} or directly from symmetry-based constructions \cite{deBoer:2017ing, Freidel:2022bai}. This has led to Carrollian formulations of constitutive relations, equilibrium partition functions, and conservation laws. It also provides an efficient framework for relating relativistic, Galilean, and Carrollian fluid dynamics within a common language.\\

Carrollian ideas have also been developed in the context of Carroll gravity. Here the goal is not merely to use Carrollian geometry as background data, but to formulate gravitational theories whose natural kinematical language is Carrollian from the outset \cite{Bergshoeff:2017btm, Henneaux:2021yzg}. This includes Carrollian limits of general relativity \cite{Hansen:2021fxi}, Carrollian versions of lower-dimensional gravity theories \cite{Grumiller:2020elf}, and Carroll black holes and geodesics \cite{Ecker:2023uwm}, all models in which Carrollian geometry governs the dynamics of the gravitational sector itself. These constructions further underline that Carrollian geometry can be studied independently of null infinity or of any one specific holographic application.\\

Carrollian ideas have also found interesting realizations in pure field theory and condensed matter systems. This includes scalar models, whose electric and magnetic versions provide simple laboratories for ultralocal and ultra-relativistic dynamics \cite{Baiguera:2022lsw, Rivera-Betancour:2022lkc}, and quantization of Carrollian field theories \cite{Cotler:2024xhb}. In condensed matter, the appearance of Carrollian kinematics in theories of fractons and related subsystem-symmetric systems, where restricted mobility and effective ultra-local behavior naturally evoke Carrollian structures, has been appreciated in \cite{Bidussi:2021nmp}. Another example is the study of fermions and flat bands \cite{Bagchi:2022eui}, where Carrollian symmetry has been proposed as an organizing principle for the effective low-energy dynamics.\\

Another direction concerns cosmology and strong-gravity regimes. Carrollian ideas have been invoked in the study of dark energy and inflationary kinematics, where the effective causal structure can become naturally Carrollian \cite{deBoer:2021jej}. They also arise in the analysis of strong-gravity limits and cosmological singularities, in particular in relation to BKL-like regimes, where ultra-locality becomes a dominant feature of the dynamics \cite{Oling:2024vmq}.\\

Carrollian symmetry also appears in the study of plane gravitational waves and other backgrounds admitting null or ultra-relativistic structures \cite{Duval:2017els}. In such cases, the Carrollian viewpoint provides a useful geometric interpretation of the background symmetries and of the associated reduced dynamics, linking with numerical simulations of gravitational waves profiling.\\

Finally, one should mention string-theoretic appearances of Carrollian symmetry \cite{Cardona:2016ytk}, and strings in near-horizon or ultra-relativistic regimes \cite{Bagchi:2023cfp}. These examples reinforce the broader point that Carrollian symmetry is not tied to a single corner of gravitational physics, but rather appears as a recurring organizing principle whenever null or effectively ultra-local structures dominate the problem.\\

The overview given above is necessarily selective and cannot do full justice to the breadth of the subject. Nevertheless, it should make clear that Carrollian geometry and Carrollian symmetry have by now developed into a broad and structurally rich field of research, lying at the intersection of null geometry, field theory, gravity, hydrodynamics, condensed-matter systems, and holography. Beyond its original appearance as an ultra-relativistic limit or as the intrinsic geometry of null hypersurfaces, Carrollian physics now provides a common conceptual and mathematical language in a remarkably wide range of problems. In this sense, even a concise survey such as the present one already reveals that Carrollian structures are no longer confined to a specialized corner of mathematical physics, but have become a unifying framework connecting several active areas of contemporary research.\\

\newpage

\section{Conclusions}

In this review we developed a unified intrinsic description of null manifolds using the framework of Carrollian geometry. Our goal was to organize this null geometry into a coherent structure, paralleling the familiar pseudo-Riemannian narrative based on metric, connection, and curvature. By collecting and systematizing results that are otherwise dispersed across the literature, we presented a covariant geometric toolkit for the study of Carrollian manifolds. 

The review began by revisiting the algebraic origin of Carrollian physics. We described the Carroll algebra as the ultra-relativistic contraction of the Poincaré group and discussed its conformal extensions. In particular, we showed how the infinite-dimensional conformal Carroll algebra arises naturally and how, for appropriate choices of spatial manifold, it becomes isomorphic to the BMS algebra governing asymptotic symmetries of asymptotically flat spacetimes. This algebraic perspective provides the historical starting point of Carrollian physics and anticipates its deep connection with the geometry of null hypersurfaces. \\

The core of the review developed the intrinsic geometry of Carrollian manifolds. We introduced Carrollian structures as the geometric data characterizing manifolds endowed with a degenerate metric and a distinguished vector field spanning its kernel. This formulation clarifies why null manifolds require additional geometric structure beyond a metric alone. Using a fully covariant approach, we organized the discussion in direct analogy with the standard pseudo-Riemannian narrative. We first defined the general Carrollian structure and its internal symmetries, then introduced the associated geometric quantities such as acceleration, vorticity, and expansion. These elements establish the basic kinematical framework required to describe null manifolds intrinsically.

We then addressed the problem of connections and curvature in Carrollian geometry. A central result is the breakdown of the Levi-Civita theorem: in the presence of a degenerate metric, no unique torsionless metric-compatible connection exists. We therefore constructed the most general intrinsic Carrollian connection and identified a special torsionless connection -- named the standard Carrollian connection. The properties of this connection were analyzed in detail, including its associated covariant derivative and curvature tensors. The resulting Carrollian curvature framework clarifies the relationships among the various curvature constructions appearing in the literature and provides a systematic language for describing geometric dynamics on null manifolds.\\

The second part of the review embedded these intrinsic constructions into an ambient spacetime perspective. Using the rigging technique, we showed how a null hypersurface embedded in a pseudo-Riemannian manifold naturally inherits a Carrollian structure. In this setting the rigged connection induced from the ambient Levi-Civita connection reproduces precisely the standard intrinsic Carrollian connection previously derived. A posteriori, this provides a rationale for singling out the standard Carrollian connection. Within this framework we obtained the Gauss and Codazzi-Mainardi equations for null hypersurfaces and demonstrated how the latter leads to the null Brown-York stress tensor, whose conservation law corresponds to the projection of Einstein’s equations onto the hypersurface.

We then extended the Carrollian description to generic hypersurfaces of arbitrary causal character. This construction shows that Carrollian structures provide a unified geometric language capable of describing spacelike, timelike, and null hypersurfaces within a single framework, with the null case emerging as a smooth limit. In this way the Carrollian formalism offers a universal description of hypersurface geometry and clarifies the intrinsic structure underlying null physics.\\

Our presentation did not follow the historical development of the subject. As discussed in the final part of the review, this choice reflects a conceptual lesson that emerges from the material presented here. Many key insights about null manifolds were historically obtained through their embedding in an ambient spacetime. However, as the preceding sections illustrate, the geometric structure of null manifolds possesses a number of subtleties that become difficult to disentangle when treated exclusively from this perspective. While the ambient construction remains indispensable in important contexts -- most notably in the study of null infinity in asymptotically flat spacetimes -- it also imposes strong constraints on the intrinsic data. The development of an intrinsic Carrollian formulation was therefore essential for isolating the universal geometric structures of null manifolds and making them applicable beyond the specific setting in which they were originally discovered. In this sense, the intrinsic approach does not replace the ambient one, but clarifies and generalizes it, providing a framework that can be deployed in a wide range of physical situations.

\subsubsection*{Acknowledgments} LC warmly thanks Stephan Stieberger for suggesting the preparation of this review for Physics Reports. We are grateful to Laurent Freidel and José Senovilla for insightful discussions and for their contributions at the early stages of this project. We have also greatly benefited from exchanges with Martina Adamo, Glenn Barnich, Cris Corral, Felipe Diaz, Florian Girelli, Temple He, Marc Henneaux, Michael Imseis, Marc Klinger, Luis Lehner, Etera Livine, Sucheta Majumdar, Rob Myers, Dominik Neuenfeld, Sabrina Pasterski, Tasso Petkou, Leo Sanhueza, Simone Speziale, Tom Wetzstein, Pengming Zhang, and Kathryn Zurek. We thank an anonymous reviewer for helpful comments and suggestions that improved the manuscript. Parts of this review were presented by LC in seminars and lectures at the HolographyCL Farewell Meeting in Viña del Mar, the GGI program From Asymptotic Symmetries to Flat Holography: Theoretical Aspects and Observable Consequences in Florence, and the workshop Holography in and beyond the AdS Paradigm in Montevideo. LC sincerely thanks the organizers of these events for their kind invitations and stimulating environments. Research at Perimeter Institute is supported in part by the Government of Canada through the Department of Innovation, Science and Economic Development Canada and by the Province of Ontario through the Ministry of Colleges and Universities. LC is supported by the Simons Collaboration on Celestial Holography.

Finally, we wish to dedicate a special thought to Rob Leigh, whose intellectual depth, generosity, and lasting influence have profoundly shaped this line of research. His clarity of vision and spirit of collaboration continue to inspire this topic and our scientific journey.

%%%%%%%%%%%%%%%%%%%%%%
\bibliographystyle{uiuchept}
\bibliography{CarrollReview.bib}

@article{Ciambelli:2023xqk,
    author = "Ciambelli, Luca",
    title = "{Dynamics of Carrollian scalar fields}",
    eprint = "2311.04113",
    archivePrefix = "arXiv",
    primaryClass = "hep-th",
    doi = "10.1088/1361-6382/ad5bb5",
    journal = "Class. Quant. Grav.",
    volume = "41",
    number = "16",
    pages = "165011",
    year = "2024"
}

@article{deBoer:2023fnj,
    author = "de Boer, Jan and Hartong, Jelle and Obers, Niels A. and Sybesma, Watse and Vandoren, Stefan",
    title = "{Carroll stories}",
    eprint = "2307.06827",
    archivePrefix = "arXiv",
    primaryClass = "hep-th",
    reportNumber = "NORDITA-2023-036",
    doi = "10.1007/JHEP09(2023)148",
    journal = "JHEP",
    volume = "09",
    pages = "148",
    year = "2023"
}

@article{Bagchi:2025vri,
    author = "Bagchi, Arjun and Banerjee, Aritra and Dhivakar, Prateksh and Mondal, Saikat and Shukla, Ashish",
    title = "{The Carrollian Kaleidoscope}",
    eprint = "2506.16164",
    archivePrefix = "arXiv",
    primaryClass = "hep-th",
    month = "6",
    year = "2025"
}

@article{Ciambelli:2023mir,
    author = "Ciambelli, Luca and Freidel, Laurent and Leigh, Robert G.",
    title = "{Null Raychaudhuri: canonical structure and the dressing time}",
    eprint = "2309.03932",
    archivePrefix = "arXiv",
    primaryClass = "hep-th",
    doi = "10.1007/JHEP01(2024)166",
    journal = "JHEP",
    volume = "01",
    pages = "166",
    year = "2024"
}

@article{Ashtekar:1981hw,
    author = "Ashtekar, A.",
    title = "{Radiative Degrees of Freedom of the Gravitational Field in Exact General Relativity}",
    doi = "10.1063/1.525169",
    journal = "J. Math. Phys.",
    volume = "22",
    pages = "2885--2895",
    year = "1981"
}

@article{Ashtekar:2024bpi,
    author = "Ashtekar, Abhay and Speziale, Simone",
    title = "{Null infinity as a weakly isolated horizon}",
    eprint = "2402.17977",
    archivePrefix = "arXiv",
    primaryClass = "hep-th",
    doi = "10.1103/PhysRevD.110.044048",
    journal = "Phys. Rev. D",
    volume = "110",
    number = "4",
    pages = "044048",
    year = "2024"
}

@article{Manzano:2023wxx,
    author = "Manzano, Miguel and Mars, Marc",
    title = "{Abstract formulation of the spacetime matching problem and null thin shells}",
    eprint = "2309.14874",
    archivePrefix = "arXiv",
    primaryClass = "gr-qc",
    doi = "10.1103/PhysRevD.109.044050",
    journal = "Phys. Rev. D",
    volume = "109",
    number = "4",
    pages = "044050",
    year = "2024"
}

@article{Hartong:2015xda,
    author = "Hartong, Jelle",
    title = "{Gauging the Carroll Algebra and Ultra-Relativistic Gravity}",
    eprint = "1505.05011",
    archivePrefix = "arXiv",
    primaryClass = "hep-th",
    doi = "10.1007/JHEP08(2015)069",
    journal = "JHEP",
    volume = "08",
    pages = "069",
    year = "2015"
}

@article{Afshar:2024llh,
    author = "Afshar, Hamid and Bekaert, Xavier and Najafizadeh, Mojtaba",
    title = "{Classification of conformal carroll algebras}",
    eprint = "2409.19953",
    archivePrefix = "arXiv",
    primaryClass = "hep-th",
    reportNumber = "IPM/P-2024/31",
    doi = "10.1007/JHEP12(2024)148",
    journal = "JHEP",
    volume = "12",
    pages = "148",
    year = "2024"
}

@article{Henneaux:1979vn,
    author = "Henneaux, Marc",
    title = "{Geometry of Zero Signature Space-times}",
    reportNumber = "PRINT-79-0606 (PRINCETON)",
    journal = "Bull. Soc. Math. Belg.",
    volume = "31",
    pages = "47--63",
    year = "1979"
}

@article{Donnay:2019jiz,
    author = "Donnay, Laura and Marteau, Charles",
    title = "{Carrollian Physics at the Black Hole Horizon}",
    eprint = "1903.09654",
    archivePrefix = "arXiv",
    primaryClass = "hep-th",
    doi = "10.1088/1361-6382/ab2fd5",
    journal = "Class. Quant. Grav.",
    volume = "36",
    number = "16",
    pages = "165002",
    year = "2019"
}

@article{Ciambelli:2018wre,
    author = "Ciambelli, Luca and Marteau, Charles and Petkou, Anastasios C. and Petropoulos, P. Marios and Siampos, Konstantinos",
    title = "{Flat holography and Carrollian fluids}",
    eprint = "1802.06809",
    archivePrefix = "arXiv",
    primaryClass = "hep-th",
    reportNumber = "CPHT-RR049.082017, CERN-TH-2017-229",
    doi = "10.1007/JHEP07(2018)165",
    journal = "JHEP",
    volume = "07",
    pages = "165",
    year = "2018"
}

@article{Gourgoulhon:2005ng,
    author = "Gourgoulhon, Eric and Jaramillo, Jose Luis",
    title = "{A 3+1 perspective on null hypersurfaces and isolated horizons}",
    eprint = "gr-qc/0503113",
    archivePrefix = "arXiv",
    doi = "10.1016/j.physrep.2005.10.005",
    journal = "Phys. Rept.",
    volume = "423",
    pages = "159--294",
    year = "2006"
}

@article{Duval:2014uoa,
    author = "Duval, C. and Gibbons, G. W. and Horvathy, P. A. and Zhang, P. M.",
    title = "{Carroll versus Newton and Galilei: two dual non-Einsteinian concepts of time}",
    eprint = "1402.0657",
    archivePrefix = "arXiv",
    primaryClass = "gr-qc",
    doi = "10.1088/0264-9381/31/8/085016",
    journal = "Class. Quant. Grav.",
    volume = "31",
    pages = "085016",
    year = "2014"
}

@article{Bekaert:2015xua,
    author = "Bekaert, Xavier and Morand, Kevin",
    title = "{Connections and dynamical trajectories in generalised Newton-Cartan gravity II. An ambient perspective}",
    eprint = "1505.03739",
    archivePrefix = "arXiv",
    primaryClass = "hep-th",
    doi = "10.1063/1.5030328",
    journal = "J. Math. Phys.",
    volume = "59",
    number = "7",
    pages = "072503",
    year = "2018"
}

@article{Chandrasekaran:2021hxc,
    author = "Chandrasekaran, Venkatesa and Flanagan, Eanna E. and Shehzad, Ibrahim and Speranza, Antony J.",
    title = "{Brown-York charges at null boundaries}",
    eprint = "2109.11567",
    archivePrefix = "arXiv",
    primaryClass = "hep-th",
    doi = "10.1007/JHEP01(2022)029",
    journal = "JHEP",
    volume = "01",
    pages = "029",
    year = "2022"
}

@article{Ciambelli:2018xat,
    author = "Ciambelli, Luca and Marteau, Charles and Petkou, Anastasios C. and Petropoulos, P. Marios and Siampos, Konstantinos",
    title = "{Covariant Galilean versus Carrollian hydrodynamics from relativistic fluids}",
    eprint = "1802.05286",
    archivePrefix = "arXiv",
    primaryClass = "hep-th",
    reportNumber = "CPHT-RR048.082017, CERN-TH-2017-228",
    doi = "10.1088/1361-6382/aacf1a",
    journal = "Class. Quant. Grav.",
    volume = "35",
    number = "16",
    pages = "165001",
    year = "2018"
}

@article{Ciambelli:2019lap,
    author = "Ciambelli, Luca and Leigh, Robert G. and Marteau, Charles and Petropoulos, P. Marios",
    title = "{Carroll Structures, Null Geometry and Conformal Isometries}",
    eprint = "1905.02221",
    archivePrefix = "arXiv",
    primaryClass = "hep-th",
    reportNumber = "CPHT-RR025.052019, CPHT-RR010.022019",
    doi = "10.1103/PhysRevD.100.046010",
    journal = "Phys. Rev. D",
    volume = "100",
    number = "4",
    pages = "046010",
    year = "2019"
}

@article{Mars:1993mj,
    author = "Mars, Marc and Senovilla, Jose M. M.",
    title = "{Geometry of general hypersurfaces in space-time: Junction conditions}",
    eprint = "gr-qc/0201054",
    archivePrefix = "arXiv",
    doi = "10.1088/0264-9381/10/9/026",
    journal = "Class. Quant. Grav.",
    volume = "10",
    pages = "1865--1897",
    year = "1993"
}

@article{Duval:2014uva,
    author = "Duval, C. and Gibbons, G. W. and Horvathy, P. A.",
    title = "{Conformal Carroll groups and BMS symmetry}",
    eprint = "1402.5894",
    archivePrefix = "arXiv",
    primaryClass = "gr-qc",
    doi = "10.1088/0264-9381/31/9/092001",
    journal = "Class. Quant. Grav.",
    volume = "31",
    pages = "092001",
    year = "2014"
}

@article{Freidel:2022vjq,
    author = "Freidel, Laurent and Jai-akson, Puttarak",
    title = "{Carrollian hydrodynamics and symplectic structure on stretched horizons}",
    eprint = "2211.06415",
    archivePrefix = "arXiv",
    primaryClass = "gr-qc",
    reportNumber = "RIKEN-iTHEMS-Report-22",
    doi = "10.1007/JHEP05(2024)135",
    journal = "JHEP",
    volume = "05",
    pages = "135",
    year = "2024"
}

@article{Bagchi:2019xfx,
    author = "Bagchi, Arjun and Mehra, Aditya and Nandi, Poulami",
    title = "{Field Theories with Conformal Carrollian Symmetry}",
    eprint = "1901.10147",
    archivePrefix = "arXiv",
    primaryClass = "hep-th",
    doi = "10.1007/JHEP05(2019)108",
    journal = "JHEP",
    volume = "05",
    pages = "108",
    year = "2019"
}

@article{Bergshoeff:2017btm,
    author = "Bergshoeff, Eric and Gomis, Joaquim and Rollier, Blaise and Rosseel, Jan and ter Veldhuis, Tonnis",
    title = "{Carroll versus Galilei Gravity}",
    eprint = "1701.06156",
    archivePrefix = "arXiv",
    primaryClass = "hep-th",
    doi = "10.1007/JHEP03(2017)165",
    journal = "JHEP",
    volume = "03",
    pages = "165",
    year = "2017"
}

@article{Alday:2024yyj,
    author = "Alday, Luis F. and Nocchi, Maria and Ruzziconi, Romain and Yelleshpur Srikant, Akshay",
    title = "{Carrollian amplitudes from holographic correlators}",
    eprint = "2406.19343",
    archivePrefix = "arXiv",
    primaryClass = "hep-th",
    doi = "10.1007/JHEP03(2025)158",
    journal = "JHEP",
    volume = "03",
    pages = "158",
    year = "2025"
}

@article{Figueroa-OFarrill:2018ilb,
    author = "Figueroa-O'Farrill, Jos{\'e} and Prohazka, Stefan",
    title = "{Spatially isotropic homogeneous spacetimes}",
    eprint = "1809.01224",
    archivePrefix = "arXiv",
    primaryClass = "hep-th",
    reportNumber = "EMPG-18-01",
    doi = "10.1007/JHEP01(2019)229",
    journal = "JHEP",
    volume = "01",
    pages = "229",
    year = "2019"
}

@article{Rivera-Betancour:2022lkc,
    author = "Rivera-Betancour, David and Vilatte, Matthieu",
    title = "{Revisiting the Carrollian scalar field}",
    eprint = "2207.01647",
    archivePrefix = "arXiv",
    primaryClass = "hep-th",
    reportNumber = "CPHT-RR022.042022",
    doi = "10.1103/PhysRevD.106.085004",
    journal = "Phys. Rev. D",
    volume = "106",
    number = "8",
    pages = "085004",
    year = "2022"
}

@article{Newman:1961qr,
    author = "Newman, Ezra and Penrose, Roger",
    title = "{An Approach to gravitational radiation by a method of spin coefficients}",
    doi = "10.1063/1.1724257",
    journal = "J. Math. Phys.",
    volume = "3",
    pages = "566--578",
    year = "1962"
}

@article{Ashtekar:1978zz,
    author = "Ashtekar, A. and Hansen, R. O.",
    title = "{A unified treatment of null and spatial infinity in general relativity. I - Universal structure, asymptotic symmetries, and conserved quantities at spatial infinity}",
    doi = "10.1063/1.523863",
    journal = "J. Math. Phys.",
    volume = "19",
    pages = "1542--1566",
    year = "1978"
}

@inproceedings{Geroch:1977big,
    author = "Geroch, Robert",
    title = "{Asymptotic Structure of Space-Time}",
    booktitle = "{Symposium on Asymptotic Structure of Space-Time}",
    doi = "10.1007/978-1-4684-2343-3_1",
    year = "1977"
}

@article{Tamburino:1966zz,
    author = "Tamburino, Louis A. and Winicour, Jeffrey H.",
    title = "{Gravitational Fields in Finite and Conformal Bondi Frames}",
    doi = "10.1103/PhysRev.150.1039",
    journal = "Phys. Rev.",
    volume = "150",
    pages = "1039--1053",
    year = "1966"
}

@article{Penrose:1965am,
    author = "Penrose, R.",
    title = "{Zero rest mass fields including gravitation: Asymptotic behavior}",
    doi = "10.1098/rspa.1965.0058",
    journal = "Proc. Roy. Soc. Lond. A",
    volume = "284",
    pages = "159",
    year = "1965"
}

@article{Penrose:1962ij,
    author = "Penrose, Roger",
    title = "{Asymptotic properties of fields and space-times}",
    doi = "10.1103/PhysRevLett.10.66",
    journal = "Phys. Rev. Lett.",
    volume = "10",
    pages = "66--68",
    year = "1963"
}

@article{Cotler:2024xhb,
    author = "Cotler, Jordan and Jensen, Kristan and Prohazka, Stefan and Raz, Amir and Riegler, Max and Salzer, Jakob",
    title = "{Quantizing Carrollian field theories}",
    eprint = "2407.11971",
    archivePrefix = "arXiv",
    primaryClass = "hep-th",
    doi = "10.1007/JHEP10(2024)049",
    journal = "JHEP",
    volume = "10",
    pages = "049",
    year = "2024"
}

@article{Baiguera:2022lsw,
    author = "Baiguera, Stefano and Oling, Gerben and Sybesma, Watse and S{\o}gaard, Benjamin T.",
    title = "{Conformal Carroll scalars with boosts}",
    eprint = "2207.03468",
    archivePrefix = "arXiv",
    primaryClass = "hep-th",
    reportNumber = "NORDITA 2022-047",
    doi = "10.21468/SciPostPhys.14.4.086",
    journal = "SciPost Phys.",
    volume = "14",
    number = "4",
    pages = "086",
    year = "2023"
}

@article{Dautcourt:1997hb,
    author = "Dautcourt, G.",
    editor = "Demianski, M. and Kopczynski, W.",
    title = "{On the ultrarelativistic limit of general relativity}",
    eprint = "gr-qc/9801093",
    archivePrefix = "arXiv",
    journal = "Acta Phys. Polon. B",
    volume = "29",
    pages = "1047--1055",
    year = "1998"
}

@article{Figueroa-OFarrill:2020gpr,
    author = "Figueroa-O'Farrill, Jos{\'e}",
    title = "{On the intrinsic torsion of spacetime structures}",
    eprint = "2009.01948",
    archivePrefix = "arXiv",
    primaryClass = "hep-th",
    reportNumber = "EMPG-20-14",
    month = "9",
    year = "2020"
}

@article{Figueroa-OFarrill:2022mcy,
    author = "Figueroa-O'Farrill, Jos{\'e} and Have, Emil and Prohazka, Stefan and Salzer, Jakob",
    title = "{The gauging procedure and carrollian gravity}",
    eprint = "2206.14178",
    archivePrefix = "arXiv",
    primaryClass = "hep-th",
    reportNumber = "EMPG-22-10",
    doi = "10.1007/JHEP09(2022)243",
    journal = "JHEP",
    volume = "09",
    pages = "243",
    year = "2022"
}

@article{Bagchi:2022emh,
    author = "Bagchi, Arjun and Banerjee, Shamik and Basu, Rudranil and Dutta, Sudipta",
    title = "{Scattering Amplitudes: Celestial and Carrollian}",
    eprint = "2202.08438",
    archivePrefix = "arXiv",
    primaryClass = "hep-th",
    doi = "10.1103/PhysRevLett.128.241601",
    journal = "Phys. Rev. Lett.",
    volume = "128",
    number = "24",
    pages = "241601",
    year = "2022"
}

@article{Bidussi:2021nmp,
    author = "Bidussi, Leo and Hartong, Jelle and Have, Emil and Musaeus, J{\o}rgen and Prohazka, Stefan",
    title = "{Fractons, dipole symmetries and curved spacetime}",
    eprint = "2111.03668",
    archivePrefix = "arXiv",
    primaryClass = "hep-th",
    doi = "10.21468/SciPostPhys.12.6.205",
    journal = "SciPost Phys.",
    volume = "12",
    number = "6",
    pages = "205",
    year = "2022"
}

@article{Figueroa-OFarrill:2017sfs,
    author = "Figueroa-O'Farrill, Jos{\'e}",
    title = "{Classification of kinematical Lie algebras}",
    eprint = "1711.05676",
    archivePrefix = "arXiv",
    primaryClass = "hep-th",
    reportNumber = "EMPG-17-19",
    month = "11",
    year = "2017"
}

@article{Chen:2021xkw,
    author = "Chen, Bin and Liu, Reiko and Zheng, Yu-fan",
    title = "{On higher-dimensional Carrollian and Galilean conformal field theories}",
    eprint = "2112.10514",
    archivePrefix = "arXiv",
    primaryClass = "hep-th",
    doi = "10.21468/SciPostPhys.14.5.088",
    journal = "SciPost Phys.",
    volume = "14",
    number = "5",
    pages = "088",
    year = "2023"
}

@article{deBoer:2017ing,
    author = "de Boer, Jan and Hartong, Jelle and Obers, Niels A. and Sybesma, Watse and Vandoren, Stefan",
    title = "{Perfect Fluids}",
    eprint = "1710.04708",
    archivePrefix = "arXiv",
    primaryClass = "hep-th",
    doi = "10.21468/SciPostPhys.5.1.003",
    journal = "SciPost Phys.",
    volume = "5",
    number = "1",
    pages = "003",
    year = "2018"
}

@article{Ecker:2023uwm,
    author = "Ecker, Florian and Grumiller, Daniel and Hartong, Jelle and P{\'e}rez, Alfredo and Prohazka, Stefan and Troncoso, Ricardo",
    title = "{Carroll black holes}",
    eprint = "2308.10947",
    archivePrefix = "arXiv",
    primaryClass = "hep-th",
    reportNumber = "TUW-23-03",
    doi = "10.21468/SciPostPhys.15.6.245",
    journal = "SciPost Phys.",
    volume = "15",
    number = "6",
    pages = "245",
    year = "2023"
}

@article{Henneaux:2021yzg,
    author = "Henneaux, Marc and Salgado-Rebolledo, Patricio",
    title = "{Carroll contractions of Lorentz-invariant theories}",
    eprint = "2109.06708",
    archivePrefix = "arXiv",
    primaryClass = "hep-th",
    doi = "10.1007/JHEP11(2021)180",
    journal = "JHEP",
    volume = "11",
    pages = "180",
    year = "2021"
}

@article{Grumiller:2020elf,
    author = "Grumiller, Daniel and Hartong, Jelle and Prohazka, Stefan and Salzer, Jakob",
    title = "{Limits of JT gravity}",
    eprint = "2011.13870",
    archivePrefix = "arXiv",
    primaryClass = "hep-th",
    reportNumber = "TUW--20--05",
    doi = "10.1007/JHEP02(2021)134",
    journal = "JHEP",
    volume = "02",
    pages = "134",
    year = "2021"
}

@article{Hansen:2021fxi,
    author = "Hansen, Dennis and Obers, Niels A. and Oling, Gerben and S{\o}gaard, Benjamin T.",
    title = "{Carroll Expansion of General Relativity}",
    eprint = "2112.12684",
    archivePrefix = "arXiv",
    primaryClass = "hep-th",
    reportNumber = "NORDITA 2021-156",
    doi = "10.21468/SciPostPhys.13.3.055",
    journal = "SciPost Phys.",
    volume = "13",
    number = "3",
    pages = "055",
    year = "2022"
}

@article{Bagchi:2022eui,
    author = "Bagchi, Arjun and Banerjee, Aritra and Basu, Rudranil and Islam, Minhajul and Mondal, Saikat",
    title = "{Magic fermions: Carroll and flat bands}",
    eprint = "2211.11640",
    archivePrefix = "arXiv",
    primaryClass = "hep-th",
    doi = "10.1007/JHEP03(2023)227",
    journal = "JHEP",
    volume = "03",
    pages = "227",
    year = "2023"
}

@article{Bergshoeff:2022eog,
    author = "Bergshoeff, Eric and Figueroa-O'Farrill, Jos{\'e} and Gomis, Joaquim",
    title = "{A non-lorentzian primer}",
    eprint = "2206.12177",
    archivePrefix = "arXiv",
    primaryClass = "hep-th",
    reportNumber = "EMPG-22-08",
    doi = "10.21468/SciPostPhysLectNotes.69",
    journal = "SciPost Phys. Lect. Notes",
    volume = "69",
    pages = "1",
    year = "2023"
}

@article{Bagchi:2023cfp,
    author = "Bagchi, Arjun and Banerjee, Aritra and Hartong, Jelle and Have, Emil and Kolekar, Kedar S. and Mandlik, Mangesh",
    title = "{Strings near black holes are Carrollian}",
    eprint = "2312.14240",
    archivePrefix = "arXiv",
    primaryClass = "hep-th",
    doi = "10.1103/PhysRevD.110.086009",
    journal = "Phys. Rev. D",
    volume = "110",
    number = "8",
    pages = "086009",
    year = "2024"
}

@article{Cardona:2016ytk,
    author = "Cardona, Biel and Gomis, Joaquim and Pons, Josep M",
    title = "{Dynamics of Carroll Strings}",
    eprint = "1605.05483",
    archivePrefix = "arXiv",
    primaryClass = "hep-th",
    reportNumber = "ICCUB-16-018",
    doi = "10.1007/JHEP07(2016)050",
    journal = "JHEP",
    volume = "07",
    pages = "050",
    year = "2016"
}

@article{Duval:2017els,
    author = "Duval, C. and Gibbons, G. W. and Horvathy, P. A. and Zhang, P. -M.",
    title = "{Carroll symmetry of plane gravitational waves}",
    eprint = "1702.08284",
    archivePrefix = "arXiv",
    primaryClass = "gr-qc",
    doi = "10.1088/1361-6382/aa7f62",
    journal = "Class. Quant. Grav.",
    volume = "34",
    number = "17",
    pages = "175003",
    year = "2017"
}

@article{Oling:2024vmq,
    author = "Oling, Gerben and Pedraza, Juan F.",
    title = "{Mixmasters in Wonderland: Chaotic dynamics from Carroll limits of gravity}",
    eprint = "2409.05836",
    archivePrefix = "arXiv",
    primaryClass = "hep-th",
    reportNumber = "IFT-UAM/CSIC-24-127",
    doi = "10.21468/SciPostPhysCore.8.1.025",
    journal = "SciPost Phys. Core",
    volume = "8",
    pages = "025",
    year = "2025"
}

@article{deBoer:2021jej,
    author = "de Boer, Jan and Hartong, Jelle and Obers, Niels A. and Sybesma, Watse and Vandoren, Stefan",
    title = "{Carroll Symmetry, Dark Energy and Inflation}",
    eprint = "2110.02319",
    archivePrefix = "arXiv",
    primaryClass = "hep-th",
    reportNumber = "NORDITA 2021-086",
    doi = "10.3389/fphy.2022.810405",
    journal = "Front. in Phys.",
    volume = "10",
    pages = "810405",
    year = "2022"
}

@article{Freidel:2022bai,
    author = "Freidel, Laurent and Jai-akson, Puttarak",
    title = "{Carrollian hydrodynamics from symmetries}",
    eprint = "2209.03328",
    archivePrefix = "arXiv",
    primaryClass = "hep-th",
    doi = "10.1088/1361-6382/acb194",
    journal = "Class. Quant. Grav.",
    volume = "40",
    number = "5",
    pages = "055009",
    year = "2023"
}

@article{Bondi,
author = {Bondi, Hermann  and Van der Burg, M. G. J.  and Metzner, A. W. K. },
title = {Gravitational waves in general relativity, VII. Waves from axi-symmetric isolated system},
journal = {Proceedings of the Royal Society of London. Series A. Mathematical and Physical Sciences},
volume = {269},
number = {1336},
pages = {21-52},
year = {1962},
doi = {10.1098/rspa.1962.0161},
}

@article{Petkou:2022bmz,
    author = "Petkou, Anastasios C. and Petropoulos, P. Marios and Betancour, David Rivera and Siampos, Konstantinos",
    title = "{Relativistic fluids, hydrodynamic frames and their Galilean versus Carrollian avatars}",
    eprint = "2205.09142",
    archivePrefix = "arXiv",
    primaryClass = "hep-th",
    reportNumber = "CPHT-RR021.042022",
    doi = "10.1007/JHEP09(2022)162",
    journal = "JHEP",
    volume = "09",
    pages = "162",
    year = "2022"
}

@article{Penna:2018gfx,
    author = "Penna, Robert F.",
    title = "{Near-horizon Carroll symmetry and black hole Love numbers}",
    eprint = "1812.05643",
    archivePrefix = "arXiv",
    primaryClass = "hep-th",
    month = "12",
    year = "2018"
}

@article{Sachs:1961zz,
    author = "Sachs, R. K.",
    title = "{Gravitational waves in general relativity. 6. The outgoing radiation condition}",
    doi = "10.1098/rspa.1961.0202",
    journal = "Proc. Roy. Soc. Lond. A",
    volume = "264",
    pages = "309--338",
    year = "1961"
}

@article{Ciambelli:2018ojf,
	archiveprefix = {arXiv},
	author = {Ciambelli, Luca and Marteau, Charles},
	doi = {10.1088/1361-6382/ab0d37},
	eprint = {1810.11037},
	journal = {Class. Quant. Grav.},
	number = {8},
	pages = {085004},
	primaryclass = {hep-th},
	reportnumber = {CPHT-RR101.102018},
	title = {{Carrollian conservation laws and Ricci-flat gravity}},
	volume = {36},
	year = {2019}}

@article{Bacry:1968zf,
    author = "Bacry, H. and Levy-Leblond, J.",
    title = "{Possible kinematics}",
    doi = "10.1063/1.1664490",
    journal = "J. Math. Phys.",
    volume = "9",
    pages = "1605--1614",
    year = "1968"
}

@article{Bagchi:2016bcd,
    author = "Bagchi, Arjun and Basu, Rudranil and Kakkar, Ashish and Mehra, Aditya",
    title = "{Flat Holography: Aspects of the dual field theory}",
    eprint = "1609.06203",
    archivePrefix = "arXiv",
    primaryClass = "hep-th",
    doi = "10.1007/JHEP12(2016)147",
    journal = "JHEP",
    volume = "12",
    pages = "147",
    year = "2016"
}

@article{SenGupta:1966qer,
    author = "Sen Gupta, N. D.",
    title = "{On an analogue of the Galilei group}",
    doi = "10.1007/BF02740871",
    journal = "Nuovo Cim. A",
    volume = "44",
    number = "2",
    pages = "512--517",
    year = "1966"
}

@article{LevyLeblond1965,
	author = {L\'evy-Leblond, J.-M.},
	journal = {A. Inst. Henri Poincar\'e III 1},
	title = {{Une nouvelle limite non-relativiste du groupe de Poincar\'e}},
	year = {1965}}

@article{Donnay:2022aba,
	archiveprefix = {arXiv},
	author = {Donnay, Laura and Fiorucci, Adrien and Herfray, Yannick and Ruzziconi, Romain},
	eprint = {2202.04702},
	month = {2},
	primaryclass = {hep-th},
	title = {{A Carrollian Perspective on Celestial Holography}},
	year = {2022}}

\end{document}